\documentclass[a4paper,fleqn]{cas-sc}

\usepackage[style=authoryear,backend=biber,natbib=true]{biblatex}

\newcommand*\dif{\mathop{}\!\mathrm{d}}

\usepackage[dvipsnames]{xcolor}

\usepackage{graphicx}

\usepackage{import}
\usepackage{needspace}

\usepackage{amsthm}
\theoremstyle{definition}
\newtheorem{definition}{Definition}[section]

\newtheorem{theorem}{Theorem}

\newcommand{\mat}[1]{{%
  \mathbf{#1}
}}

\definecolor{blue1}{HTML}{30336b}
\definecolor{green1}{HTML}{00b894}
\definecolor{red1}{HTML}{d63031}
\definecolor{yellow1}{HTML}{7F6000}
\definecolor{pink1}{HTML}{7030a0}
\definecolor{light1}{HTML}{f1f2f6}

\DeclareMathOperator*{\argmin}{argmin}

\usepackage[most]{tcolorbox}

\tcbset{textmarker/.style={%
        enhanced,
        parbox=false,boxrule=0mm,boxsep=0mm,arc=0mm,
        outer arc=0mm,left=6mm,right=3mm,top=7pt,bottom=7pt,
        toptitle=1mm,bottomtitle=1mm,oversize}}

\newtcolorbox{hintBox}{textmarker,
    borderline west={6pt}{0pt}{yellow},
    colback=yellow!10!white}
\newtcolorbox{importantBox}{textmarker,
    borderline west={6pt}{0pt}{red},
    colback=red!10!white}
\newtcolorbox{noteBox}{textmarker,
    borderline west={6pt}{0pt}{green},
    colback=green!10!white}
\newtcolorbox{blueBox}{textmarker,
    borderline west={6pt}{0pt}{blue1},
    colback=blue1!10!white}
\newtcolorbox{pinkBox}{textmarker,
    borderline west={6pt}{0pt}{pink1},
    colback=pink1!10!white}
\newtcolorbox{lightBox}{textmarker,
    borderline west={6pt}{0pt}{light1},
    colback=light1!10!white}

\newtcolorbox{greyBox}{textmarker,
    borderline west={6pt}{0pt}{blue1},
    colback=gray!10!white}

\newcommand{\greenbox}[1]{\begin{noteBox} #1 \end{noteBox}}

\newcommand{\yellowbox}[1]{\begin{hintBox} #1 \end{hintBox}}

\newcommand{\bluebox}[1]{\begin{blueBox} #1 \end{blueBox}}

\usepackage{parskip}
\usepackage{algorithm}
\usepackage{algpseudocode}

\def\tsc#1{\csdef{#1}{\textsc{\lowercase{#1}}\xspace}}
\tsc{WGM}
\tsc{QE}
\tsc{EP}
\tsc{PMS}
\tsc{BEC}
\tsc{DE}

\begin{document}
\let\WriteBookmarks\relax
\renewcommand{\textfraction}{.05}
\renewcommand{\floatpagefraction}{.8}
\shorttitle{Astronomix}
\shortauthors{Storcks et~al.}

\title [mode = title]{Per Astronomix ad Astra: High-Order Differentiable (Magneto)hydrodynamics with Energy-Conserving Self-Gravity}

\author[1]{Leonard Storcks}[orcid=0000-0002-3542-7632]
\cormark[1]
\cortext[cor1]{Corresponding author}
\ead{leonard.storcks@pm.me}
\credit{method development, implementation, testing, writing}

\author[2]{Nils Thuerey}
\credit{supervision}

\author[1]{Tobias Buck}
\credit{supervision}

\affiliation[1]{organization={Heidelberg University},
                addressline={Im Neuenheimer Feld 225}, 
                city={Heidelberg},
                country={Germany}}

\affiliation[2]{organization={Technical University of Munich},
                addressline={Boltzmannstraße 3}, 
                city={Garching bei München},
                country={Germany}}

\begin{abstract}
We present \texttt{astronomix}, a performant differentiable (magneto)hydrodynamics
simulator written in \texttt{Python}/\texttt{JAX}. We demonstrate how automatic differentiation, 
validated against hand-derived analytical functional derivatives and finite differences, enables 
inverse modeling over millions of parameters and allows for sensitivity and stability analysis as 
well as correct eigenmode initialization. The differentiability of \texttt{astronomix} furthermore 
enables training machine-learning models inside the simulator. On a single GPU at a given
resolution, \texttt{astronomix} has runtimes of the same order of magnitude
as the GPU-optimized code AthenaPK but reaches far lower errors on smooth problems due to its higher order.
\texttt{astronomix} scales to multiple GPUs ($\sim 6.5$
strong scaling speedup on $8$ GPUs) and multiple nodes ($\sim 76\%$ weak
scaling efficiency on $16$ GPUs over $4$ nodes). We also present a novel
fourth-order self-gravity scheme which
complements the fifth-order finite difference constrained transport
magnetohydrodynamics scheme implemented in \texttt{astronomix}. To maximize 
performance, we created an agentic skill that generates and 
validates custom Pallas GPU kernels from our \texttt{JAX} reference code and test 
suite. The simulator is available 
at \url{https://github.com/leo1200/astronomix}.
\end{abstract}

\begin{graphicalabstract}
 \includegraphics[width=1.0\textwidth]{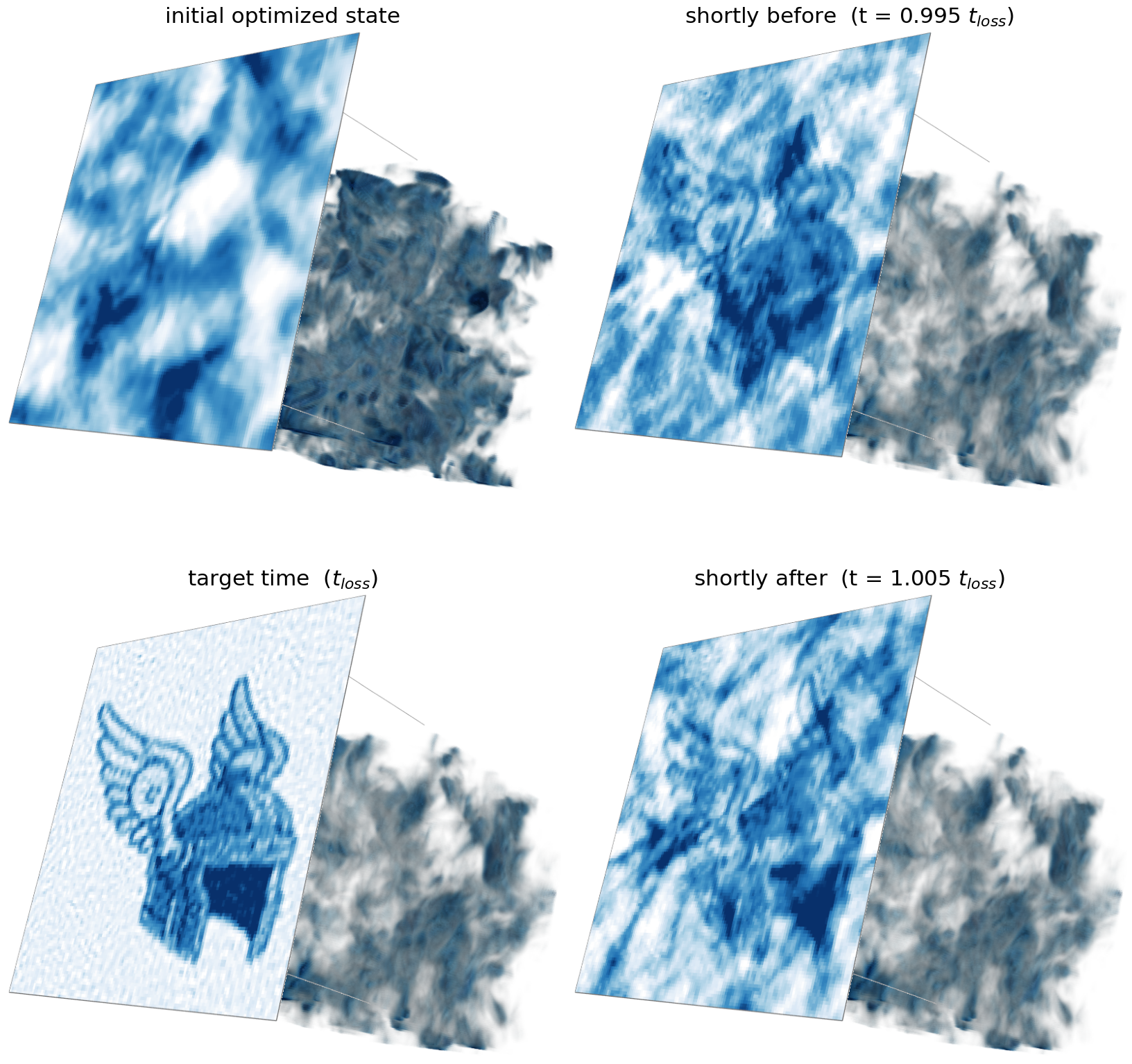}
\end{graphicalabstract}

\begin{highlights}
\item fifth-order finite difference magnetohydrodynamics with an energy-conserving self-gravity scheme and improved stability mechanisms
\item lower single-GPU error per runtime than AthenaPK on smooth problems (based on the higher convergence order), with efficient scaling to multiple GPUs and multiple nodes
\item fully differentiable: opening the door for PDE-constrained optimization, inference, sensitivity analysis and solver-in-the-loop applications in astrophysical fluid dynamics
\end{highlights}


\maketitle

\section{Introduction}

Starting in the 1970s \citep{lucy77}, 
fluid simulators became essential research tools
in astrophysics \citep{leer18}. Conversely, astrophysicists like Bram van Leer had a lasting impact on computational fluid
dynamics as a whole \citep{hirsch97}.

Since then, astrophysical fluid codes such as \texttt{AREPO}, \texttt{ATHENA} and \texttt{RAMSES} \citep{Springel10,Stone08,Teyssier2002}
have evolved into mature research tools. In recent years the
High-Performance Computing (HPC) landscape has shifted towards GPUs, hand-in-hand with advancements in machine 
learning \citep{jouppi17}. This shift has prompted adaptations
of classical low-level simulation codes based on vendor-specific GPU-programming language extensions like CUDA \citep{cuda}, abstraction layers like \texttt{SYCL} \citep{sycl} or a framework like \texttt{kokkos} \citep{kokkos} (see \cite{zier24, grete20, athenaK}). These adaptations have further complicated already highly complex and difficult to maintain codebases. Method development, in contrast, would benefit from high-level code allowing for implementations close to the
mathematical formulation. Additionally, in large simulations one would often like to perform data analysis on-the-fly directly on the GPU. However, integrating data analysis scripts, typically written in high-level Python, into such classical simulators tends to be difficult.

In addition, there is a rich field of applications which require derivatives through simulations. One of the most classical setups is the optimization of airfoils through aerodynamic simulations \citep{Jameson1988}, an example of PDE-constrained optimization \citep{Herzog2010}. Whether we want to do gradient-based inverse modeling on parameters of the simulation \citep{hmc, holzschuh}, perform sensitivity analysis or more recently integrate machine learning models into the simulator \citep{um21, wei25}, we need the simulator to be differentiable. While finite differencing is always available, gradients are inexact and high-dimensional gradients are too expensive \citep{baydin18}. Classically, derivatives through simulations are therefore taken by solving an adjoint system backwards in time \citep{wilcox13}.
However, deriving a properly discretized adjoint system and corresponding numerical solver is tedious (see \cite{wilcox13}), practically infeasible for an astrophysical simulator with rich physics ranging from magnetic fields through self-gravity to cosmic rays. Therefore, none of the classical astrophysical fluid codes are differentiable and, in spite of the rich observational data, PDE-constrained optimization in astrophysics has been limited to a few specialized examples \citep{hanasoge11}.

Building on the same numerical frameworks that are used for 
machine learning such as PyTorch \citep{pytorch} and 
JAX \citep{jax}, a new field of automatically differentiable 
numerical solvers has emerged
\citep{diffrax,jaxfluids2, jax_fem, jax_sph, diffSPH}. 
But while there are differentiable fluid codes that scale to 
multiple GPUs, like \texttt{jaxfluids} \citep{jaxfluids2}, 
\texttt{XLB} \citep{ataei2024xlb} or \texttt{janc} \citep{janc},
they lack important physics, such as
magnetic fields interacting with the astrophysical plasma and
the gravitational self-interaction of the fluid,
necessary for much of astrophysical research, from
studying wind-blown bubbles \citep{lachlan24} to star
formation \citep{grudi21} and magnetohydrodynamical turbulence \citep{beattie2024}. Additionally, automatically differentiable 
fluid simulators are still a young field, with canonical gradient 
verification tests yet to be developed, theoretical insights on 
the usefulness and conditioning of the gradients to be made or 
transferred from adjacent fields (e.g. for differentiation 
through chaotic simulations 
\citep{least_squares_shadowing, Pires1996}) and further 
applications to be discovered.

From the above points we conclude that there is immense potential in developing a modern differentiable astrophysical fluid simulator. Such a simulator could open the door for bringing together the rich astrophysical data, advanced simulation techniques and PDE-constrained optimization literature for inverse modeling. It could allow for training subgrid machine learning models inside the simulator to bridge the vast astronomical scales. It could empower a broad community of astrophysicists to write modules and on-the-fly analysis tools in high-level Python.

We therefore present \texttt{astronomix}, a differentiable astrophysical fluid simulator written in \texttt{JAX}, featuring

\begin{itemize}
    \item 1D, 2D and 3D simulations on fixed Cartesian grids, plus radially symmetric simulations following \cite{euler_spherical} (see \cite{storcks24})
    \item a fifth-order finite difference WENO constrained transport scheme following \cite{how_mhd}, 
    which we complement with a novel fourth-order self-gravity scheme, additional stability 
    measures and optionally a low-memory Runge-Kutta integrator (Sec. \ref{sec:fd})
    \item a second-order finite volume scheme \citep{vanleer11} with a simple constrained transport 
    MHD scheme following \cite{pang24} (Sec. \ref{app:finite_volume})
    \item differentiation through simulations with adaptive time stepping via \textit{online checkpointing} 
    \citep{stumm10} and 
    checkpointed while loops~\citep{kidger2025}
    (Sec. \ref{sec:checkpointing})
    \item an optional backend in Pallas with optimized GPU kernels, generated from our 
    high-level JAX code via an agentic skill (Sec. \ref{sec:pallas})
\end{itemize}

In the following, we 

\begin{itemize}
    \item show that the forward performance of \texttt{astronomix} on a single GPU is
    competitive with AthenaPK \citep{athenapk}, an open source performance portable MHD code based on the 
    adaptive mesh refinement library Parthenon \citep{parthenon},
    and that \texttt{astronomix} scales decently to
    multiple GPUs and multiple nodes (Sec. \ref{sec:performance})
    \item discuss the performance of reverse-mode automatic differentiation 
    and the choice of the number of checkpoints \citep{stumm10} (Sec. \ref{sec:checkpointing})
    \item introduce validation tests of gradients against analytical functional derivatives in 
    smooth settings (Sec. \ref{sec:analytical_gradients}) and against finite differencing in a 
    shock tube setup (Sec. \ref{sec:shock_gradients})
    \item demonstrate different use cases of the solver's differentiability, namely
    \begin{itemize}
        \item proper eigenmode initialization of a Kelvin--Helmholtz instability (Sec. \ref{sec:khi_eigen})
        \item field-level inference of initial conditions through a turbulent simulation (Sec. \ref{sec:inverse_example})
    \end{itemize}
    \item discuss gradient and optimization pathologies and how multiple shooting can alleviate them (Sec. \ref{sec:multiple_shooting})
\end{itemize}

\section{Governing equations}

While astrophysical flows can typically be assumed to be inviscid, an astrophysical fluid simulator 
must incorporate physics beyond the Euler equations: astrophysical flows are typically magnetized, 
self-gravitating, radiating and feature complex source terms like supernovae or stellar winds driving 
supersonic, compressible, reactive conditions \citep[Sec. 1]{teyssier15}.

At its core, \texttt{astronomix} implements the equations of ideal magnetohydrodynamics (MHD) for a perfectly conducting plasma \citep{freidberg14}, as given by the continuity equation

\begin{equation}
    \partial_t \rho + \vec{\nabla} \cdot (\rho \vec{v}) = 0,
\end{equation}

with density $\rho$ and velocity $\vec{v}$ and the momentum balance equation

\begin{equation}
    \partial_t (\rho \vec{v}) + \vec{\nabla} \cdot \left(\rho \vec{v}\vec{v}^T + \left(P + \frac{B^2}{2}\right)\mathbb{I} - \vec{B}\vec{B}^T\right) = 0,
\end{equation}

where $P$ is the gas pressure and $\vec{B}$ is the magnetic field vector. Throughout we
adopt units with $\mu_0 = 1$ (the factor $4\pi$ absorbed into $\vec{B}$), so that the
magnetic pressure is $B^2/2$ and the Alfv\'en speed is $v_A = B/\sqrt{\rho}$.

The gas pressure $P$ follows either from the energy equation

\begin{equation}
    \partial_t \left(\rho e + \frac{B^2}{2}\right) + \vec{\nabla} \cdot \left[(\rho e + P)\vec{v} - (\vec{v} \times \vec{B}) \times \vec{B} \right] = 0,
\end{equation}

with specific total energy $e = e_{th} + \frac{1}{2} \vec{v}^2$, where $e_{th}$ is the specific thermal energy, and ideal gas equation of state

\begin{equation}
    P = (\gamma - 1) \rho e_{th}, \quad \gamma = \text{ specific heat ratio } = \frac{c_p}{c_v}, \quad \text{ for a monatomic gas } \gamma = \frac{f+2}{f} = \frac{5}{3}
\end{equation}

--- the whole system is then called the adiabatic MHD equations --- or under an isothermal assumption from $P = c_s^2 \rho$, where $c_s$ is the sound speed --- the isothermal MHD equations. We will mostly focus on the adiabatic MHD equations.

The magnetic field evolves in time according to the induction equation, here for the case of perfect conductivity:

\begin{equation}
    \partial_t \vec{B} = \vec{\nabla} \times (\vec{v} \times \vec{B}).
\end{equation}

More details on the governing equations and their eigen-structure can be found in Appendix \ref{app:equations}.

The addition of self-gravity is described in Sec. \ref{sec:self_gravity}.

\section{Methodology and forward verification}

The core physics we focus on in this work are hydrodynamics,
magnetohydrodynamics and self-gravity. While many astrophysical
plasmas have relatively weak magnetic fields \citep[Ch. 1]{bosch20}, magnetic fields
are thought to take part in regulating star formation and
influence angular momentum in accretion disks
(see \cite{arepo_mhd} and references therein). Magnetic
fields also have a significant impact on the evolution of
wind-blown bubbles \citep{lachlan24}, influence core-collapse
supernovae \citep{arepo_mhd} and are core to the dynamics
of Earth's magnetosphere, coronal magnetic fields and the
solar wind \citep{marsch91,parker58}. Self-gravity is important for structure formation;
specifically, it creates the overdensities from which stars
are born \citep{McKee2007}. Together with our turbulence driving module,
we therefore cover the key dynamical processes involved
in star formation -- turbulence, magnetic fields, and self-gravity
\citep{McKee2007}.

One of the most fundamental choices in designing a numerical scheme for solving a partial differential equation 
is the representation of the solution. In a finite volume scheme, the fluid state is represented in terms of 
volume averages over cells, e.g. the average density. In a finite difference scheme, the solution is represented 
in terms of point values. 
This distinction has fundamental consequences:

\begin{itemize}
    \item consider both the finite volume and finite difference scheme are formulated in a conservative flux form, in the finite difference scheme the discretely conserved quantity (the sum of the point values) equals the physical integral of the conserved quantity to the order of the scheme, while the equivalence is exact for the finite volume scheme
    \item for the inclusion of nonlinear source terms $S$ which are functions of the fluid state $W$ the direct evaluation $S(W)$ is correct in the finite difference scheme, while in a finite volume scheme, one must account for the fact that on each cell, $\langle S(W) \rangle \neq S(\langle W \rangle)$, i.e. naively evaluating source terms with the cell-averaged quantities 
    introduces errors, which can possibly degrade the order of the overall scheme
\end{itemize}

As we are interested in studying the behavior of high-order schemes and non-linear source terms such as self-gravity are ubiquitous in astrophysics, we focus on a fifth-order finite difference scheme as the workhorse in \texttt{astronomix} and implement a second-order finite volume scheme mostly for comparison's sake.

For both the finite volume and the finite difference 
scheme the time integration for solving the hyperbolic system

\begin{equation}
    \partial_t \vec{U} + \vec{\nabla}_{\vec{x}} \cdot \mat{F}(\vec{U}) = 0, \quad \text{state vector } \vec{U}, \quad \text{flux matrix } \mat{F}
\end{equation}

is based on the discrete right-hand
side in flux-form (here written for one dimension, $x$)

\begin{equation}
    {\partial_t \vec{U}}_i=\frac{1}{\Delta x}\left[\vec{F}^*_{i-\frac{1}{2}}-\vec{F}^*_{i+\frac{1}{2}}\right].
\end{equation}

In the finite volume scheme $\vec{F}^*_{i\pm\frac{1}{2}}$ can be understood 
as true interface fluxes, for the interpretation in the finite difference 
scheme, see Sec. \ref{sec:self_gravity_scheme}.

Another fundamental decision is how to handle multiple spatial dimensions.
One might either apply operator splitting, consecutively applying updates
along different dimensions, or use an unsplit scheme where the spatial operators
are added up and handled together in the time integration scheme. 
We will focus on unsplit schemes here, i.e. the right-hand side in three dimensions
reads

\begin{equation}
    {\partial_t \vec{U}}_{i,j,k}
    =
    \frac{1}{\Delta x}
    \left[
        \vec{F}^{x*}_{i-\frac{1}{2},j,k}
        -
        \vec{F}^{x*}_{i+\frac{1}{2},j,k}
    \right]
    +
    \frac{1}{\Delta y}
    \left[
        \vec{F}^{y*}_{i,j-\frac{1}{2},k}
        -
        \vec{F}^{y*}_{i,j+\frac{1}{2},k}
    \right]
    +
    \frac{1}{\Delta z}
    \left[
        \vec{F}^{z*}_{i,j,k-\frac{1}{2}}
        -
        \vec{F}^{z*}_{i,j,k+\frac{1}{2}}
    \right].
\end{equation}

A Runge-Kutta scheme is applied for time stepping. Before each time step, $\Delta t$ is
provided by a time step estimator based on the CFL criterion. In the coming sections
we therefore have to show how each scheme provides $\vec{F}^*_{i\pm\frac{1}{2}}$ and
the timestep estimate $\Delta t$.

Regarding the inclusion of magnetic fields, one core challenge of numerical MHD is keeping the magnetic
field divergence-free, $\vec{\nabla} \cdot \vec{B} = 0$.
If this is not achieved, simulation results quickly 
become unphysical. From projection methods over hyperbolic divergence
cleaning methods \citep{dedner02} to additional
terms diffusing divergence errors \citep{powell99} and constrained transport methods
\citep{evans88}, a plethora of methods have been developed
(see \cite{toth00} and \cite{helzel11} for an overview).
Here, we employ constrained transport methods.
While there exist many variants of constrained transport methods 
(e.g. with staggered or unstaggered magnetic
and electric fields) they commonly introduce the electric field, 
and the magnetic field update must satisfy the Maxwell-Faraday equation

\begin{equation}
   \partial_t \vec{B} = -\vec{\nabla} \times \vec{E},
\end{equation}

which retains $\vec{\nabla} \cdot \vec{B} = 0$ as the divergence of a curl vanishes.
The handling of the magnetic fields differs between our finite volume and finite difference
implementations and is described in the respective sections.

The general structure of the unsplit time integration in \texttt{astronomix}
is given in simplified pseudocode in Algorithm \ref{algs:astronomix}. Additional
physical modules such as stellar wind or cooling might either be included as right-hand
side terms evaluated at each Runge-Kutta substage or as updates to the state at each
time step.

The finite volume scheme serves as a simple baseline method and lets us make runtime 
comparisons to other simple second-order finite volume codes. However, the flagship method
of \texttt{astronomix} is the finite difference scheme. The detailed description
of the finite volume scheme is given in Sec. \ref{app:finite_volume}, the description
of the finite difference scheme follows here.

\begin{algorithm}
\caption{Conserved-variable unsplit time evolution}
\label{algs:astronomix}
\begin{algorithmic}[1]
\State $Q \gets Q_{\mathrm{init}}$ \Comment{cell-centered conserved state}
\State $\mathcal{A} \gets \mathcal{A}_{\mathrm{init}}$ \Comment{auxiliary evolved fields, e.g.\ staggered $B$ components}
\State $U \gets (Q, \mathcal{A})$ \Comment{full evolved state}
\While{$t < t_{\mathrm{end}}$}
    \State $\Delta t \gets \Call{TimestepEstimator}{U}$ \Comment{estimate time step}
    \State $U \gets \Call{IntegrationLevelModuleUpdates}{U}$ \Comment{integration-level module effects}
    \Statex
    \Function{RHS}{$U$} \Comment{set up RHS function}
        \State $(Q, \mathcal{A}) \gets U$
        \State $Q \gets \Call{SyncFromAux}{Q, \mathcal{A}}$ \Comment{reconstruct dependent fields, e.g.\ cell-centered $B$ from staggered}
        \State $F_x \gets \Call{FluxesX}{Q}$ \Comment{hydro or MHD fluxes from FD or FV scheme}
        \State $\dot{Q}_{\mathrm{flux},x} \gets \dfrac{1}{\Delta x}\bigl(F_{x,\,i-1/2} - F_{x,\,i+1/2}\bigr)$ \Comment{$i$: index along $x$}
        \State $\quad\vdots$ \Comment{analogously for $y$, $z$}
        \State $\dot{Q}_{\mathrm{flux}} \gets \dot{Q}_{\mathrm{flux},x} + \dot{Q}_{\mathrm{flux},y} + \dot{Q}_{\mathrm{flux},z}$
        \State $\dot{\mathcal{A}} \gets \Call{AuxRHS}{Q, (F_x, F_y, F_z)}$ \Comment{own RHS from shared fluxes, e.g.\ constrained transport}
        \State $\dot{S}_{\mathrm{mod}} \gets \Call{ModuleSources}{Q, (F_x, F_y, F_z)}$ \Comment{source-level module effects}
        \State $\dot{Q} \gets \dot{Q}_{\mathrm{flux}} + \dot{S}_{\mathrm{mod}}$
        \State \Return $(\dot{Q}, \dot{\mathcal{A}})$ \Comment{time derivatives for every component of $U$}
    \EndFunction
    \Statex
    \State $\textsc{preStage} \gets$ pre-stage hook \Comment{boundary handling, positivity clipping on all of $U$}
    \State $\textsc{finalize} \gets$ finalize hook \Comment{merge $\mathcal{A}$ into $Q$ (\Call{SyncFromAux}{}), clip}
    \State $U \gets \Call{RKStep}{U, \Delta t, \textsc{rhs}, \textsc{preStage}, \textsc{finalize}}$ \Comment{advances $U$ component-wise}
    \State $t \gets t + \Delta t$
\EndWhile
\end{algorithmic}
\end{algorithm}

\subsection{Finite difference scheme}
\label{sec:fd}

As the flagship numerical method, we implemented a
fifth-order weighted-essentially-non-oscillatory (WENO) finite difference
scheme with constrained-transport handling
of the magnetic field following \citet{how_mhd}, keeping
the divergence of the magnetic field zero to numerical precision.

We expand on the \cite{how_mhd} scheme in the following ways:
we complement it with a high-order energy-conserving self-gravity module (see Sec. \ref{sec:self_gravity}),
we implement stability-improving measures such as a positivity-preserving
flux limiter \citep{Hu2013}, and finally we provide a low-storage fourth-order Runge-Kutta
scheme in addition to the default strong-stability-preserving Runge-Kutta
scheme. The available Runge-Kutta schemes are presented in Sec. \ref{app:rk_integrators}.

In the following we discuss 

\begin{itemize}
    \item the calculation of the fluxes which enter into the right-hand side of the Runge-Kutta integrators (Sec. \ref{sec:weno_fluxes})
    \item the constrained transport scheme for divergence-free magnetohydrodynamics (Sec. \ref{sec:constrained_transport})
    \item the calculation of the time step (Sec. \ref{sec:fd_time_step})
    \item measures for improved stability (Sec. \ref{sec:stability})
\end{itemize}

\subsubsection{Calculation of the WENO fluxes}
\label{sec:weno_fluxes}

The cell-centered fluxes and eigensystem of the flux Jacobian are 
given in Sec. \ref{app:equations} for adiabatic and isothermal hydrodynamics
and magnetohydrodynamics.

The idea of WENO is to find interface fluxes by interpolating the
cell-centered fluxes using several stencils, and then weighting the stencils
based on their smoothness \citep{shu99}.
 
The reconstruction is done in characteristic variables to better capture the
underlying wave structure. At each interface, we compute the eigenstructure
(evaluated at the average of the left and right states), and project all
stencils into characteristic space.
 
Consider the interface at $i + \tfrac{1}{2}$. 
For the example of adiabatic MHD our vector of conserved variables is
\begin{equation}
  \vec{q} = (\rho,\ \rho v_x,\ \rho v_y,\ \rho v_z,\ B_x,\ B_y,\ B_z,\ E)^{T},
\end{equation}
with $N_\mathrm{vars} = 8$ variables. In the eigenstructure of the MHD
equations, we have $N_\mathrm{char} = 7$ characteristic waves.
 
We calculate the flux as follows:
\begin{enumerate}
  \item We retrieve the eigenstructure given by the right and left eigenvector
    matrices $\vec{R}_{i+1/2} \in \mathbb{R}^{N_\mathrm{vars} \times N_\mathrm{char}}$
    and $\vec{L}_{i+1/2} \in \mathbb{R}^{N_\mathrm{char} \times N_\mathrm{vars}}$,
    as well as the eigenvalues $\lambda$ at
    $\vec{q}_{i+1/2} \approx \tfrac{1}{2}(\vec{q}_i + \vec{q}_{i+1})$.
 
  \item In the stencil $m = i-2, \dots, i+3$, we project the fluxes $\vec{F}_m$
    and conserved variables $\vec{q}_m$ into characteristic space,
    \begin{equation}
      F_{s,m} = \vec{L}^s_{i+1/2}\, \vec{F}_m, \qquad
      q_{s,m} = \vec{L}^s_{i+1/2}\, \vec{q}_m,
    \end{equation}
    where $\vec{L}^s_{i+1/2}$ is the $s$-th row of $\vec{L}$, so $F_{s,m}$
    and $q_{s,m}$ are scalar fields. All fluxes and conserved variables in the
    stencil $m = i-2, \dots, i+3$ are projected using the same
    $\vec{L}^s_{i+1/2}$ at the interface $i + \tfrac{1}{2}$.
 
  \item We compute the differences
    \begin{equation}
      \Delta F_{s,\,m+1/2} = F_{s,m+1} - F_{s,m}, \qquad
      \Delta q_{s,\,m+1/2} = q_{s,m+1} - q_{s,m}
    \end{equation}
    for $m = i-2, \dots, i+2$.
 
  \item We use local Lax--Friedrichs flux splitting to split the fluxes into
    $F_s^+$ and $F_s^-$ such that $\partial_u F_s^+$ only has non-negative
    eigenvalues, and $\partial_u F_s^-$ only has non-positive eigenvalues. Both
    can then be properly upwinded with skewed stencils (for $F_s^+$ we use a
    left-biased stencil, for $F_s^-$ we use a right-biased stencil, see step~5),
    \begin{align}
      \Delta F^{+}_{s,\,m+1/2} &= \tfrac{1}{2}\left(\Delta F_{s,\,m+1/2} + \alpha^s\, \Delta q_{s,\,m+1/2}\right),
        & m &= i-2, \dots, i+1, \\
      \Delta F^{-}_{s,\,m+1/2} &= \tfrac{1}{2}\left(\Delta F_{s,\,m+1/2} - \alpha^s\, \Delta q_{s,\,m+1/2}\right),
        & m &= i-1, \dots, i+2,
    \end{align}
    where $\alpha^s = \max\!\left(\lvert \lambda^s_m \rvert\right)$ over the
    stencil $m = i-2, \dots, i+3$.
 
  \item We can compactly write the WENO flux reconstruction as
    \begin{equation}
    \begin{aligned}
      \vec{F}_{i+1/2} = {}& \tfrac{1}{12}\left(-\vec{F}_{i-1} + 7\vec{F}_i + 7\vec{F}_{i+1} - \vec{F}_{i+2}\right) \\
      & + \sum_{s=1}^{N_\mathrm{char}} \Big[
        -\phi\!\left(\Delta F^{+}_{s,\,i-3/2},\ \Delta F^{+}_{s,\,i-1/2},\ \Delta F^{+}_{s,\,i+1/2},\ \Delta F^{+}_{s,\,i+3/2}\right) \\
      & \qquad\quad\ \,
        + \phi\!\left(\Delta F^{-}_{s,\,i+5/2},\ \Delta F^{-}_{s,\,i+3/2},\ \Delta F^{-}_{s,\,i+1/2},\ \Delta F^{-}_{s,\,i-1/2}\right)
        \Big]\, \vec{R}^s_{i+1/2},
    \end{aligned}
    \end{equation}
    where $\vec{R}^s_{i+1/2}$ is the $s$-th column of $\vec{R}$ at the
    interface $i + \tfrac{1}{2}$, and $\phi$ is the WENO interpolant function
    given by
    \begin{equation}
      \phi(a, b, c, d) = \tfrac{1}{3}\,\omega_0\,(a - 2b + c)
        + \tfrac{1}{6}\left(\omega_2 - \tfrac{1}{2}\right)(b - 2c + d),
    \end{equation}
    with weight functions
    \begin{gather}
      \omega_0 = \frac{\alpha_0}{\alpha_0 + \alpha_1 + \alpha_2}, \qquad
      \omega_2 = \frac{\alpha_2}{\alpha_0 + \alpha_1 + \alpha_2}, \\[4pt]
      \alpha_0 = \frac{1}{(\varepsilon + \mathrm{IS}_0)^2}, \qquad
      \alpha_1 = \frac{6}{(\varepsilon + \mathrm{IS}_1)^2}, \qquad
      \alpha_2 = \frac{3}{(\varepsilon + \mathrm{IS}_2)^2},
    \end{gather}
    and smoothness indicators
    \begin{align}
      \mathrm{IS}_0 &= 13\,(a - b)^2 + 3\,(a - 3b)^2, \\
      \mathrm{IS}_1 &= 13\,(b - c)^2 + 3\,(b + c)^2, \\
      \mathrm{IS}_2 &= 13\,(c - d)^2 + 3\,(3c - d)^2.
    \end{align}
    Here $\varepsilon$ is a small parameter to avoid division by zero, here
    taken as $10^{-7}$. $\varepsilon$ is dimensional, so this assumes code
    units of order one.
\end{enumerate}

\subsubsection{Constrained-transport right-hand side for the interface magnetic fields}
\label{sec:constrained_transport}

Following the algorithm of \citet{how_mhd} interface magnetic fields are evolved with the same
Runge--Kutta method as the cell-centered fluid state based on a constrained
transport right-hand side. At the
end of each Runge--Kutta time step, the cell-centered magnetic fields are
updated based on the interface magnetic fields.

We denote the cell-centered fields by $\rho$, $\vec{v} = (v_x, v_y, v_z)$ and
$\vec{B} = (B_x, B_y, B_z)$, all located at cell centers $(i, j, k)$, and the
interface magnetic fields by $b_x$, $b_y$ and $b_z$, located respectively at the
$x$-, $y$- and $z$-interfaces $(i+\tfrac12, j, k)$, $(i, j+\tfrac12, k)$ and
$(i, j, k+\tfrac12)$. We make use of the 4th-order center-to-face and 6th-order
face-to-center interpolation operators
\begin{align}
  \mathcal{I}^{cf}_x f\big|_{i+1/2} &= \tfrac{1}{16}\left(-f_{i-1} + 9 f_i + 9 f_{i+1} - f_{i+2}\right), \\
  \mathcal{I}^{fc}_x f\big|_{i} &= \tfrac{1}{256}\left(3 f_{i-5/2} - 25 f_{i-3/2} + 150 f_{i-1/2} + 150 f_{i+1/2} - 25 f_{i+3/2} + 3 f_{i+5/2}\right),
\end{align}
(and analogously along $y$ and $z$). Note that $\mathcal{I}^{cf}$ shifts a field
by half a cell along the given axis, so it serves both to move cell-centered
values to interfaces and to move interface values to edges. We further use the
6th-order interface-to-center difference operator
\begin{equation}
  \delta_x f\big|_i = \tfrac{75}{64}\left(f_{i+1/2} - f_{i-1/2}\right) - \tfrac{25}{384}\left(f_{i+3/2} - f_{i-3/2}\right) + \tfrac{3}{640}\left(f_{i+5/2} - f_{i-5/2}\right), \qquad \partial_x f \approx \frac{\delta_x f}{\Delta x},
\end{equation}
and the point-value to cell-average correction following
\citet{Buchmuller2014}, along one axis $a$ given by
\begin{equation}
  \bar q = q + \frac{1}{24}\,\delta^2_a q + \dots, \qquad
  \delta^2_a q\big|_i = q_{i+1} - 2 q_i + q_{i-1} \ \text{along axis } a,
\end{equation}
the discrete form of the averaging operator
$\operatorname{avg}_a = 1 + \tfrac{\Delta a^2}{24}\,\partial_a^2 + O(\Delta a^4)$,
applied per axis and composed where a quantity is averaged along two axes.

At the beginning of a simulation, the interface magnetic fields might be
initialized from the cell-centered magnetic fields via center-to-face
interpolation along each field's own direction,
\begin{equation}
  b_x = \mathcal{I}^{cf}_x B_x, \qquad b_y = \mathcal{I}^{cf}_y B_y, \qquad b_z = \mathcal{I}^{cf}_z B_z.
\end{equation}
Vice versa, the cell-centered magnetic fields are calculated from the interface
magnetic fields via 6th-order face-to-center interpolation,
\begin{equation}
  B_x = \mathcal{I}^{fc}_x b_x, \qquad B_y = \mathcal{I}^{fc}_y b_y, \qquad B_z = \mathcal{I}^{fc}_z b_z,
\end{equation}
after which, for the ideal gas equation of state, the total energy is updated by
the change in cell-centered magnetic energy,
$E \leftarrow E + \tfrac12\left(\lvert\vec{B}_\mathrm{new}\rvert^2 - \lvert\vec{B}_\mathrm{old}\rvert^2\right)$,
so that the thermal pressure is left unchanged (no correction is needed for the
isothermal equation of state). Keeping the thermal pressure fixed this way
protects positivity at the cost of changes to the total energy at the spatial
truncation order of the scheme. We measured $|E(t_\mathrm{end}) - E(0)|/|E(0)|$
for the Alfvén-wave test (Sec. \ref{sec:alfven_test}); it falls from $1.3\cdot10^{-3}$ at $N = 16$ to
$1.6\cdot10^{-6}$ at $N = 64$.

The construction goal for the right-hand side is to achieve high-order 
accuracy while preserving zero divergence. This is achieved by constructing high-order 
point values of the electric field on cell edges and 
applying a point-value discrete curl. The right-hand side is based on the 
WENO interface magnetic fluxes $\mathcal{F}^{d}_{B_c}$ (the flux of 
$B_c$ through the $d$-interface, see Sec.~\ref{sec:weno_fluxes}). Note
that these WENO fluxes are not pointwise evaluations of the magnetic fluxes
at the interfaces. For something of the form 
$\frac{\hat{f}_{i+\frac{1}{2}} - \hat{f}_{i-\frac{1}{2}}}{\Delta x}$ to be a high-order
approximation for $f'(x_i)$, as in the flux formulation of the WENO scheme, 
$\hat{f}_{i+\frac{1}{2}}$ must be the interface evaluation of the 
deconvolved flux function $\tilde{f}$ (see Sec. \ref{sec:self_gravity_scheme} 
for more details). Concretely, in smooth regions the WENO flux satisfies
\begin{equation}
\left.\mathcal{F}^{x}_{B_y}\right|_{i+\frac12}
= \left.\operatorname{avg}_x^{-1}\!\left(B_y v_x - B_x v_y\right)\right|_{i+\frac12}
+ O(\Delta^5),
\qquad B_y v_x - B_x v_y = -E_z,
\end{equation}
where $\operatorname{avg}_x = 1 + \tfrac{\Delta x^2}{24}\,\partial_x^2
+ O(\Delta^4)$ is the averaging operator along $x$
(cf.\ the point-value to cell-average correction above), ignoring the
dissipation terms present in non-smooth regions. Written out in terms of
interface evaluations,
\begin{equation}
\left.\mathcal{F}^{x}_{B_y}\right|_{i+\frac12}
= \left.\left(B_y v_x - B_x v_y\right)\right|_{i+\frac12}
- \frac{\Delta x^2}{24}\,
  \left.\partial_x^2\left(B_y v_x - B_x v_y\right)\right|_{i+\frac12}
+ O(\Delta^4).
\end{equation}

The right-hand side is calculated with the following steps:

\begin{enumerate}
  \item We form modified magnetic fluxes by augmenting the WENO interface
    magnetic fluxes $\mathcal{F}^{d}_{B_c}$ with the cell-centered
    products $B_d v_c$, interpolated to the interface as a single product
    (rather than interpolating the factors separately and multiplying),
    \begin{align}
      \mathcal{F}^{x,*}_{B_y} &= \mathcal{F}^{x}_{B_y} + \mathcal{I}^{cf}_x(B_x v_y), &
      \mathcal{F}^{x,*}_{B_z} &= \mathcal{F}^{x}_{B_z} + \mathcal{I}^{cf}_x(B_x v_z), \\
      \mathcal{F}^{y,*}_{B_x} &= \mathcal{F}^{y}_{B_x} + \mathcal{I}^{cf}_y(B_y v_x), &
      \mathcal{F}^{y,*}_{B_z} &= \mathcal{F}^{y}_{B_z} + \mathcal{I}^{cf}_y(B_y v_z), \\
      \mathcal{F}^{z,*}_{B_x} &= \mathcal{F}^{z}_{B_x} + \mathcal{I}^{cf}_z(B_z v_x), &
      \mathcal{F}^{z,*}_{B_y} &= \mathcal{F}^{z}_{B_y} + \mathcal{I}^{cf}_z(B_z v_y).
    \end{align}
    Since $\mathcal{I}^{cf}_x(B_x v_y)\big|_{i+\frac12}
    = (B_x v_y)\big|_{i+\frac12} + O(\Delta x^4)$ is a point-value
    interpolation, adding it to the expansion above gives, to leading order,
    \begin{equation}
      \mathcal{F}^{x,*}_{B_y} \approx B_y v_x
    - \tfrac{\Delta x^2}{24}\,\partial_x^2 (B_y v_x)
    + \tfrac{\Delta x^2}{24}\,\partial_x^2 (B_x v_y),
    \end{equation}
    where the electric-field contribution that is upwinded along the wrong
    direction for the corresponding edge update (step 2) has been
    neutralized to leading order.

  \item We combine the modified fluxes into the edge-centered electric field
    (electromotive force) components $\Omega_x$, $\Omega_y$ and $\Omega_z$, each
    interpolated from its two adjacent interfaces onto their shared edge,
    \begin{align}
      \Omega_x &= \mathcal{I}^{cf}_y \mathcal{F}^{z,*}_{B_y} - \mathcal{I}^{cf}_z \mathcal{F}^{y,*}_{B_z}, & &\text{at } (i, j+\tfrac12, k+\tfrac12), \\
      \Omega_y &= \mathcal{I}^{cf}_z \mathcal{F}^{x,*}_{B_z} - \mathcal{I}^{cf}_x \mathcal{F}^{z,*}_{B_x}, & &\text{at } (i+\tfrac12, j, k+\tfrac12), \\
      \Omega_z &= \mathcal{I}^{cf}_x \mathcal{F}^{y,*}_{B_x} - \mathcal{I}^{cf}_y \mathcal{F}^{x,*}_{B_y}, & &\text{at } (i+\tfrac12, j+\tfrac12, k).
    \end{align}
    To leading order we get
    \begin{equation}
    \Omega_z \approx
    \left(1 - \tfrac{\Delta x^2}{24}\,\partial_x^2\right)
    \left(1 - \tfrac{\Delta y^2}{24}\,\partial_y^2\right) E_z,
    \end{equation}
    i.e.\ $\Omega$ is not yet a point value of the electric field but
    rather its double deconvolution along the two axes spanning the
    edge plane.

  \item This double deconvolution is cancelled by applying the
    point-value to cell-average correction of \citet{Buchmuller2014}
    along the two axes spanning each edge plane. Since averaging is the
    inverse of deconvolution, this yields high-order point values of the
    electric field at the edges, $\bar\Omega_x$, $\bar\Omega_y$ and
    $\bar\Omega_z$; e.g.\ $\bar\Omega_z \approx E_z$ at
    $(i+\tfrac12, j+\tfrac12, k)$.

  \item With genuine point values in hand, the point-value difference
    operator $\delta$ now yields a high-order pointwise approximation of
    the curl at the interface centers. We update the interface fields by the discrete curl
    $\partial_t \vec{b} = -\nabla\times\vec{\Omega}$,
    \begin{align}
      \mathrm{rhs}_{b_x} &= -\frac{1}{\Delta y}\,\delta_y \bar\Omega_z + \frac{1}{\Delta z}\,\delta_z \bar\Omega_y, \\
      \mathrm{rhs}_{b_y} &= -\frac{1}{\Delta z}\,\delta_z \bar\Omega_x + \frac{1}{\Delta x}\,\delta_x \bar\Omega_z, \\
      \mathrm{rhs}_{b_z} &= -\frac{1}{\Delta x}\,\delta_x \bar\Omega_y + \frac{1}{\Delta y}\,\delta_y \bar\Omega_x,
    \end{align}
    where in reduced dimensionality the terms involving derivatives along
    inactive axes drop out.
\end{enumerate}

Because the interface fields are advanced by the discrete curl of edge-centered
electric fields, and the discrete divergence
$\nabla\cdot\vec{b} = \delta_x b_x/\Delta x + \delta_y b_y/\Delta y + \delta_z b_z/\Delta z$
is built from the same operator $\delta$, the mixed differences commute and
$\nabla\cdot\vec{b} = 0$ is preserved by construction. The interpolations and the 
average correction are formally fourth-order
accurate in the transverse directions (cf. \cite{Buchmuller2014}) but in practice
the WENO truncation error dominates at attainable resolutions and fifth-order
convergence is observed \citep{how_mhd}, matching the behavior of the underlying scheme.

\subsubsection{Time step estimation}
\label{sec:fd_time_step}
 
The CFL time step is given by
\begin{equation}
  \Delta t_\mathrm{adv} = C_\mathrm{CFL}\,\frac{\Delta x}{\lambda_x + \lambda_y + \lambda_z},
\end{equation}
where $\lambda_d$ is the largest absolute eigenvalue of the flux Jacobian along $d$ (see Sec. \ref{app:equations}).
The strong-stability-preserving fourth-order Runge-Kutta scheme
is stable up to $C_\mathrm{CFL} = 1.5$ (see Sec.~\ref{app:rk:ssprk4}).

\subsubsection{Measures for improved stability}
\label{sec:stability}

Violent flows such as strong gravitational collapses or high-Mach turbulence
can result in negative densities or pressures, or high-Mach
overshoots in the immediate neighborhood of a deep void. This kind of behavior
is not specific to the WENO scheme applied here but a general challenge in numerical
hydrodynamics of the extreme conditions one encounters in astrophysics.

We extend the \cite{how_mhd} method by two fairly standard stability mechanisms:
blending the interface flux toward a diffusive first-order flux at trouble cells,
and enforcing positivity of the updated conserved state after each Runge--Kutta stage.
 
\emph{Flux blending toward Lax--Friedrichs.} 
At each interface the high-order
WENO flux is optionally blended toward the first-order local Lax--Friedrichs (Rusanov) flux
with a per-interface weight $w \in [0, 1]$,
\begin{equation}
  \hat{\vec{F}}_{i+1/2} = (1 - w_{i+1/2})\,\vec{F}^\mathrm{WENO}_{i+1/2} + w_{i+1/2}\,\vec{F}^\mathrm{LLF}_{i+1/2},
\end{equation}
where the Lax--Friedrichs flux is built from the physical fluxes $\vec{F}_L$,
$\vec{F}_R$ and conserved states $\vec{q}_L = \vec{q}_i$,
$\vec{q}_R = \vec{q}_{i+1}$ of the two cells adjacent to the interface,
\begin{equation}
  \vec{F}^\mathrm{LLF}_{i+1/2} = \tfrac12(\vec{F}_L + \vec{F}_R) - \tfrac12\,\alpha\,(\vec{q}_R - \vec{q}_L), \qquad
  \alpha = \max\left(\lvert v_{d,L}\rvert + c_L,\ \lvert v_{d,R}\rvert + c_R\right),
\end{equation}
with $d$ the sweep axis and $c$ the fast magnetosonic speed (MHD) or sound speed
(hydro). The weight is set by one or both of two activation paths, when both are 
active the stronger weight is taken, $w = \max(w_\mathrm{void}, w_\mathrm{pp})$:
\begin{enumerate}
  \item \emph{Deep-void protection.} To dampen the high-Mach overshoot near a deep void,
    $w$ ramps linearly from $1$ at the density floor to $0$ at $\beta\rho_\mathrm{min}$,
    using the smaller of the two adjacent densities $\rho_f = \min(\rho_L, \rho_R)$,
    \begin{equation}
      w_\mathrm{void} = \mathrm{clip}\!\left(\frac{\beta\rho_\mathrm{min} - \rho_f}{(\beta - 1)\,\rho_\mathrm{min}},\ 0,\ 1\right),
    \end{equation}
    with $\beta > 1$ the blend factor.
\item \emph{Positivity-preserving flux.} Negative densities and pressures from
    over-depleting fluxes are avoided with the weight $w_\mathrm{pp} = 1 - \theta_\mathrm{keep}$,
    where $\theta_{\mathrm{keep},\,i+1/2} \in [0, 1]$ is the fraction of the antidiffusive
    flux $\vec{A}_{i+1/2} = \vec{F}^\mathrm{WENO}_{i+1/2} - \vec{F}^\mathrm{LLF}_{i+1/2}$
    that is retained, so the limited flux is
    $\vec{F}^\mathrm{LLF}_{i+1/2} + \theta_{\mathrm{keep},\,i+1/2}\,\vec{A}_{i+1/2}$.
    It factorizes into a density and a pressure fraction,
    $\theta_\mathrm{keep} = \theta_\rho\,\theta_p$, applied sequentially: each is the
    largest fraction that would keep a single Euler step with the limited flux above
    the respective floor. Writing $\sigma = \Delta t/\Delta x$
    for the current stage, the first-order (Lax--Friedrichs) update is
    $\vec{U}^\mathrm{LF}_i = \vec{U}_i - \sigma\,(\vec{F}^\mathrm{LLF}_{i+1/2} - \vec{F}^\mathrm{LLF}_{i-1/2})$,
    with density component $\rho^\mathrm{LF}_i$.

    The density fraction $\theta_{\rho,\,i+1/2}$ keeping the density above the floor
    $\rho_\mathrm{min}$ is computed from the antidiffusive mass depletion of cell $i$ and its
    headroom above the floor,
    \begin{equation}
      P_i = \sigma\left[\max(0, A_{\rho,\,i+1/2}) + \max(0, -A_{\rho,\,i-1/2})\right], \quad
      Q_i = \max\!\left(0,\ \rho^\mathrm{LF}_i - \rho_\mathrm{min}\right), \quad
      R_i = \min\!\left(1,\ Q_i / P_i\right),
    \end{equation}
    with $R_i = 1$ when $P_i = 0$ (no depleting fluxes). Every interface is limited by
    the cell it depletes,
    \begin{equation}
      \theta_{\rho,\,i+1/2} = \begin{cases} R_i, & A_{\rho,\,i+1/2} \geq 0, \\ R_{i+1}, & A_{\rho,\,i+1/2} < 0, \end{cases}
    \end{equation}
    such that for a first-order Euler update
    \begin{equation}
      \rho^\mathrm{new}_i
        = \rho^\mathrm{LF}_i
          - \sigma\left(\theta_{\rho,\,i+1/2}\,A_{\rho,\,i+1/2}
          - \theta_{\rho,\,i-1/2}\,A_{\rho,\,i-1/2}\right)
        \;\geq\; \rho^\mathrm{LF}_i - R_i\,P_i
        \;\geq\; \rho^\mathrm{LF}_i - Q_i
        \;\geq\; \rho_\mathrm{min},
    \end{equation}
    since only depleting fluxes reduce $\rho_i$, and each carries a fraction
    $\theta_\rho \leq R_i$ by construction. The bound is preserved by the subsequent
    pressure limiting, since $\theta_\mathrm{keep} \leq \theta_\rho$.
    This approach can be viewed as a reduction (the lower-bound branch) of the
    \citet{Zalesak1979} methodology, which in its full form also prevents new extrema
    by constructing local bounds from neighboring cells.

    The pressure fraction (ideal gas) follows \citet{Hu2013}. With the internal-energy
    residual, written in conserved variables with momentum $\vec{m} = \rho\vec{v}$,
    \begin{equation}
      q(\vec{U}) = \rho\,(E - e_\mathrm{floor}) - \tfrac12\lvert\vec{m}\rvert^2 - \tfrac12\rho\lvert\vec{B}\rvert^2 \ \geq 0 \iff p \geq p_\mathrm{min}
      \quad (\text{for } \rho > 0), \qquad e_\mathrm{floor} = \frac{p_\mathrm{min}}{\gamma - 1},
    \end{equation}
    and the density-limited antidiffusive increment
    $\delta\vec{U}_i = -\sigma\left[(\theta_\rho\vec{A})_{i+1/2} - (\theta_\rho\vec{A})_{i-1/2}\right]$,
    the per-cell admissible fraction
    \begin{equation}
      s_i = \max\left\{\, s \in [0, 1] : q(\vec{U}^\mathrm{LF}_i + s\,\delta\vec{U}_i) \geq 0 \,\right\}
    \end{equation}
    is found by bisection ($q$ is quadratic in $s$ for hydro and cubic for MHD, and
    $s_i = 0$ when already $q(\vec{U}^\mathrm{LF}_i) < 0$). In contrast to the density
    fraction, each interface takes the minimum of its two adjacent cells,
    $\theta_{p,\,i+1/2} = \min(s_i,\, s_{i+1})$.
\end{enumerate}
The blended flux also enters the constrained transport function, which
does not need to be adapted otherwise.
 
\emph{Positivity enforcement.} \texttt{astronomix} also features the following
positivity enforcement options, which are applied after each Runge--Kutta stage
\begin{enumerate}
  \item \emph{Hard floor.} Densities below $\rho_\mathrm{min}$ are raised to the
    floor and, for an ideal gas, the pressure recovered from the total energy is
    floored at $p_\mathrm{min}$ and the energy redefined accordingly.
  \item \emph{Neighbor redistribution.} Each cell with $\rho \leq \rho_\mathrm{min}$
    has its density, momentum and (ideal gas) total energy replaced by the average
    over the valid ($\rho > \rho_\mathrm{min}$) cells in its $3^d$ neighborhood,
    with the patched velocity taken as the mass-weighted neighbor mean and clipped
    to $v_\mathrm{max}$ (specified by the user). Isolated vacuum cells fall back to the floor. 
    This replaces the sharp floor cell of the hard-floor option with a smoother, physically
    plausible value. The (rare) patched cells can cause the overall mass conservation to be
    slightly violated. This protection is also implemented in the Fortran code of the
    \cite{how_mhd} method but not described in the paper.
\item \emph{Conservative internal-energy redistribution.} Where the internal
    energy approaches its floor, it is pulled in from higher internal energy neighbors through an
    antisymmetric face flux
    \begin{equation}
      f_{i+1/2} = C_{i+1/2}\,(e_i - e_{i+1}), \qquad
      C_{i+1/2} = C_0\,\mathbb{1}\!\left[\min(e_i,\, e_{i+1}) < \varepsilon_\mathrm{act}\, e_\mathrm{floor}\right],
    \end{equation}
    where $e$ is the internal-energy density, $e_\mathrm{floor} = p_\mathrm{min}/(\gamma - 1)$,
    $C_0 \ll 1$ a user-set diffusion coefficient, and $\varepsilon_\mathrm{act}$ an
    activation factor, so the flux is non-zero only at faces touching a near-floor
    cell. The transfer is applied over a small user-set number of passes, with the
    activation mask and fluxes recomputed from the corrected internal energy after
    each pass. Because $f$ is a face flux the total energy is conserved exactly.
\end{enumerate}

Should invalid values still occur, there is an option to reset them to the
floor values. Additionally, there is a vacuum-rest option that zeroes the
momentum of below-floor cells (so the recovered velocity is $0$ rather than a
spurious $\vec{m}/\rho_\mathrm{min}$ that runs away in a deep void). For the
redistribution and conservative modes, one can set a velocity cap that clips
$\lvert\vec{v}\rvert$ to $v_\mathrm{max}$.

\subsection{JAX as a numerical specification language for optimized GPU kernels}
\label{sec:pallas}

Key metrics for a simulation code are performance, memory efficiency and readability.
However, it is typically difficult to optimize these objectives simultaneously in high-level languages like \texttt{JAX}.
To provide a specific example, naive finite difference implementations can lead to a $20\times$ increase in temporary memory. 
Optimizing for a lower temporary memory
footprint, e.g., by doing the projection onto the eigenvectors one at a time rather
than by a big einsum with all eigenvectors realized simultaneously, can come at the cost
of performance.
Additionally, such optimizations quickly worsen readability of an implementation, and obstruct the underlying mathematical formulation.
High-level languages like \texttt{JAX} often support kernel languages in which low-level GPU kernels can be linked into the high-level code.
Such optimized GPU kernels have the potential to unlock both higher performance and
memory efficiency, e.g. by smartly fusing operations. On the other hand, developing these
GPU kernels by hand is typically tedious and error-prone. 

The availability of capable coding agents enables
a different paradigm: Native high-level \texttt{JAX} code acts as a numerical
specification language from which coding agents can create an optimized GPU 
backend of the simulator, \texttt{Pallas} in this case. 
\footnote{The
agent-assisted code development and the Pallas-backend generation described below
used Claude Code with the Claude Opus 4.8 model.}
Native \texttt{JAX} hereby has advantages
over pseudocode: it is unambiguous and executable; outputs of different
parts of the code can directly be compared. It is a middle ground between
a high-level description and a low-level implementation: precise but readable.
The \texttt{Pallas} backend lives in separate files from the native \texttt{JAX}
code, can be activated in the simulation configuration and is never touched by
the human developer. An agentic skill guides the agent in writing
an efficient \texttt{Pallas} backend. Even if one wants to drive the numerical 
code development itself with an agentic loop, it is more token efficient for the agent to 
iterate on native \texttt{JAX} code and only translate to the \texttt{Pallas} backend
when the numerical details are worked out.

The agentic skill guides the agent to translate one native \texttt{JAX} kernel at a
time and to validate the result against the native reference before accepting it.
This validation encompasses checks of both final results and intermediate kernel outputs of the \texttt{Pallas}
backend compared to the native backend on the forward and gradient test suite presented
in the paper (the convergence, shock-tube, blast and turbulence tests of
Sec.~\ref{sec:forward_tests} and the gradient-verification tests of Sec.~\ref{sec:analytical_gradients}
and~\ref{sec:shock_gradients}). Kernel correctness is maintained by
regenerating the \texttt{Pallas} backend from the native code whenever the latter changes. 

In the following we will use FD/FV (JAX) as a shorthand for the finite difference / finite volume
scheme with native JAX backend, and FD (Pallas) for the finite difference scheme with Pallas backend.

\subsection{Forward-tests}
\label{sec:forward_tests}

In this section we present results from a range of standard hydrodynamical
and magnetohydrodynamical validation tests, comparing different
aspects of the solver configurations implemented in
\texttt{astronomix}. As more advanced \textit{astrophysical} tests,
we discuss an MHD jet (Sec. \ref{sec:mhd_jet}) as well as
driven isothermal turbulence in the high-Mach high magnetic field
as well as the low-Mach low magnetic field regime
(Sec. \ref{sec:iso_turb}).

\subsubsection{3D Hydrodynamical convergence test}

We first verify the spatial accuracy of the hydrodynamics solvers on a smooth,
small-amplitude acoustic wave (an analytical solution to the linearized Euler equations)
in three dimensions.

A grid-aligned wave would only test the solver along one axis. We therefore set up initial wave
conditions parallel to the unit vector $\vec{\hat{n}} = \left(\frac{1}{3}, \frac{2}{3}, \frac{2}{3}\right)^T$
($x_\parallel := \vec{\hat{n}} \cdot \vec{x}$) given by
\begin{equation}
  \begin{aligned}
    \rho &= \rho_0 + \epsilon\,\rho_0\,\cos(k_\parallel x_\parallel - \omega t), \\
    \vec{v} &= \vec{\hat{n}} \, \epsilon\,c_s\,\cos(k_\parallel x_\parallel - \omega t), \\
    p &= p_0 + \epsilon\,\rho_0 c_s^2\,\cos(k_\parallel x_\parallel - \omega t),
  \end{aligned}
\end{equation}
with $\omega = c_s|\vec{k}|, k_\parallel = 2\pi, \epsilon = 10^{-6}$
and background chosen such 
that $c_s = \sqrt{\gamma p_0/\rho_0} = 1$. The computational
domain is periodic along all axes with uniform grid spacing 
and size $\vec{L} = (3.0, 1.5, 1.5)^T$ such that 
the wave wraps periodically in all dimensions. The state is
evolved for five periods.

\begin{figure}
 \centering
 \includegraphics[width=1.0\textwidth]{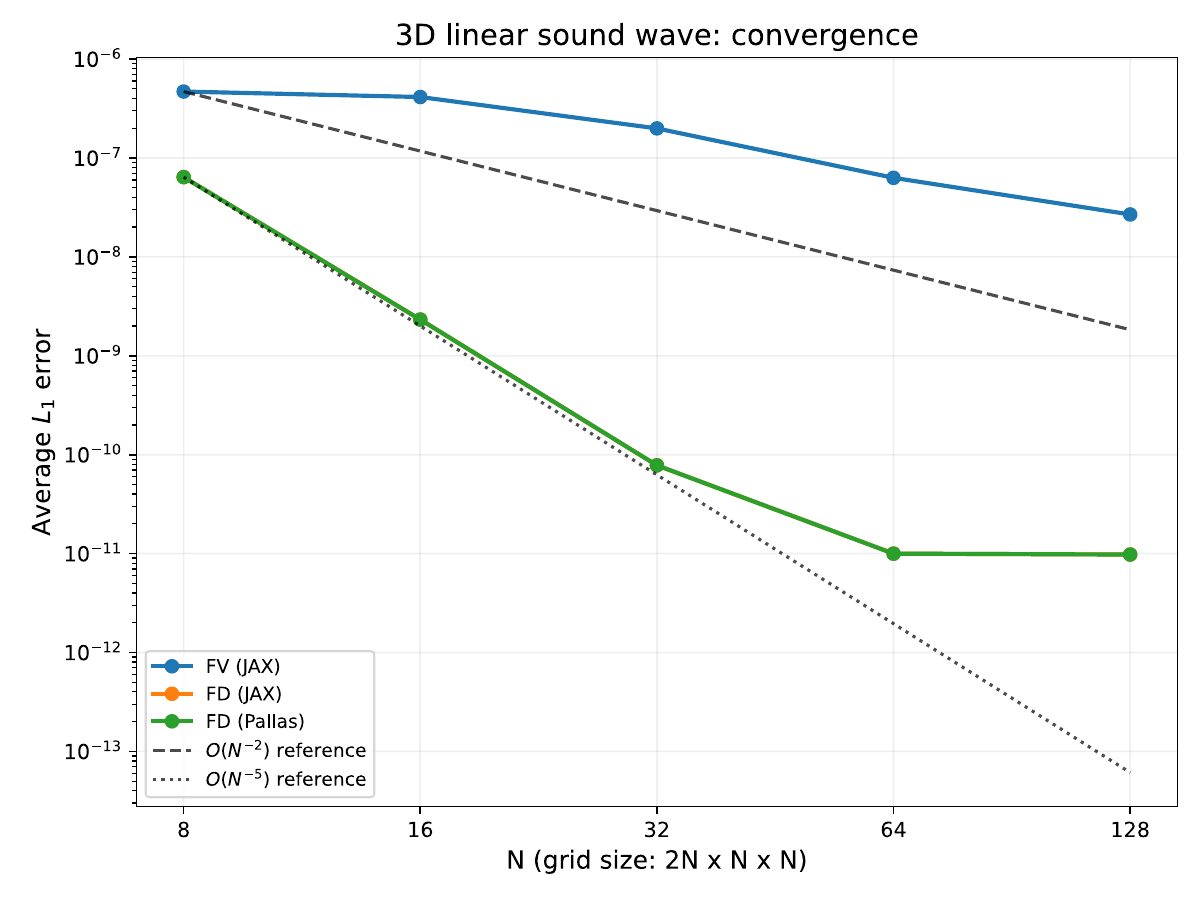}
 \caption{Convergence of the 3D linear sound-wave test: $L_1$ error of
  the solution versus linear resolution $N$ (grid $(2N, N, N)$) at fixed
  integration time, for the finite-volume scheme (FV) and the finite-difference
  scheme on the native-JAX and Pallas backends (FD, the two backends
  overlap exactly). Dashed and dotted lines mark $O(N^{-2})$ and $O(N^{-5})$
  reference slopes.}
 \label{fig:hydro_convergence}
\end{figure}

The mean $L_1$ error across the primitive variables as a function 
of resolution is shown in Fig. \ref{fig:hydro_convergence} for the 
finite volume scheme with default settings (Minmod limiter 
and HLL Riemann solver) and the finite difference solver. The finite
difference solver converges at fifth order, the finite volume solver
at second order as expected. We also used this test to check the Pallas
backend of the finite difference solver against the native JAX
backend and found perfect agreement.

\subsubsection{1D Shock tube}

A standard 1D shock tube test is given 
by the initial conditions

\begin{equation}
  \begin{aligned}
    \rho(x) &= 
    \begin{cases}
    1.0, & \text{if } x < 0.5 \\
    0.125, & \text{otherwise}
    \end{cases} \\
    u(x) &= 0 \\
    p(x) &= 
    \begin{cases}
    1.0, & \text{if } x < 0.5 \\
    0.1, & \text{otherwise}
    \end{cases}
  \end{aligned}
\end{equation}

on the domain $[0,1]$ with open boundaries.

\begin{figure}
 \centering
 \includegraphics[width=1.0\textwidth]{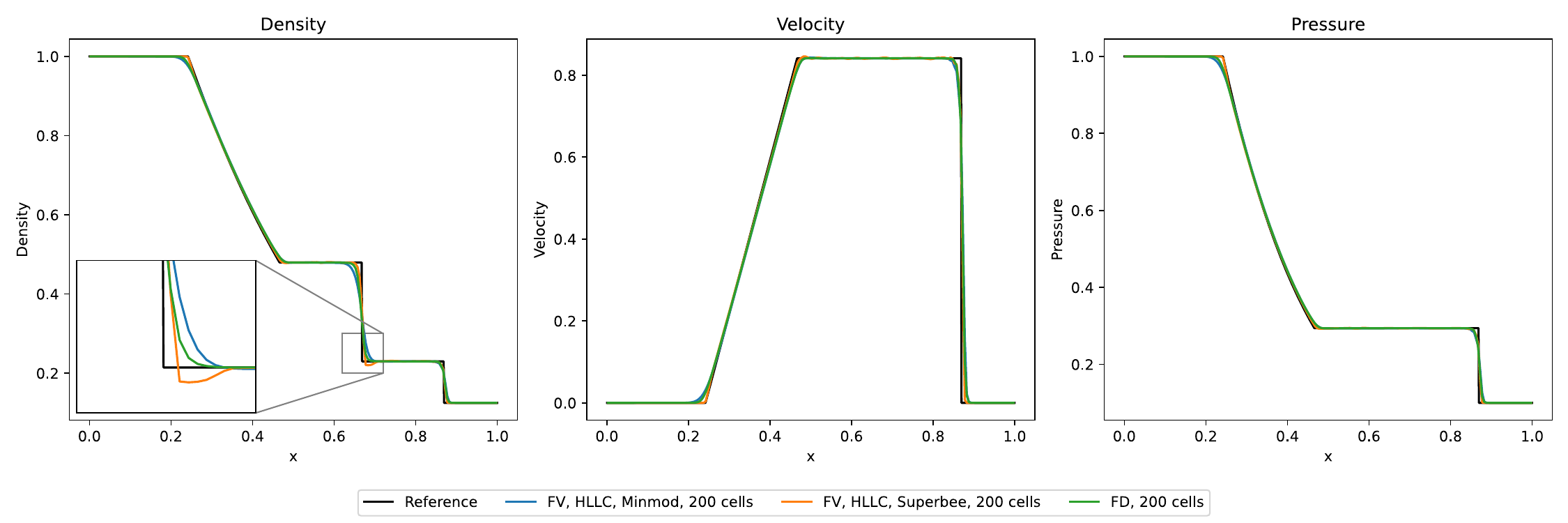}
 \caption{Sod shock tube: density, velocity and pressure at $t = 0.2$ for the
  finite-volume method with HLLC Riemann solver and Minmod and Superbee
  limiters, and the finite-difference scheme, all at $N = 200$ cells, compared
  with the analytical Riemann solution. The inset zooms in on the contact
  discontinuity, where the Superbee limiter is least diffusive but slightly
  overshoots.}
 \label{fig:shock_tube}
\end{figure}

The final density, velocity and pressure at $t = 0.2$ are
given in Fig. \ref{fig:shock_tube} for different solver configurations.
Compared are the finite volume method with HLLC Riemann solver and 
Minmod as well as Superbee limiter and the finite difference
scheme. All of these configurations closely match
the analytical reference solution.
The finite volume method with Superbee limiter 
is least diffusive in this test but slightly overshoots
at the contact discontinuity, as shown in the highlight box. 
While the Superbee limiter
is total variation diminishing for an advection equation,
our reconstruction is in the primitive, not the advected
characteristic variables, which can cause 
such slight overshoots.

\subsubsection{1D Double blast}

Next we consider the double blast problem of \cite{woodward84}.
The initial conditions are 

\begin{equation}
\begin{aligned}
\rho(x) &= 1.0, \\
u(x) &= 0.0, \\
p(x) &=
\begin{cases}
1000.0, & x < 0.1, \\
0.01,   & 0.1 \leq x \leq 0.9, \\
100.0,  & x > 0.9.
\end{cases}
\end{aligned}
\end{equation}

\begin{figure}
 \centering
 \includegraphics[width=1.0\textwidth]{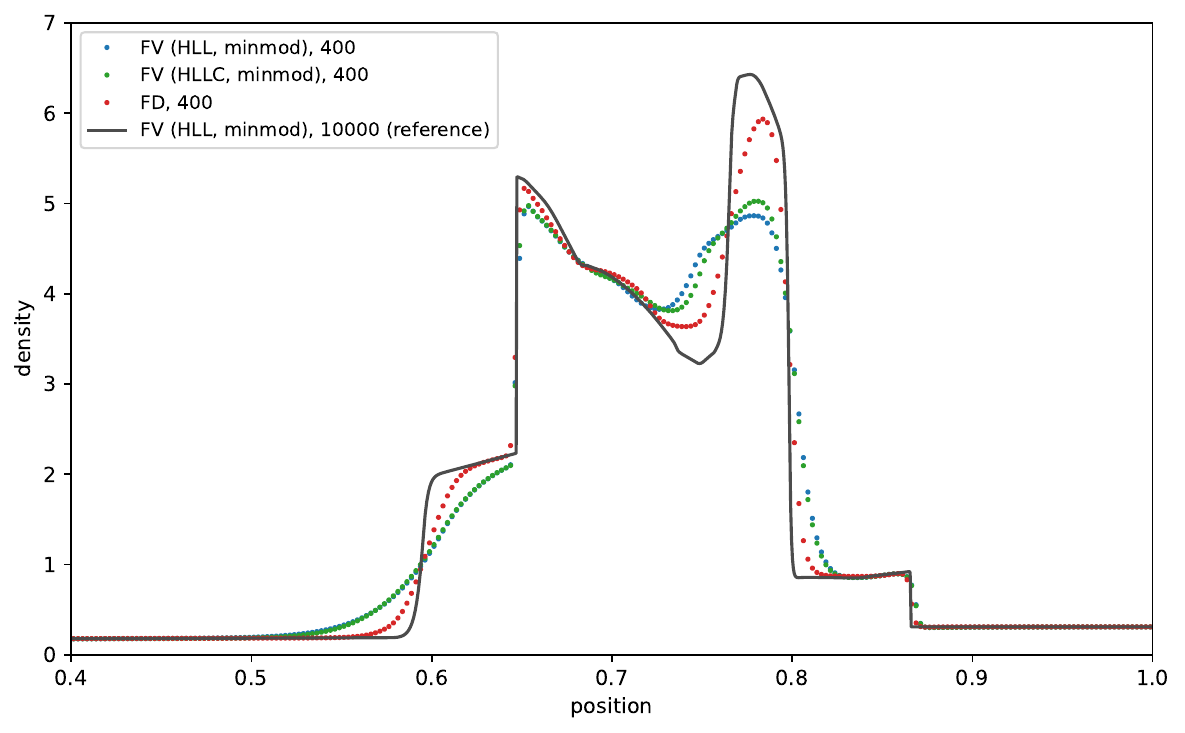}
 \caption{Double blast wave of \cite{woodward84}: density at $t = 0.038$ for the
  finite-volume method with HLL and HLLC Riemann solvers (Minmod limiter) and
  the finite-difference scheme, all at $N = 400$ cells, together with a
  converged $N = 10000$ finite-volume reference (solid line). The
  finite-difference result is closest to the reference at the central density
  peak.}
 \label{fig:double_blast}
\end{figure}

The domain $[0,1]$ has reflecting boundaries and the fluid state is evolved
until $t = 0.038$, by which time the two blast waves have collided. In Fig.
\ref{fig:double_blast} we compare the final density for
the finite volume method with HLL and HLLC Riemann solver as 
well as the finite difference method, all at a resolution of 
$N = 400$ cells. The HLL and HLLC results closely agree, the
finite difference results are closest to the high-resolution
reference, in particular at the density peak near $x \approx 0.78$
which the coarse finite-volume runs underresolve.

\subsubsection{3D Sedov test}

In the Sedov test, a fixed thermal energy $E_\mathrm{expl} = 1$ is injected
into a small spherical region of an otherwise uniform medium of density $\rho = 1$ and
ambient pressure $p_\mathrm{amb} = 10^{-4}$, with zero initial velocity.

The initial pressure is given by

\begin{equation}
  p(r) = p_\mathrm{amb} + \Delta p\,w(r), \qquad
  w(r) = \frac{1}{2}\left[\,1 - \tanh\!\left(\frac{r - r_\mathrm{expl}}{\delta}\right)\right],
  \label{eq:sedov}
\end{equation}

where we slightly smoothed the edges of the injection region to reduce grid artifacts.

$\Delta p$ is calculated such that the correct energy $E_\mathrm{expl}$ is injected

\begin{equation}
  \sum_\mathrm{cells} \frac{p(r) - p_\mathrm{amb}}{\gamma - 1}\,\Delta V = E_\mathrm{expl}
  \quad\Longrightarrow\quad
  \Delta p = \frac{E_\mathrm{expl}\,(\gamma - 1)}{\Delta V \sum_\mathrm{cells} w(r)}.
\end{equation}

\begin{figure}
 \centering
 \includegraphics[width=1.0\textwidth]{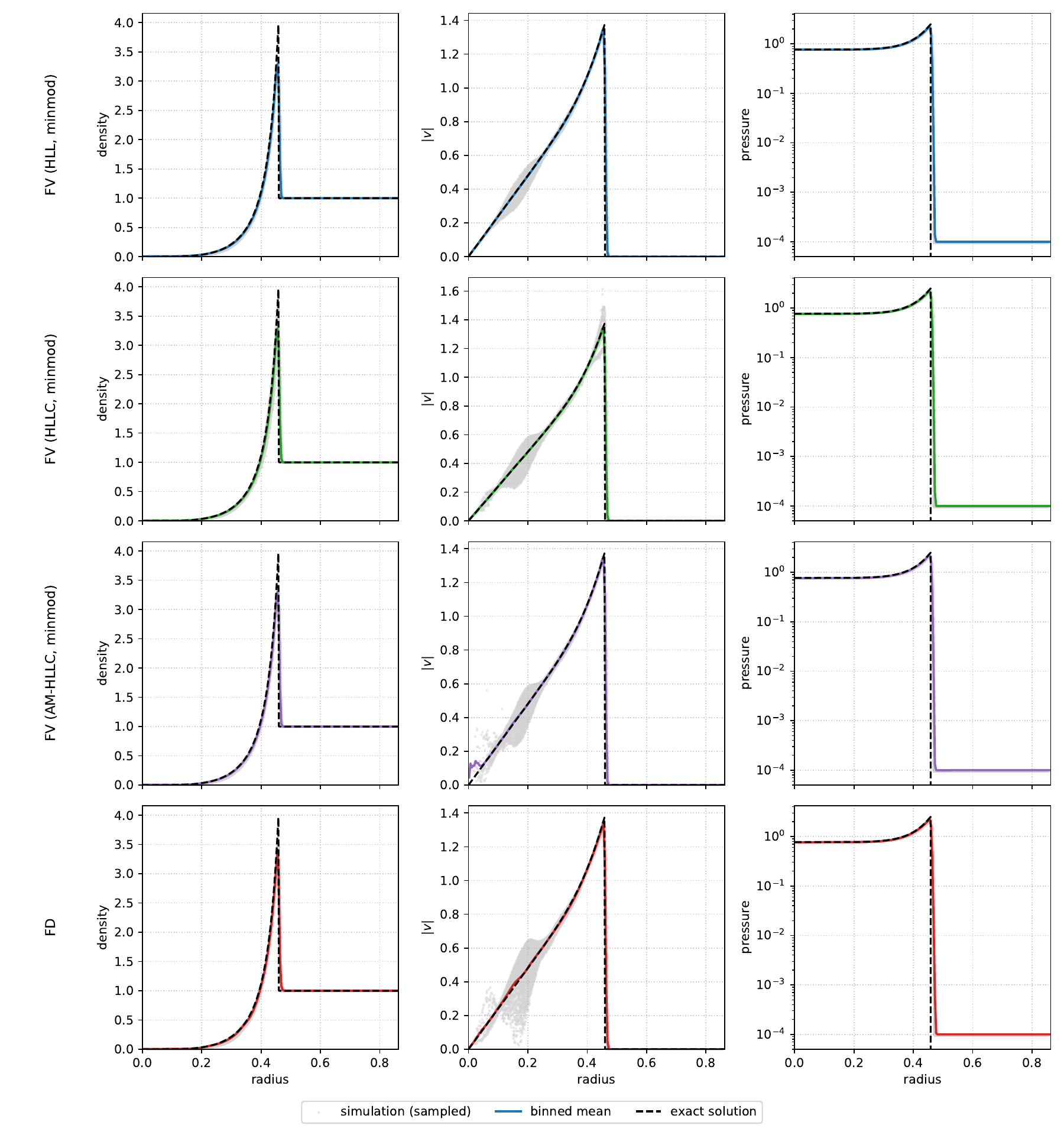}
 \caption{Sedov--Taylor blast wave at $t = 0.1$: radial profiles of density,
  velocity magnitude and pressure (columns) for the finite-volume method with
  Minmod limiter and HLL, HLLC and AM-HLLC Riemann solvers and the
  finite-difference method (rows), all at $256^3$ cells. Grey points are the
  individual cell values, the colored lines their radial binned means, and the
  dashed black line is the analytical self-similar solution.}
 \label{fig:sedov_blast}
\end{figure}

In Fig. \ref{fig:sedov_blast} we show the radial distribution of 
density, velocity magnitude and pressure at $t = 0.1$ for the finite volume method with Minmod limiter
and HLL, HLLC and AM-HLLC Riemann solver and the finite difference
method compared to the analytical
solution. All numerical results were obtained at a resolution
of $256^3$ cells. The binned means of the radial distributions
closely follow the reference solution for all configurations.
All methods show a spread in velocity distribution inside the
blast due to grid artifacts (which still remain in spite of our smoothing).
The HLLC Riemann solver additionally has a large velocity spread
at the shock front; it suffers from the \textit{carbuncle instability}
in multi-dimensional simulations with strong shocks \citep{baumgart24}.
The modern AM-HLLC variant \citep{hllc_am} alleviates this issue.

\subsubsection{3D MHD convergence test}
\label{sec:alfven_test}

We test the spatial convergence of the MHD schemes using a circularly
polarized Alfvén-wave setup, analogous to the hydrodynamical sound-wave
test. We follow Sec.~3.2 of \cite{how_mhd}. Consider a wave which, in a
coordinate system $\tilde{x}, \tilde{y}, \tilde{z}$, propagates along
$-\tilde{x}$ with the velocity and magnetic-field perturbations rotating
in the $\tilde{y}$--$\tilde{z}$ plane:

\Needspace{10\baselineskip}
\begin{equation}
  \begin{aligned}
    \rho           &= \rho_0, \\
    v_{\tilde{x}}  &= 0, \\
    v_{\tilde{y}}  &= A \sin\!\big(k_\parallel (\tilde{x} + v_A t)\big), \\
    v_{\tilde{z}}  &= A \cos\!\big(k_\parallel (\tilde{x} + v_A t)\big), \\
    B_{\tilde{x}}  &= B_\parallel, \\
    B_{\tilde{y}}  &= A \sin\!\big(k_\parallel (\tilde{x} + v_A t)\big), \\
    B_{\tilde{z}}  &= A \cos\!\big(k_\parallel (\tilde{x} + v_A t)\big), \\
    p              &= p_0,
  \end{aligned}
\end{equation}

with $A = 0.1$, $B_\parallel = 1.0$, $\rho_0 = 1.0$, $p_0 = 0.1$,
$k_\parallel = 2\pi$ and $v_A = B_\parallel/\sqrt{\rho_0} = 1$.
The wave is left-going along $\tilde{x}$ at the Alfvén speed.

To avoid alignment with the grid, we rotate this configuration by the
Euler angles $-\arctan(2/\sqrt{5})$ about the $y$-axis followed by
$\arctan(2)$ about the $z$-axis. The corresponding matrix $R$, which maps
a vector from the unrotated frame $(\tilde{x}, \tilde{y}, \tilde{z})$ into
the simulation frame $(x, y, z)$, is

\Needspace{10\baselineskip}
\begin{equation}
  R =
  \begin{pmatrix}
    \dfrac{1}{3} & -\dfrac{2}{\sqrt{5}} & -\dfrac{2}{3\sqrt{5}} \\[8pt]
    \dfrac{2}{3} &  \dfrac{1}{\sqrt{5}} & -\dfrac{4}{3\sqrt{5}} \\[8pt]
    \dfrac{2}{3} &  0                   &  \dfrac{\sqrt{5}}{3}
  \end{pmatrix},
\end{equation}
with the inverse (simulation $\to$ unrotated) rotation given by its
transpose $R^{\mathsf{T}}$. The first column of $R$ is the wave's
propagation direction in the simulation frame,
$\vec{\hat{n}} = \left(\tfrac{1}{3}, \tfrac{2}{3}, \tfrac{2}{3}\right)^{T}$,
i.e. the unrotated $\tilde{x}$-axis is mapped onto $\vec{\hat{n}}$, which is
aligned with none of the coordinate axes.

The computational domain is periodic along all axes and has size
$\vec{L} = (3.0, 1.5, 1.5)^{T}$, discretized on a grid of $(2N, N, N)$
cells so that the grid spacing is uniform along all dimensions and the
wave wraps exactly once along each axis. The state is evolved for five
periods (to $t = 5$), after which the analytic solution has returned to
its initial condition.

For a divergence-free initialization, we construct the magnetic
perturbation as the discrete curl of a vector potential: on the staggered
edge grid for the finite-difference scheme, and on the collocated
cell-centered grid for the finite-volume scheme. The uniform background
field $B_\parallel\,\vec{\hat{n}}$ is added analytically in both cases,
since no periodic vector potential exists for a uniform field.

\begin{figure}
 \centering
 \includegraphics[width=1.0\textwidth]{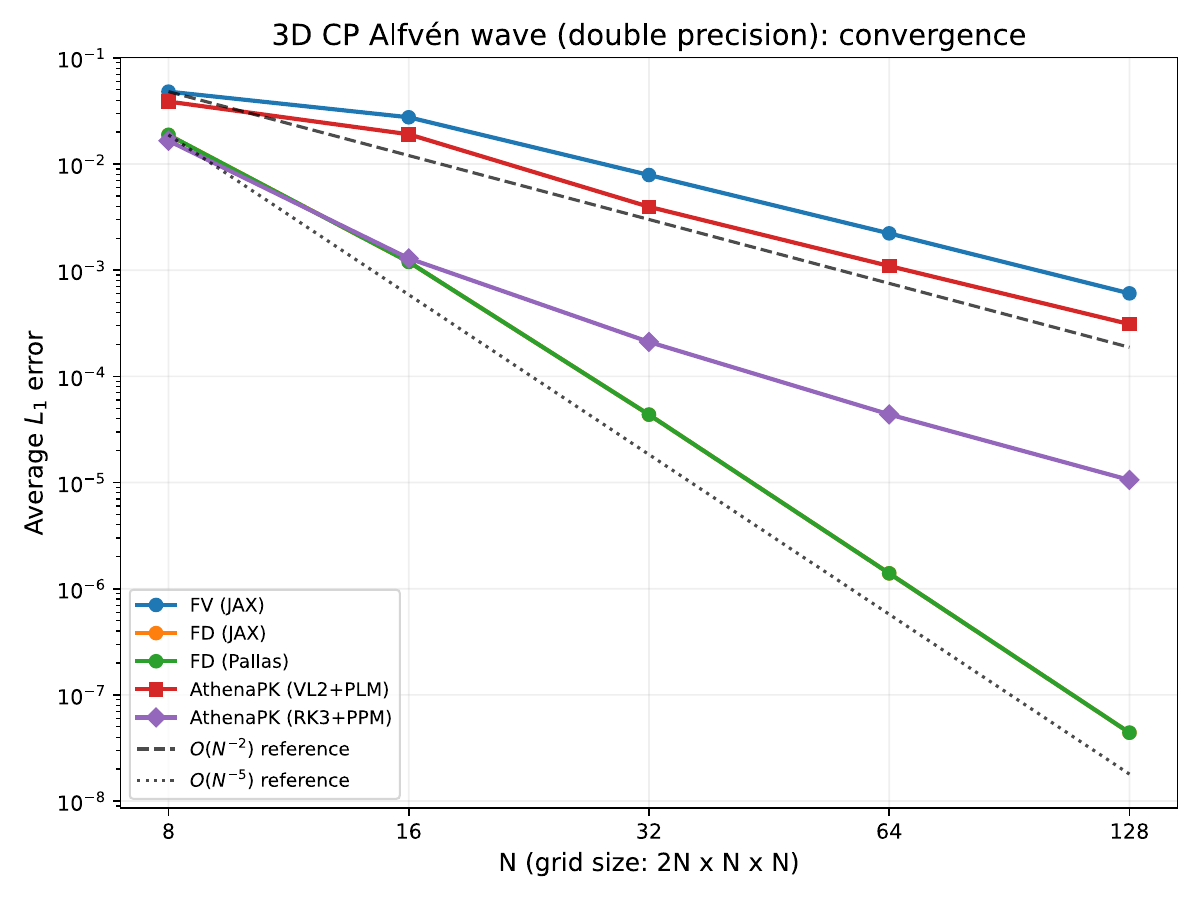}
 \caption{Convergence of the 3D circularly polarized Alfvén-wave test (double
  precision): average $L_1$ error versus linear resolution $N$ (grid
  $(2N, N, N)$) for the finite-volume (FV) and finite-difference (FD) schemes
  and the AthenaPK reference code in a second-order (second-order van Leer integrator 
  and piecewise linear reconstruction, called VL2+PLM) and third-order (third order 
  Runge-Kutta time integrator and piecewise parabolic reconstruction, called RK3+PPM) setup. 
  Dashed and dotted lines mark $O(N^{-2})$ and
  $O(N^{-5})$ reference slopes.}
 \label{fig:mhd_convergence}
\end{figure}

We quantify convergence with the $L_1$ norm of the difference between the
evolved and analytic states at $t = 5$, averaged over the primitive
variables, as in the sound-wave test. As shown in
Fig.~\ref{fig:mhd_convergence}, the finite volume scheme 
converges at second order, like the second-order AthenaPK reference finite
volume scheme (with second-order van Leer integrator and piecewise linear reconstruction, called
VL2+PLM), while the finite difference scheme converges at fifth 
order.

\subsubsection{2D Orszag-Tang vortex test}

\begin{figure}
 \centering
 \includegraphics[width=1.0\textwidth]{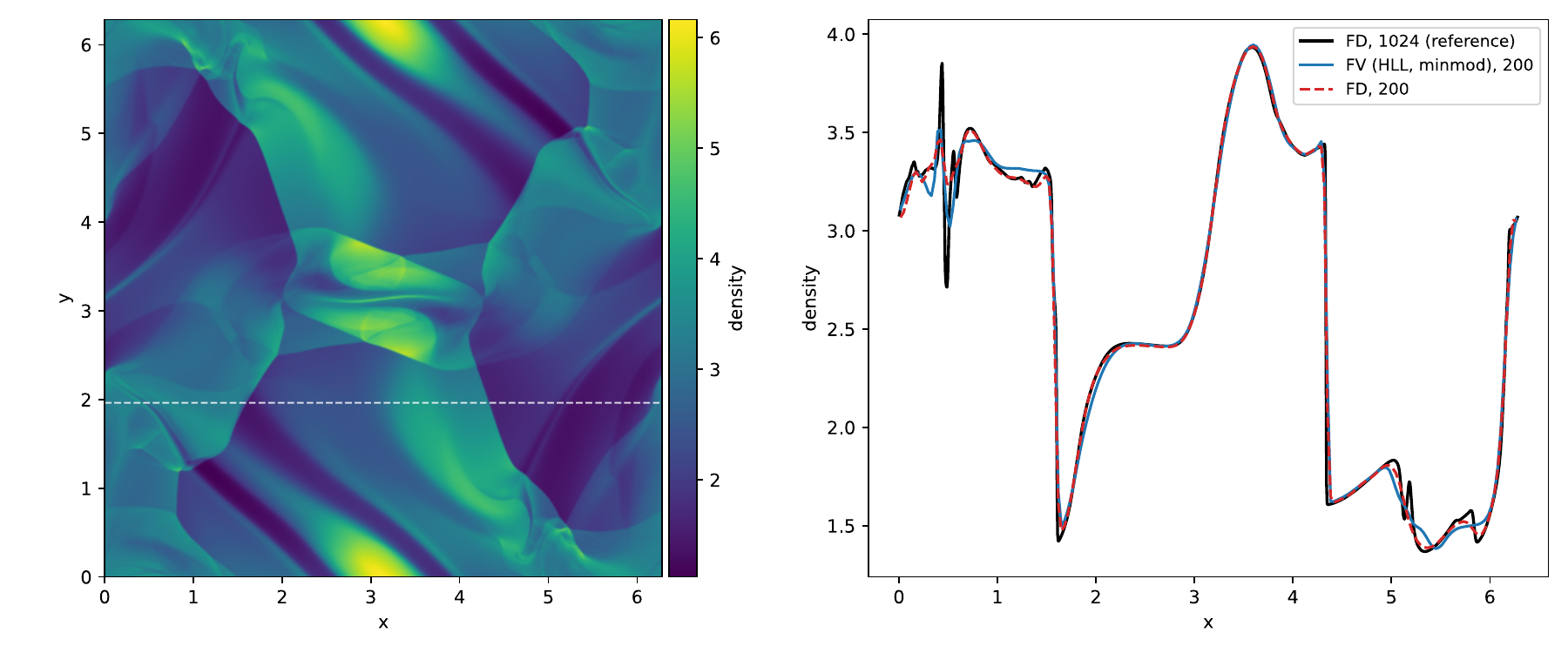}
 \caption{For the Orszag-Tang vortex with initial conditions
  given in Eq. \ref{eq:orszag_tang} the density field (left, finite difference
  at $1024^2$ cells) and a cut through it at $y = 0.625\pi$ (right) at $t = 3.0$
  are shown. The cut compares the finite-volume and finite-difference
  schemes at $200^2$ cells and the finite-difference scheme at $1024^2$ cells.
  The dashed line on the left marks the location of the cut.}
 \label{fig:orszag_tang}
\end{figure}

Consider the Orszag-Tang vortex with initial conditions

\begin{equation}
\begin{aligned}
  \rho &= \gamma^2, \\
  v_x  &= -\sin y, \\
  v_y  &= \sin x, \\
  v_z  &= 0, \\
  B_x  &= -\sin y, \\
  B_y  &= \sin(2x), \\
  B_z  &= 0, \\
  p    &= \gamma.
\end{aligned}
\label{eq:orszag_tang}
\end{equation}

with $\gamma = \frac{5}{3}$ which we simulate on the domain $[0,2\pi]^2$ with periodic boundaries until $t = 3.0$, following
Example 5.2 in \cite{pang24}. In Fig.~\ref{fig:orszag_tang} the density field
obtained with the finite-difference scheme at $1024^2$ cells is shown on the left
and profiles along a horizontal cut at $y = 0.625\pi$ for the finite difference
and finite volume scheme at $N = 200$ on the right. The finite difference
scheme at $N = 200$ is closer to the $N = 1024$ reference solution.

\subsubsection{3D MHD blast test}

Next, we consider the 3D MHD blast test following Sec. 3.7 in \cite{how_mhd}.

The background state is given by a uniform density $\rho = 1$, zero velocity and a uniform diagonal
magnetic field $\vec{B} = (B_0/\sqrt{2}, B_0/\sqrt{2}, 0)$ with $B_0 = 10$. The computational domain
has dimensions $[0,1]^3$.

Around the center of the box, we inject thermal energy via
\begin{equation}
  p(r) =
  \begin{cases}
    100, & r \leq r_0, \\
    1 + 99\,\dfrac{r_1 - r}{r_1 - r_0}, & r_0 < r \leq r_1, \\
    1, & r > r_1,
  \end{cases}
\end{equation}
with $r_0 = 0.125$ and $r_1 = 1.1\,r_0$. The state is evolved until $t = 0.02$.

\begin{figure}
 \centering
 \includegraphics[width=1.0\textwidth]{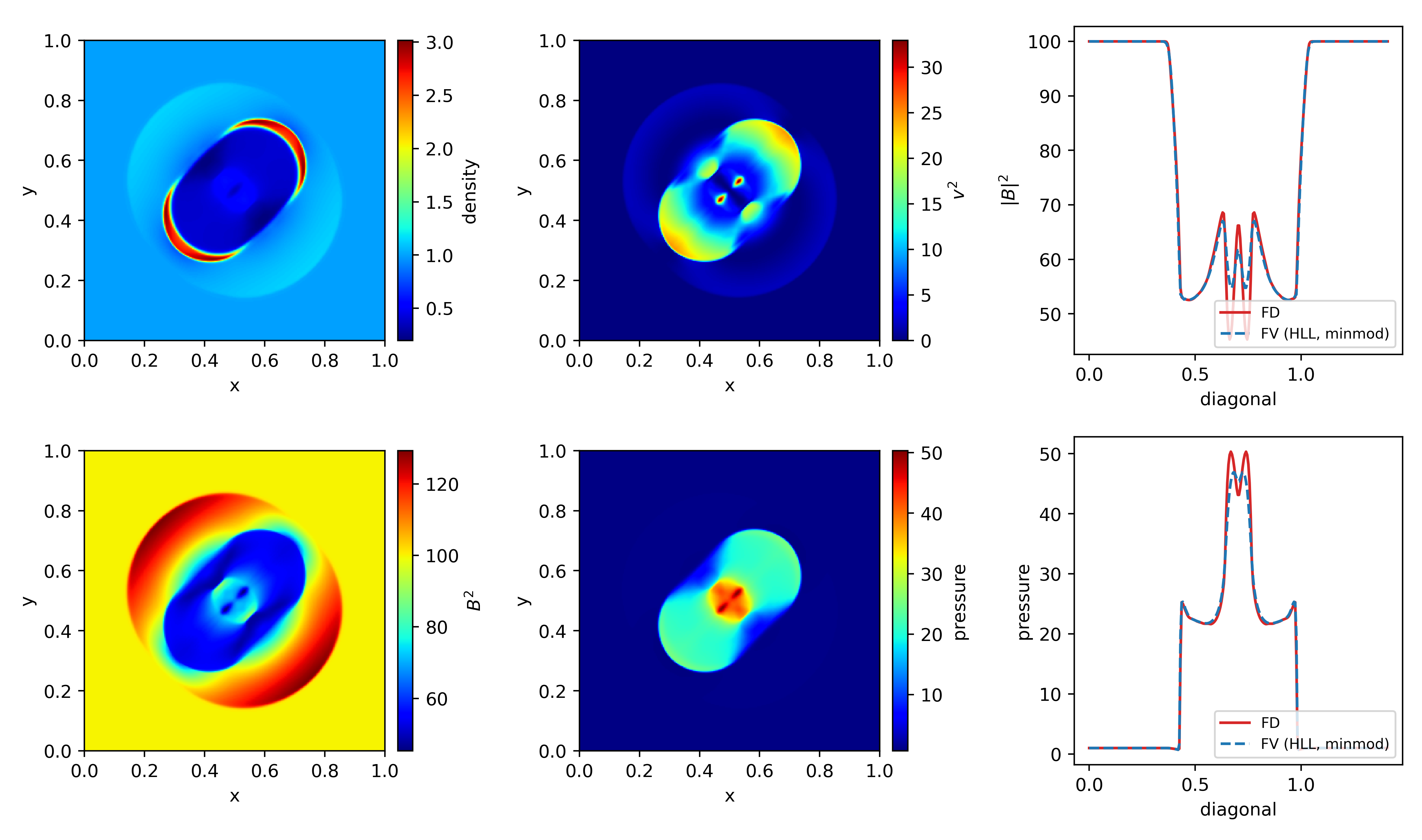}
 \caption{Strongly magnetized 3D MHD blast wave at $t = 0.02$ and $256^3$ cells.
  The four left panels show central slices of the finite-difference solution
  (density, kinetic energy $v^2$, magnetic pressure $B^2$ and thermal pressure);
  the right column shows $|B|^2$ and pressure along the box diagonal for the
  finite-difference scheme and the finite-volume scheme at the same
  resolution.}
 \label{fig:mhd_blast}
\end{figure}

The strong magnetic field results in an anisotropic blast elongated 
along the background magnetic field vector. Fig.~\ref{fig:mhd_blast} shows the
density, kinetic energy, magnetic and thermal pressure slices through $z = 0$
for the finite difference scheme at $256^3$ cells on the left and profiles
through $z = 0, x = y$ for the finite difference and finite volume schemes
on the right. The finite difference scheme produces more detailed
features.

\begin{figure}
 \centering
 \includegraphics[width=1.0\textwidth]{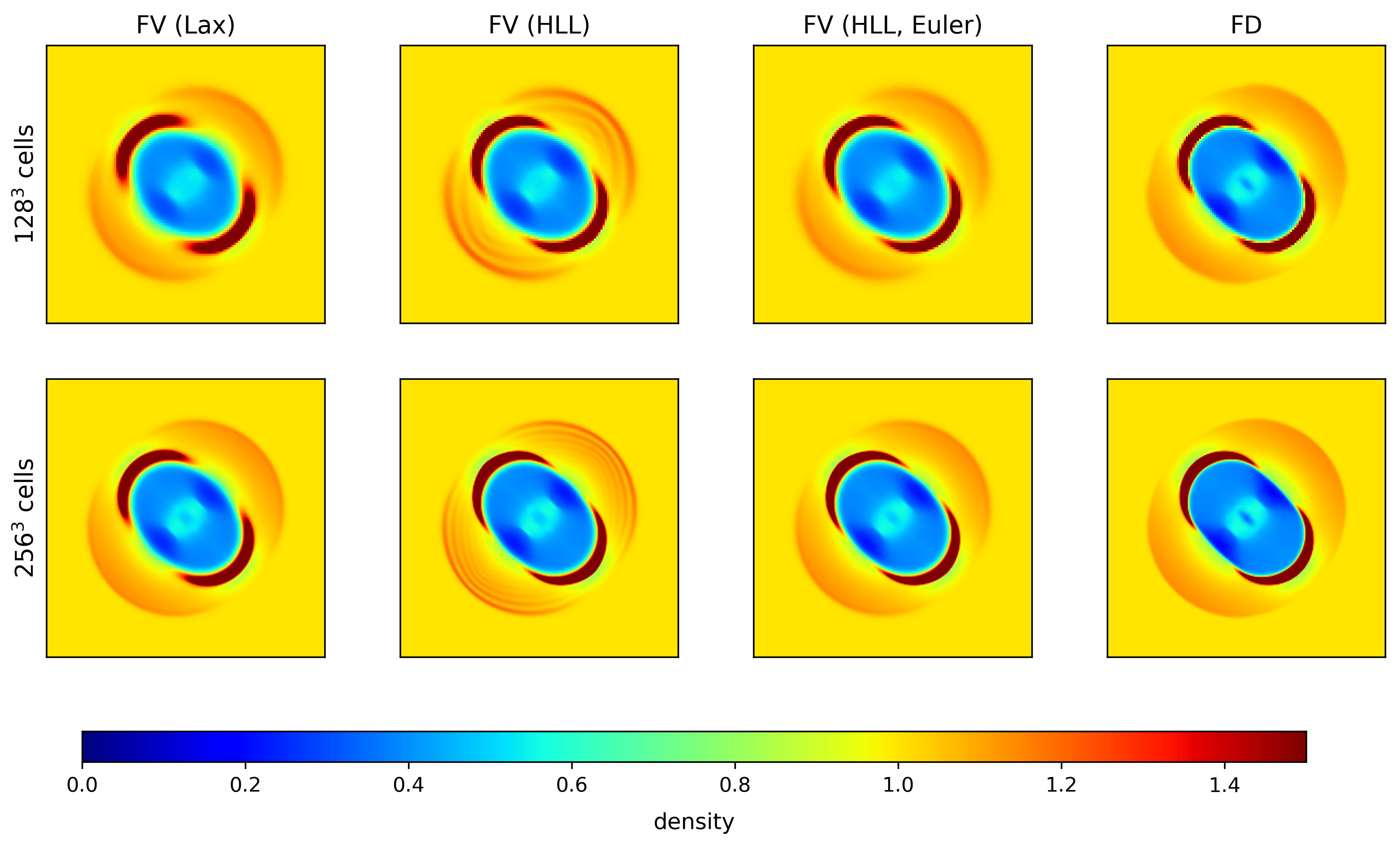}
 \caption{Central density slice of the MHD blast wave at $128^3$ (top) and
  $256^3$ (bottom) cells for the finite-volume Lax and HLL schemes (one with a second-order implicit midpoint magnetic step 
  and one with a first-order implicit Euler magnetic step) and the
  finite-difference scheme. When using an HLL solver, the finite-volume MHD scheme based on \cite{pang24} develops spurious
  post-shock oscillations that sharpen with resolution.}
 \label{fig:mhd_blast_oscillations}
\end{figure}

Fig.~\ref{fig:mhd_blast_oscillations} compares the central density
slice across solver configurations and resolutions. We found that
the \cite{pang24} MHD scheme creates oscillations around shocks in
three dimensions when paired with an HLL (or HLLC) solver in place of
the Lax--Friedrichs solver used in the paper. At the price of using a first-order
implicit Euler method instead of the second-order implicit midpoint method
for the magnetic time step, these oscillations can be removed.

As we want to maintain second-order accuracy in time and as a Lax--Friedrichs
solver is typically too diffusive for astrophysical applications such
as simulating MHD turbulence, we will therefore explore different finite volume MHD approaches
in the future, e.g. re-implementing the Athena methodology \citep{Stone08}, and
use the finite difference scheme as our flagship method.

\subsubsection{3D MHD Jet}
\label{sec:mhd_jet}

\begin{figure}
 \centering
 \includegraphics[width=1.0\textwidth]{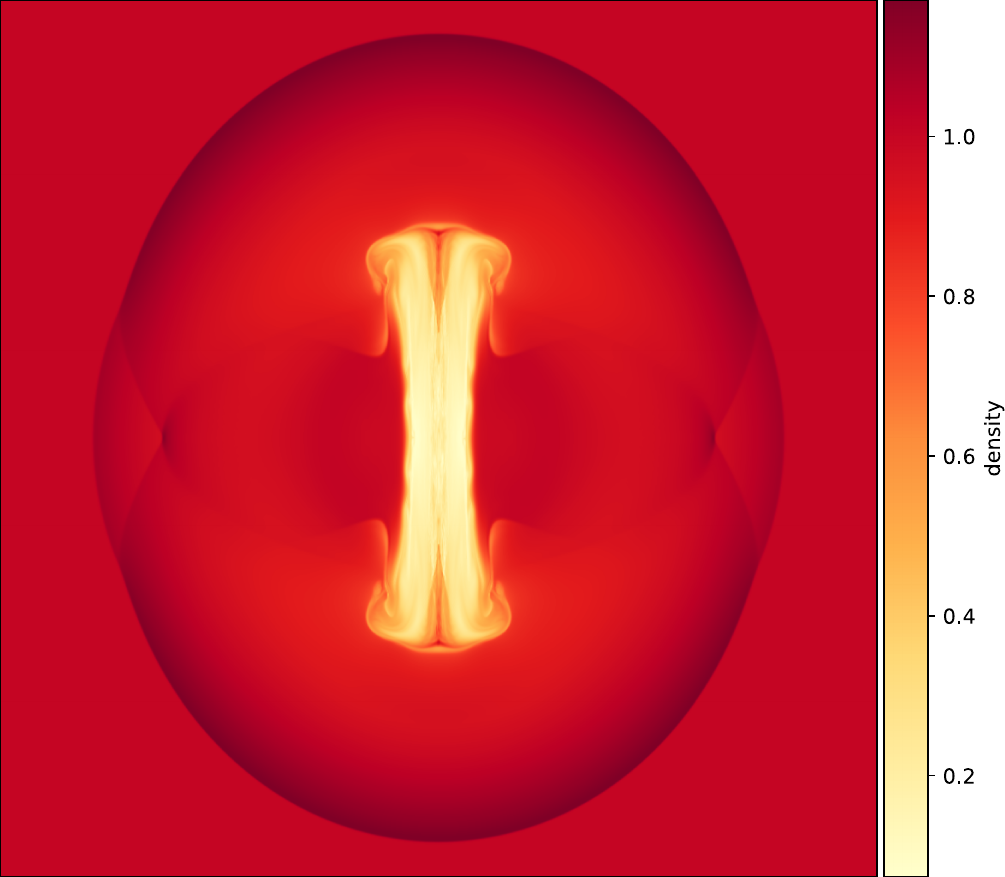}
 \caption{Magnetically driven jet at $t = 5$, density slice through the jet axis,
  computed with the finite-difference scheme at $1024^3$ cells.}
 \label{fig:mhd_jet}
\end{figure}

Next, we demonstrate a more astrophysically motivated setup: a magnetically
driven jet launched from a magnetized central region (cf. Sec. 3.8 in \cite{how_mhd}).
In a box of size $24$ with
open boundaries the medium is initially uniform ($\rho = 1$, $p = 1$, $\vec{v} =
0$), and the magnetic field is initialized from the vector potential
\begin{equation}
  \vec{A} = e^{-r^2}\left(-(y - y_c),\; (x - x_c),\; \tfrac{1}{2}A_0\right),
  \qquad A_0 = 20,
\end{equation}
where $r$ is the distance from the box center $(x_c, y_c, z_c)$. \texttt{astronomix} provides
a convenience function to easily set up the magnetic field from the vector potential.
Fig.~\ref{fig:mhd_jet} shows a
density slice through the jet axis at $t = 5$ obtained with the finite-difference
scheme at $1024^3$ cells. The jet and its surrounding
bow shock are well-resolved.

\subsubsection{3D isothermal MHD turbulence}
\label{sec:iso_turb}

As a final stress test we simulate statistically
stationary, driven isothermal MHD turbulence in a periodic box of size
$L_0 = 1$, following Sec.~3.9 of \cite{how_mhd}. We consider the two
regimes studied there: a supersonic, strongly magnetized ``ISM'' case
($M_\mathrm{turb} \approx 10$, $\beta_p = \frac{p_\mathrm{thermal}}{p_\mathrm{mag}} = 0.1$)
and a subsonic, weakly magnetized ``ICM'' case
($M_\mathrm{turb} \approx 0.5$, $\beta_p = 10^{6}$). The ISM case is evolved for five turbulent crossing times
and the ICM case for thirty, matching the durations from \cite{how_mhd}. We complement their $256^3$ runs with
$512^3$ runs and discuss challenges in stability and possible solutions. While
\cite{how_mhd} use white-in-time noise to drive the turbulence, we use the now
more standard Ornstein--Uhlenbeck process.

\paragraph{Ornstein--Uhlenbeck driving.}
The turbulence is sustained by a solenoidal, large-scale acceleration field
$\vec{f}(\vec{x}, t)$. Each forcing realization is built in Fourier
space from a Helmholtz-projected Gaussian field,
\begin{equation}
  \hat{\vec{w}}(\vec{k}) =
  \big(\mathbb{1} - \hat{\vec{k}}\,\hat{\vec{k}}^{\mathsf{T}}\big)\,
  \sqrt{P(k)}\;\vec{\eta}(\vec{k}),
  \qquad
  P(k) = k^{6}\,\exp\!\big(-8\,k / k_\mathrm{pk}\big),
\end{equation}
where $\vec{\eta}$ is unit-variance complex white noise, the projector
$(\mathbb{1} - \hat{\vec{k}}\hat{\vec{k}}^{\mathsf{T}})$ removes the
compressible (curl-free) component so that $\nabla\!\cdot\!\vec{w} = 0$, and the
$\vec{k} = 0$ mode is set to zero. The injection spectrum $P(k)$ matches that of
\cite{how_mhd} and peaks at $k_f = 0.75\,k_\mathrm{pk}$; we set
$k_\mathrm{pk} = 4\pi/L_0$, so the driving peaks at $k_f = 3\pi/L_0$,
corresponding to a peak forcing wavelength of $2\pi/k_f = 2L_0/3$. As the
injection scale we adopt $L_\mathrm{inj} = 2\pi/k_\mathrm{pk} = L_0/2$.
Each draw is normalized to unit rms,
$\langle |\vec{w}|^2 \rangle = 1$.

The persistent forcing field is advanced once per timestep with the exact
discretization of an OU process,
\begin{equation}
  \vec{f} \;\leftarrow\; \alpha\,\vec{f}
  + \sqrt{1 - \alpha^{2}}\;\vec{\xi},
  \qquad
  \alpha = \exp\!\big(-\Delta t / \tau_f\big),
  \label{eq:ou_update}
\end{equation}
where $\vec{\xi}$ is an independent unit-rms solenoidal draw with spectrum
$P(k)$. Equation~\eqref{eq:ou_update} keeps the field at unit rms and yields an
exponentially correlated acceleration,
$\langle \vec{f}(t)\!\cdot\!\vec{f}(t') \rangle \propto
\exp(-|t - t'|/\tau_f)$, with correlation time $\tau_f$ (we use
$\tau_f = 0.5$). The field is applied as a constant-amplitude acceleration,
\begin{equation}
  \vec{v} \;\leftarrow\; \vec{v} + F_0\,\vec{f}\,\Delta t,
\end{equation}
so that the forcing has fixed rms amplitude $F_0$ and the energy injection
adapts self-consistently to the flow rather than being prescribed. We choose
$F_0 = 3.5$, which drives a steady $v_\mathrm{rms} \approx 1$. Because the
acceleration is independent of the fluid state, the realization is exactly
reproducible for a given timestep sequence and contributes trivially to the
tangent/adjoint, preserving the end-to-end differentiability of
\texttt{astronomix}. The OU driving is parameterized by $(F_0, \tau_f, k_f)$ and
is statistically equivalent to the white-in-time prescription of \cite{how_mhd} for the diagnostics
considered here. 

\paragraph{Normalization.}
We set $\rho_0 = 1$ and the isothermal sound speed $a = 1/M_\mathrm{turb}$, so
that the driven $v_\mathrm{rms} \approx 1$ gives the target Mach number
$M_\mathrm{turb} = v_\mathrm{rms}/a$. The uniform guide field
$\vec{B}_0 = B_0\,\hat{\vec{z}}$ follows from the plasma beta,
$B_0 = \sqrt{2 P_\mathrm{th}/\beta_p}$ with $P_\mathrm{th} = a^2 \rho_0$ the
isothermal thermal pressure. The crossing time is
$t_\mathrm{cross} = L_\mathrm{inj}/v_\mathrm{rms}$.

\paragraph{Stabilization mechanisms.}
\cite{how_mhd} do not describe measures to stabilize their turbulent
simulations, yet two are present in their Fortran code, which we adopt: a
density floor $\rho_\mathrm{min}$, applied as a read-only clamp inside the
characteristic eigen-decomposition, and the neighbor-redistribution
step of Sec.~\ref{sec:stability}, there applied once per step after the forcing update.
For the turbulence runs we use $\rho_\mathrm{min} = 0.02$ and $v_\mathrm{max} = 50$.

For most runs these suffice, but for the $512^3$ high-Mach cases they do not: a
late-time blow-up develops in the supersonic ISM run ($M_\mathrm{turb} \approx
10$, $\beta_p = 0.1$), where strong rarefactions carve deep voids in which the
density sits pinned at $\rho_\mathrm{min}$ and the recovered velocity
$\vec{m}/\rho_\mathrm{min}$ spikes to hypersonic values that the high-order
reconstruction cannot tolerate. We therefore additionally activate the deep-void
measures of Sec.~\ref{sec:stability}: the deep-void flux blending
($w_\mathrm{void}$, blend factor $\beta = 8$) and the vacuum-rest option, both
evaluated within every Runge--Kutta substage. Different from 
\cite{how_mhd} we also perform the neighbor redistribution at every substage.
With these measures the $M_\mathrm{turb}
\approx 10$, $\beta_p = 0.1$ run remains stable for the full five crossing times, 
tested up to $512^3$ cells.

\begin{figure}
 \centering
 \includegraphics[width=1.0\textwidth]{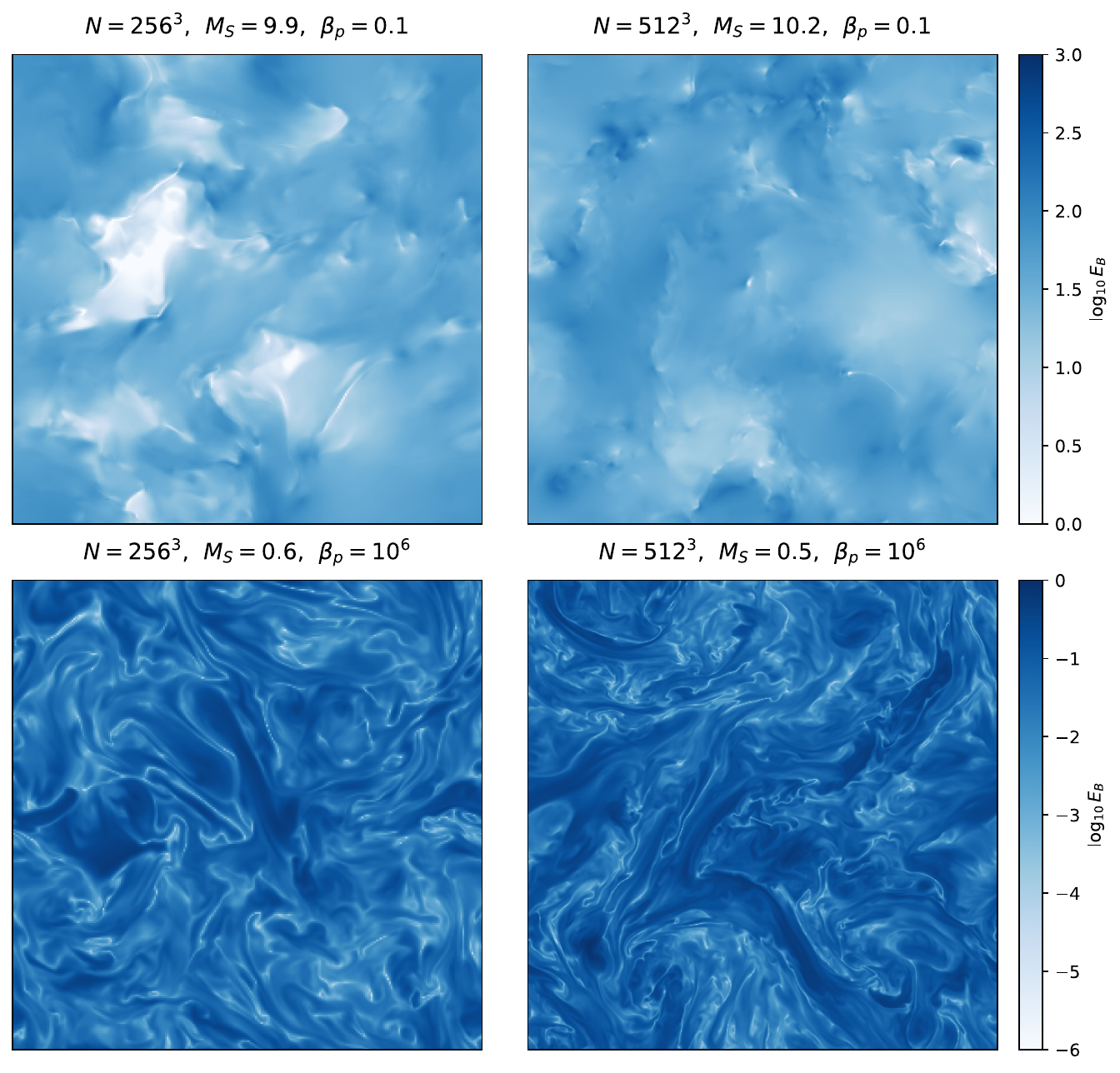}
 \caption{Driven isothermal-MHD turbulence: mid-plane slices of the magnetic
  energy $\log_{10} E_B$ for the supersonic ISM regime ($M_\mathrm{turb}\approx10$,
  $\beta_p=0.1$; top) and the subsonic ICM regime ($M_\mathrm{turb}\approx0.5$,
  $\beta_p=10^{6}$; bottom), each at $256^3$ (left) and $512^3$ (right) cells,
  computed with the finite-difference scheme. The
  ISM case is shock-dominated with deep voids; the ICM case shows flux
  ribbons amplified by a small-scale dynamo.}
 \label{fig:turbulence_slices}
\end{figure}

\begin{figure}
 \centering
 \includegraphics[width=1.0\textwidth]{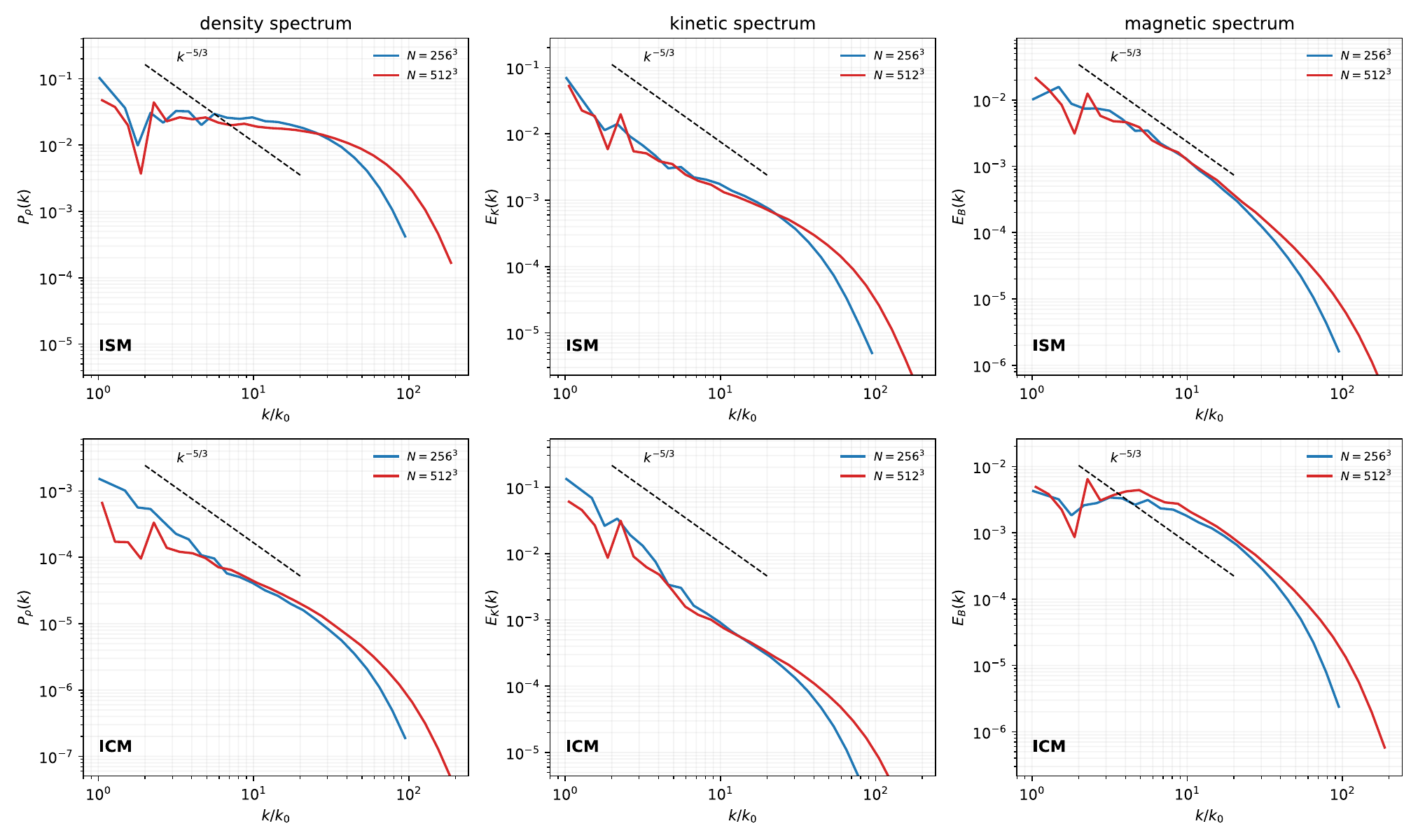}
 \caption{Saturated power spectra of the driven isothermal-MHD turbulence,
  time-averaged over the stationary phase: density $P_\rho(k)$ (left), kinetic
  $E_K(k)$ (middle), and magnetic $E_B(k)$ (right) for the ISM (top) and ICM
  (bottom) regimes at $256^3$ and $512^3$ cells. Dashed lines mark the $k^{-5/3}$
  slope. The inertial ranges extend with
  resolution; the supersonic ISM density spectrum is flatter than the Kolmogorov
  reference, consistent with highly compressible turbulence.}
 \label{fig:turbulence_spectra}
\end{figure}

\paragraph{Results.}
Fig.~\ref{fig:turbulence_slices} shows mid-plane slices of the magnetic
energy for the two regimes at $256^3$ and $512^3$. The supersonic ISM case is shock-dominated,
whereas the subsonic ICM is filled with smooth magnetic flux ribbons. Both
morphologies become more intricate with resolution. In
the ICM case the magnetic energy, seeded at $E_B \sim 10^{-6}$, is amplified by
several orders of magnitude through a small-scale dynamo. For the supersonic ISM
case it would also be interesting to look at dynamo effects in longer-duration
runs; here we retained the \cite{how_mhd} setup. Fig.~\ref{fig:turbulence_spectra}
shows the corresponding power spectra, time-averaged over the statistically
stationary phase. In the subsonic ICM case the kinetic and magnetic spectra
develop inertial ranges broadly consistent with the Kolmogorov $k^{-5/3}$
reference. Increasing the resolution from $256^3$ to $512^3$
extends the inertial range to higher wavenumbers.

\subsection{High-order self-gravity with improved energy conservation}
\label{sec:self_gravity}

There has been significant numerical work on properly coupling hydrodynamics and
self-gravity \citep{mullen21,muller95,hanawa25,jiang13}. For instance, \citet{hanawa25}
have recently proposed a fourth-order self-gravity scheme for finite volume
hydrodynamics with linear momentum conservation. However, integrating such a
nonlinear source term into a finite volume scheme with high-order accuracy requires
complicated correction terms, and ensuring numerical stability can be challenging.
By nature, finite difference schemes such as the one implemented in \texttt{astronomix}
are better suited for the integration of nonlinear source terms and we
present a novel fourth-order accurate self-gravity scheme with improved energy conservation.
The gravitational energy flux is built from the WENO
mass flux and made high-order through an explicit deconvolution correction
(Sec.~\ref{sec:self_gravity_scheme}), yielding \emph{semi-discrete} (time-continuous)
energy conservation. In contrast to the
momentum-conserving finite-volume schemes of \citet{mullen21} and \citet{hanawa25}, we
focus on energy conservation here and defer a locally flux-conservative momentum coupling
to future work.

The Euler equations with self-gravity are given by

\begin{equation}
    \begin{gathered}
        \partial_t \rho + \vec{\nabla} \cdot (\rho \vec{v}) = 0  \\
        \partial_t (\rho \vec{v}) + \vec{\nabla} \cdot (\rho \vec{v}\vec{v}^T + P\,\mathbb{I}) = \rho \vec{g} \\
        \partial_t (\rho e) + \vec{\nabla} \cdot \left[(\rho e + P)\vec{v} \right] = \rho \vec{v} \cdot \vec{g} \\ \\
        \nabla^2 \Phi = 4 \pi G \rho \, (\text{Poisson equation}), \quad \vec{g} = -\vec{\nabla} \Phi \rightarrow \vec{\nabla} \times \vec{g} =0,
    \end{gathered}
    \label{eq:euler_grav}
\end{equation}

where, as before, $e = e_{th} + \frac{1}{2} \vec{v}^2$ with closure $P = (\gamma - 1) \rho e_{th}$. Let us define

\begin{equation}
    S_E := \rho \vec{v} \cdot \vec{g}, \quad S_m := \rho \vec{g}.
\end{equation}

The gravitational energy is given by

\begin{equation}
    E_\mathrm{grav} = \frac{1}{2} \int \rho \Phi \, \dif V,
\end{equation}

such that the total energy of the system reads

\begin{equation}
    E_{\mathrm{tot}} = \int \rho e \dif V + E_\mathrm{grav}.
\end{equation}

When the density changes with $\partial_t \rho = \dot{\rho}$, then the gravitational energy changes with

\begin{equation}
    \partial_t E_{\mathrm{grav}} = \int \dot{\rho} \Phi \, \dif V.
\end{equation}

A proof is given in Sec. \ref{app:gravitational_energy}.

The continuous energy source term fulfills energy conservation

\begin{equation}
    \label{eq:source_energy_conservation}
    \begin{aligned}
        \int S_E \, \dif V &= \int \rho \vec{v} \cdot \vec{g} \, \dif V \\
                           &= -\int \rho \vec{v} \cdot \vec{\nabla} \Phi \, \dif V \\
                           &= -\oint \Phi \, \rho \vec{v} \cdot \dif \vec{A}
                              + \int \Phi \, \vec{\nabla} \cdot \left( \rho \vec{v} \right) \dif V \\
                           &= -\int \Phi \, \dot{\rho} \, \dif V \\
                           &= -\partial_t E_{\mathrm{grav}},
    \end{aligned}
\end{equation}

where the surface term vanishes because the density falls off sufficiently fast
(or there is no mass flux through the boundary), and in the second-to-last step we
used the continuity equation $\dot{\rho} = -\vec{\nabla} \cdot \left( \rho \vec{v} \right)$.

For the discrete approximation

\begin{equation}
    E_\mathrm{grav,d} = \frac{1}{2} \sum_{i} V_i \rho_i \Phi_i = \frac{1}{2} \sum_{ij} V_i V_j \rho_i K_{ij} \rho_j,
\end{equation}

with Green's kernel $K_{ij}$, the corresponding change is

\begin{equation}
    \partial_t E_\mathrm{grav,d} = \sum_{i} V_i \dot{\rho}_i \Phi_i.
\end{equation}

Regarding energy conservation, our aim is to find an energy source term $S_{E,i}$ such that

\begin{equation}
    \sum_i V_i S_{E,i} = -\partial_t E_\mathrm{grav,d},
\end{equation}

ensuring energy conservation at the semi-discrete level, i.e. with time kept continuous. The application of a Runge-Kutta scheme
will still yield an energy error, as $E_\mathrm{grav,d}$ is quadratic in $\rho$
and explicit Runge-Kutta schemes do not exactly conserve quadratic invariants. One can easily see this
for an Euler step

\begin{equation}
    \begin{aligned}
        \Delta E_\mathrm{grav,d} &= \frac{1}{2} \sum_{ij} V_i V_j (\rho_i + \Delta t \dot{\rho}_i) K_{ij} (\rho_j + \Delta t \dot{\rho}_j) - \frac{1}{2} \sum_{ij} V_i V_j \rho_i K_{ij} \rho_j \\
                                 &= \Delta t \sum_{ij} V_i V_j \dot{\rho}_i K_{ij} \rho_j + \frac{1}{2} \Delta t^2 \sum_{ij} V_i V_j \dot{\rho}_i K_{ij} \dot{\rho}_j \\
                                 &= \Delta t \, \partial_t E_\mathrm{grav,d} + \mathcal{O}(\Delta t^2).
    \end{aligned}
\end{equation}

This is a second-order error per Euler step, yielding a first-order error $\mathcal{O}(\Delta t)$
for the first-order Euler scheme globally. Similarly, the fourth-order Runge-Kutta
scheme actually applied for time integration yields a global energy error of
$\mathcal{O}(\Delta t^4)$. 

Note that Eq. \ref{eq:source_energy_conservation} does not hold discretely, essentially
because the mass flux applied by the numerical scheme (which determines $\partial_t \rho$)
contains not only the bulk flux $\rho \vec{v}$ but also dissipative contributions; these
also do work against the potential, which a simple source term scheme misses.

To incorporate self-gravity into \texttt{astronomix} we have to cover the following two aspects

\begin{itemize}
    \item solve the Poisson problem $\nabla^2 \Phi = 4 \pi G \rho$ to obtain the potential from the density field
    \item couple the gravitational potential and the hydrodynamics
\end{itemize}

Both multigrid methods and fast Fourier methods can be used for solving the Poisson problem \citep{tomida23}. We resort to Fast Fourier methods, as Fast Fourier Transforms are already efficiently implemented in \texttt{JAX}. In Fourier space, solving the Poisson equation becomes a simple multiplication. We need to transform the density field to Fourier space and the solution back. In $d$ dimensions and for a grid with $N$ grid points per dimension,
the cost of the Fast Fourier Transform (FFT) scales with $\mathcal{O}(N^d \log N)$. Fast Fourier methods are most efficient for periodic boundaries and are not suited for simulations with adaptive mesh refinement \citep{tomida23}. Open boundaries can be realized via the Hockney--Eastwood method \citep[see][]{eastwood79} or via James's method \citep{james77}.

For fully periodic problems (the Jeans and slab tests
below) we use the periodic spectral Green's function, applying the Jeans swindle
by subtracting the mean density and setting the $\vec{k} = 0$ mode to zero. The
Evrard collapse is an isolated system; there we use the Hockney--Eastwood open-boundary
method \citep{eastwood79}, zero-padding the density to twice the domain size along
each axis and convolving with the isolated-system Green's function, so that periodic
image masses do not contaminate the solution.

For the finite volume scheme, the Fourier solve returns the cell-averaged potential $\bar{\Phi}_{i,j,k}$
accurate to the given cell-averaged density field $\bar{\rho}_{i,j,k}$, as averaging commutes
with $\nabla^2$.\footnote{The box-filter that maps point values to cell averages is a convolution
and hence diagonal in Fourier space, so it commutes with the (also Fourier-diagonal) inverse Laplacian.} For the finite
difference scheme, the Fourier solve returns the cell-centered
potential $\Phi_{i,j,k}$ accurate to the cell-centered density field $\rho_{i,j,k}$.

We now turn to the coupling of the gravitational potential to the hydrodynamics. An ideal numerical coupling scheme would

\begin{itemize}
    \item conserve energy and linear and angular momentum
    \item have zero rotation of the gravitational acceleration field
    \item be numerically stable, ideally positivity preserving
    \item have the same spatial order as the underlying hydrodynamical scheme
\end{itemize}

The self-gravity coupling variants we consider here are all implemented as parts
of the $\Call{ModuleSources}{Q, (F_x, F_y, F_z)}$ function in 
Algorithm \ref{algs:astronomix}, handled as right-hand side terms
in the main Runge-Kutta time integration scheme. In practice one might
not need to recompute the potential at every Runge-Kutta stage, but we
do not consider such optimizations here. Also, all coupling variants
only couple to the momentum and energy equations,
not the continuity equation.

\subsubsection{Simple source term scheme}

Let us start with the simplest discretization of the gravitational source terms for the 
finite difference scheme. From the fifth-order accurate cell-centered potential $\Phi_{i,j,k}$ 
we obtain a fifth-order cell-centered acceleration via sixth-order finite differencing

\begin{equation}
    \label{eq:grav_acceleration}
    (g_x)_{i,j,k}
=
-\frac{1}{60\,\Delta x}
\left(
\Phi_{i+3,j,k}
-9\Phi_{i+2,j,k}
+45\Phi_{i+1,j,k}
-45\Phi_{i-1,j,k}
+9\Phi_{i-2,j,k}
-\Phi_{i-3,j,k}
\right)
\end{equation}

($g_y$ and $g_z$ analogously) and can then evaluate the source terms via

\begin{equation}
    S_m = (\rho \vec{g})_{i,j,k} = \rho_{i,j,k} \vec{g}_{i,j,k}, \quad S_E = (\rho \vec{v} \cdot \vec{g})_{i,j,k} = \rho_{i,j,k} \vec{v}_{i,j,k} \cdot \vec{g}_{i,j,k}.
\end{equation}

As we are dealing with point values in a finite difference scheme, the multiplication 
retains fifth-order accuracy. This is different from a finite volume scheme, 
where $\langle \rho \vec{g}\rangle \neq \langle \rho \rangle \langle \vec{g} \rangle$ and 
correction terms are necessary for high-order accuracy \citep{hanawa25}.

While the time discretization yields a spurious coupling of the
gravitational source terms to the pressure (note that in the primitive equations gravity only couples to the velocity),
it is of the order of the time integration scheme, limiting the risk of negative pressures.

Additionally, the simple $\vec{g}_{i,j,k}$ in the momentum source is curl-free with respect to
the discrete curl operator built from the same sixth-order central differences as
Eq. \ref{eq:grav_acceleration}. This momentum source is shared by all three coupling
schemes discussed below, which differ only in the energy source. Note that global (total) linear
momentum conservation generally holds for periodic boundary conditions: for a symmetric Poisson
solve and antisymmetric central difference acceleration calculation, the net self-force vanishes.

\subsubsection{Fourth-order scheme with semi-discrete (time-continuous) energy conservation}
\label{sec:self_gravity_scheme}

The deficiency of the simple source term scheme is that it couples to the bulk flux
$\rho \vec{v}$ rather than the actual numerical flux, which also includes dissipative fluxes.

Consider the energy source term

\begin{equation}
\begin{aligned}
S_E &= \rho \vec{v} \cdot \vec{g} \\
    &= -\rho \vec{v} \cdot \vec{\nabla} \Phi \\
    &= -\vec{\nabla} \cdot \left(\rho \vec{v} \Phi \right) + \Phi \vec{\nabla} \cdot \left( \rho \vec{v} \right) \\
\end{aligned}
\end{equation}

now apply the continuity equation $\vec{\nabla} \cdot \left( \rho \vec{v} \right) = -\partial_t \rho$
to obtain

\begin{equation}
\begin{aligned}
S_E &= -\vec{\nabla} \cdot \left(\rho \vec{v} \Phi \right) - \Phi \partial_t \rho \\
\end{aligned}
\end{equation}

The term $\Phi \partial_t \rho$ can be evaluated to high order based on $\Phi$
from the Poisson solve and $\partial_t \rho$ from the FD scheme (note that there
is no source term in the density equation). $\vec{\nabla} \cdot \left(\rho \vec{v} \Phi \right)$
is more difficult to properly handle. One could integrate it directly into the WENO flux,
but this would be tedious.

For discrete energy conservation, we need the expression for
$\left(\vec{\nabla} \cdot \left(\rho \vec{v} \Phi \right)\right)_i$ to telescope out
when summing (as $\Phi \partial_t \rho$ guarantees energy conservation). Therefore,
the form of the discretization must be (shown for the $x$-term, with mass flux $f = \rho v_x$)

\begin{equation}
\left(\partial_x \left(f \Phi \right)\right)_i = \frac{q_{i+\frac{1}{2}} - q_{i-\frac{1}{2}}}{\Delta x},
\end{equation}

with a numerical flux $q_{i+\frac{1}{2}}$ for $q = \Phi f$ that is single-valued at each face. The classical approach

\begin{equation}
\left(\partial_x \left(f \Phi \right)\right)_i \approx \frac{\Phi_{i+\frac{1}{2}}\hat{f}_{i+\frac{1}{2}} - \Phi_{i-\frac{1}{2}}\hat{f}_{i-\frac{1}{2}}}{\Delta x},
\end{equation}

where $\hat{f}_{i\pm\frac{1}{2}}$ is the numerical mass flux of the underlying WENO scheme,
the same flux which determines $(\partial_t \rho)_i$ (accounting for dissipative flux contributions),
and the face potential is obtained from the cell-centered potential
by the sixth-order symmetric interpolation

\begin{equation}
    \label{eq:phi_face_interpolation}
    \Phi_{i+\frac{1}{2}} = \frac{1}{256}\left[ 3\left(\Phi_{i-2} + \Phi_{i+3}\right) - 25\left(\Phi_{i-1} + \Phi_{i+2}\right) + 150\left(\Phi_{i} + \Phi_{i+1}\right) \right],
\end{equation}

telescopes exactly and thus conserves energy, but is not high-order. 
To understand why the approach is not high-order,
we must go back a bit to understand how high-order flux difference formulations
work, i.e. how something of the form
$\frac{\hat{f}_{i+\frac{1}{2}} - \hat{f}_{i-\frac{1}{2}}}{\Delta x}$ can be a high-order
approximation for $f'(x_i)$. To see this, define the box-filter convolution

\begin{equation}
    \bar{f}(x) = \frac{1}{\Delta x} \int_{x - \frac{\Delta x}{2}}^{x + \frac{\Delta x}{2}} f(\xi) \, d\xi = f + \frac{\Delta x^2}{24} f'' + \mathcal{O}(\Delta x^4),
\end{equation}

and the deconvolution $\tilde{f}$ such that $\bar{\tilde{f}} = f$, formally only asymptotically defined via

\begin{equation}
    \tilde{f} = f - \frac{\Delta x^2}{24} f'' + \mathcal{O}(\Delta x^4).
\end{equation}

We can then write

\begin{equation}
    f(x) = \frac{1}{\Delta x} \int_{x - \frac{\Delta x}{2}}^{x + \frac{\Delta x}{2}} \tilde{f}(\xi) \, d\xi.
\end{equation}

Differentiation of both sides yields

\begin{equation}
    f'(x) = \frac{1}{\Delta x} \left( \tilde{f}\left(x+\frac{\Delta x}{2}\right) - \tilde{f}\left(x-\frac{\Delta x}{2}\right) \right).
\end{equation}

Hence, for a high-order finite difference flux formula, the face values we use must
actually be high-order face values of the deconvolved flux function $\tilde{f}$, so

\begin{equation}
    \hat{f}_{i\pm\frac{1}{2}} = \tilde{f}\left(x_i\pm\frac{\Delta x}{2}\right).
\end{equation}

However, note that the order of $\hat{f}$ automatically degrades at discontinuities, so the
argument we make here is cleanest for smooth flows.

But while $\hat{f}_{i\pm\frac{1}{2}}$ is the high-order face evaluation of the
deconvolution of $f$, $\Phi_{i+\frac{1}{2}}\hat{f}_{i+\frac{1}{2}}$ is not the high-order
face evaluation of the deconvolved product $\widetilde{\Phi f}$. Deconvolution does not
commute with multiplication, $\Phi \tilde{f} \neq \widetilde{\Phi f}$.

To restore high-order we might either compute $\widehat{(\Phi f)}_{i\pm \frac{1}{2}}$
with the same reconstruction scheme we use for the hydrodynamics or calculate an explicit
correction term

\begin{equation}
\begin{aligned}
    \hat{q}_{i+\frac{1}{2}} &= q_{i+\frac{1}{2}} - \frac{\Delta x^2}{24} q''_{i+\frac{1}{2}} + \mathcal{O}(\Delta x^4) \\
                            &= \Phi_{i+\frac{1}{2}}\left(\hat{f}_{i+\frac{1}{2}} + \frac{\Delta x^2}{24} \hat{f}''_{i+\frac{1}{2}}\right) - \frac{\Delta x^2}{24} \left( \Phi''_{i+\frac{1}{2}} f_{i+\frac{1}{2}} + 2f'_{i+\frac{1}{2}} \Phi'_{i+\frac{1}{2}} + \Phi_{i+\frac{1}{2}} f''_{i+\frac{1}{2}} \right) + \mathcal{O}(\Delta x^4) \\
                            &= \Phi_{i+\frac{1}{2}} \hat{f}_{i+\frac{1}{2}} - \frac{\Delta x^2}{24} \left( \Phi''_{i+\frac{1}{2}} \hat{f}_{i+\frac{1}{2}} + 2\hat{f}'_{i+\frac{1}{2}} \Phi'_{i+\frac{1}{2}} \right) + \mathcal{O}(\Delta x^4),
\end{aligned}
\end{equation}

where $f$ and $f'$ in the correction terms could be replaced by $\hat{f}$ and $\hat{f}'$,
since $f = \hat{f} + \mathcal{O}(\Delta x^{2})$ and the corrections are scaled with $\Delta x^2$, such
that fourth-order accuracy is retained. In practice, we evaluate the correction term
at the cell centers -- with $\Phi'$ from the sixth-order gradient of Eq. \ref{eq:grav_acceleration}
and $\Phi''$ and $f'$ from the second-order centered differences
$\Phi''_i = (\Phi_{i+1} - 2\Phi_i + \Phi_{i-1})/\Delta x^2$ and
$f'_i = (f_{i+1} - f_{i-1})/(2 \Delta x)$ on the point values $f_i = (\rho v_x)_i$ --
and average it to the faces, $(\cdot)_{i+\frac{1}{2}} = \frac{1}{2}\left[(\cdot)_i + (\cdot)_{i+1}\right]$.
As the correction term is already scaled with $\Delta x^2$, these second-order evaluations
retain the overall fourth-order accuracy.

Note that the correction term does not affect the semi-discrete energy conservation argument.

In the following, we will call 

\begin{itemize}
    \item the simple source term scheme, \textbf{simple source}
    \item the time-continuous energy-conserving scheme without high-order correction, \textbf{flux-based}
    \item the time-continuous energy-conserving scheme with high-order correction, \textbf{flux-based corrected}
\end{itemize}

\subsubsection{Jeans linear waves test}

Consider a small-amplitude linear wave such that an analytical
solution to the linearized Euler equations with self-gravity
can be found and is valid (see Sec. \ref{app:jeans}).

This analytical solution is given by

\begin{equation}
    \begin{aligned}
        \rho &= \rho_B + \rho_B \epsilon \sin{\left( \vec{k} \cdot \vec{x} - \omega t\right)} \\
        \vec{v} &= \frac{\epsilon \omega \vec{k}}{k^2} \sin{\left(\vec{k} \cdot \vec{x} - \omega t\right)} \\
        P &= \frac{c_s^2 \rho_B}{\gamma} + c_s^2 \rho_B \epsilon \sin{\left( \vec{k} \cdot \vec{x} - \omega t\right)} \\
        \omega &= \sqrt{k^2 c_s^2 - 4 \pi G \rho_B},
    \end{aligned}
\end{equation}

where for $\rho_B = 1, c_s^2 = 1, \gamma = \frac{5}{3}, 4\pi G = 1, \epsilon = 10^{-6}$,
$\vec{k} = (2,4,4)^T$ and a box size $[L_x, L_y, L_z] = 2\pi [k_x^{-1}, k_y^{-1}, k_z^{-1}]$
we obtain the setup of \citet{hanawa25}, which we also adopt here. The computational volume is
resolved with $N_x \times N_y \times N_z$ cells where $N_y = N_z = \frac{N_x}{2}$. The boundary conditions 
are periodic. The simulation is initialized from the analytical solution at $t = 0$.

\begin{figure}
 \centering
 \includegraphics[width=1.0\textwidth]{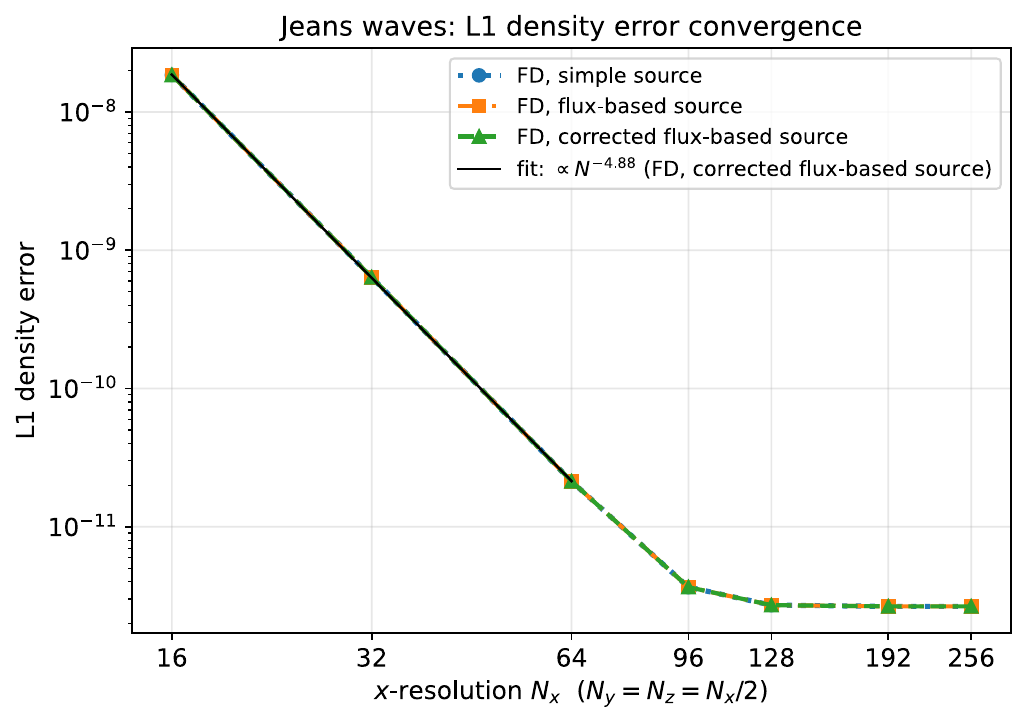}
 \caption{Convergence of the $L_1$ density error in the Jeans linear wave test for the
 three self-gravity coupling schemes. The near-identical high-order convergence of the
 uncorrected flux-based scheme here is a small-amplitude artifact (its correction term is
 second order in the perturbation and sits below the round-off floor); the schemes are
 separated by the larger-amplitude slab test of Fig.~\ref{fig:slab_convergence}.}
 \label{fig:jeans_convergence}
\end{figure}

The double-precision $L_1$ density error

\begin{equation}
    L_1 = \frac{1}{N_x N_y N_z} \sum_{ijk} \left| \rho_{ijk}(T) - \rho_{ijk}(0) \right|,
\end{equation}

measured as a function of $N_x$ at $t = \frac{2\pi}{\omega}$
for the different schemes is presented in Fig. \ref{fig:jeans_convergence}. For all schemes, the $L_1$ error converges
approximately at fifth order. The uncorrected flux-based scheme here also converges at
high order, as the correction term is second order in the perturbation amplitude and
therefore below the accumulated round-off error. Therefore, a test with larger perturbation amplitude is necessary.

\subsubsection{Slab advection test}

This test covers the advection of a large-amplitude
equilibrium slab, following \citet{hanawa25}.

The density field is given by
\begin{equation}
    \rho(\vec{x}) = \rho_0 \left[ 1 + \epsilon \cos\!\left(\vec{k}\cdot\vec{x}\right) + \frac{\epsilon^2}{3} \cos\!\left(2\,\vec{k}\cdot\vec{x}\right) \right],
\end{equation}
together with a uniform velocity field $\vec{v} = \vec{v}_0$ and
the pressure profile
\begin{equation}
\begin{aligned}
    P(\vec{x}) = p_0 + \frac{4 \pi G \epsilon \rho_0^2}{|\vec{k}|^2}
    \Bigg[
        & \left(1 - \frac{\epsilon^2}{12}\right) \cos\!\left(\vec{k}\cdot\vec{x}\right)
        + \frac{\epsilon}{3}\, \cos\!\left(2\,\vec{k}\cdot\vec{x}\right) \\
        & + \frac{\epsilon^2}{12}\, \cos\!\left(3\,\vec{k}\cdot\vec{x}\right)
        + \frac{\epsilon^3}{144}\, \cos\!\left(4\,\vec{k}\cdot\vec{x}\right)
    \Bigg].
\end{aligned}
\end{equation}
The equilibrium slab returns to its initial
state after $T = 2\pi / (\vec{k}\cdot\vec{v}_0)$.

\begin{figure}
 \centering
 \includegraphics[width=1.0\textwidth]{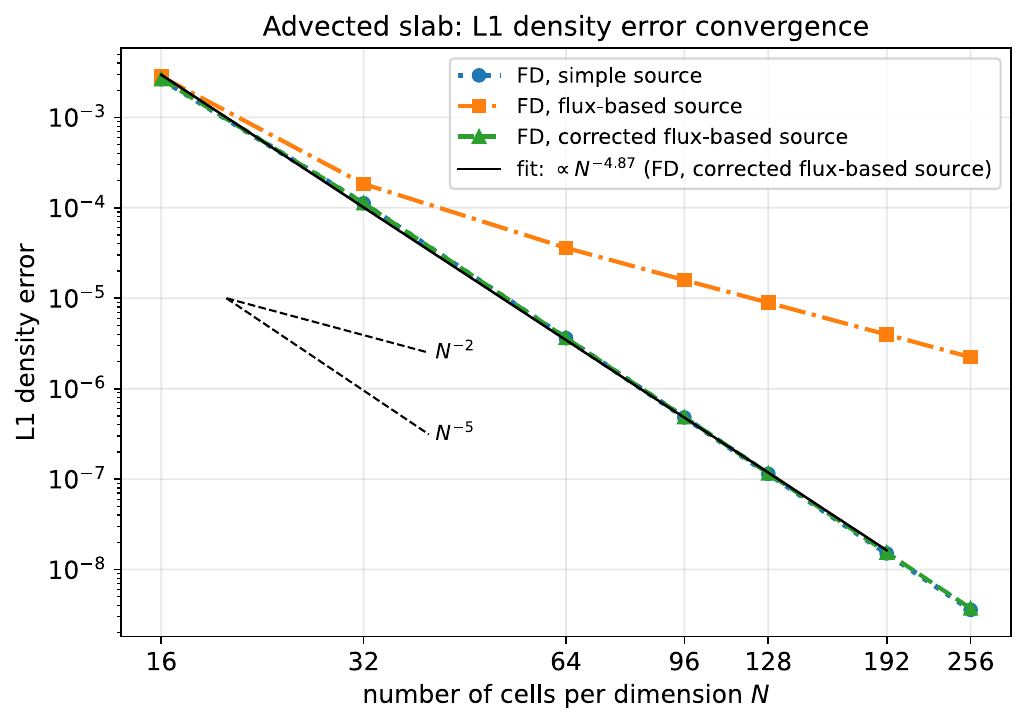}
 \caption{Convergence of the $L_1$ density error in the slab advection test for the three self-gravity coupling schemes.}
 \label{fig:slab_convergence}
\end{figure}

We adopt $G = 1/(4\pi)$, $\rho_0 = 1$, $p_0 = 6$,
$\epsilon = 0.3$, an adiabatic index $\gamma = 5/3$, a wave vector
$\vec{k} = (2/3,\, 2/3,\, 2/3)^T$, and an advection velocity
$\vec{v}_0 = (0.6,\, 0.6,\, 0.6)^T$, such that the slab moves along
the box diagonal. The cubic domain has a side length of
$L = 2\pi \max_i (1/k_i) = 3\pi$ with fully periodic boundaries. The
simulations are evolved for one period $T$ at resolutions of
$N = 16,\, 32,\, 64,\, 96$ cells per dimension, and we report the
$L_1$ error of the density, comparing the different source term formulations
in Fig. \ref{fig:slab_convergence}. The simple source term
scheme and the corrected flux-based scheme show fifth-order
convergence while, as expected, the uncorrected flux-based scheme
only converges at second order. Note that although the correction
term is only derived to $\mathcal{O}(\Delta x^4)$, the remaining
fourth-order error evidently carries a small enough constant to
stay below the fifth-order error of the underlying hydrodynamics
scheme at the resolutions considered, such that we observe
fifth-order convergence, similar to the observed fifth order
in the MHD scheme.

\subsubsection{Energy conservation test in Evrard's collapse}

Next we consider a challenging gravitational collapse problem, Evrard's collapse
\citep{Evrard1988} (see also \cite[Sec. 9.1]{Springel10}).

The initial condition is a sphere of radius $R = 1$ and mass $M = 1$ with the density
profile
\begin{equation}
    \rho(r) =
    \begin{cases}
        \dfrac{M}{2 \pi R^2 r}, & r \leq R, \\[1.5ex]
        10^{-4}, & r > R,
    \end{cases}
\end{equation}
embedded in a tenuous ambient medium. The gas is initially at rest and
nearly cold, with a uniform specific internal energy of $e_{th} = 0.05$
(in units of $G M / R$, with $G = 1$), so that the pressure follows
from $P = (\gamma - 1)\, \rho\, e_{th}$ with an adiabatic index of
$\gamma = 5/3$.

Without the necessary pressure support, the sphere collapses inside-out,
forming an accretion shock at the center that subsequently propagates outward
through the still-infalling gas until the system virializes. The evolution is
characterized by strong energy transfer between potential, kinetic and thermal energy.

\begin{figure}
 \centering
 \includegraphics[width=1.0\textwidth]{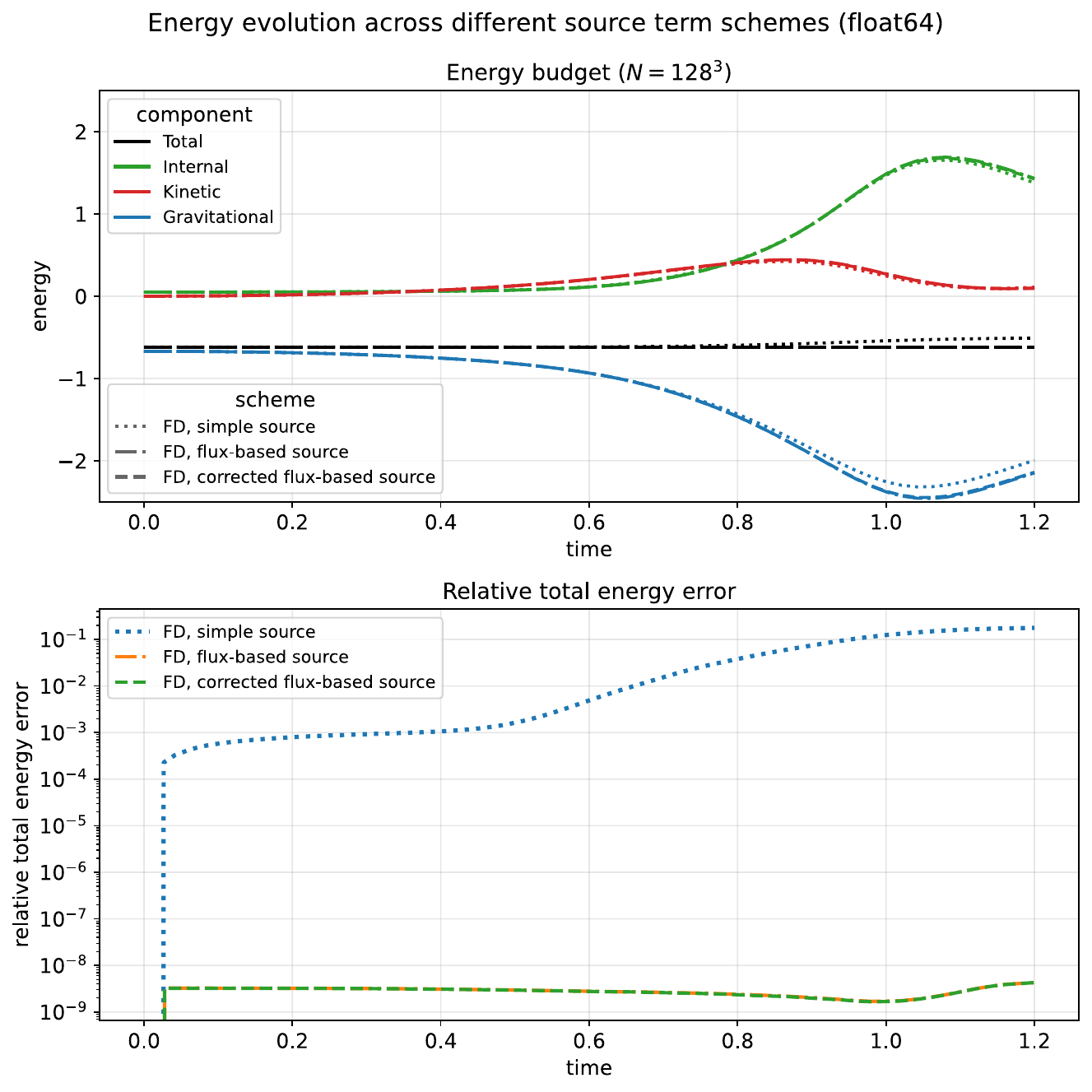}
 \caption{Evolution of the energy components and of the total energy error in Evrard's collapse for the different self-gravity coupling schemes ($128^3$ cells, double precision).}
 \label{fig:evrard_energy}
\end{figure}

The evolution of the different energy terms and the total energy error
are plotted in Fig. \ref{fig:evrard_energy} for a simulation with 
$128^3$ cells, run in double precision. While the simple source term scheme shows an energy
error of up to $18\%$, for the flux-based schemes we observe energy 
errors below $10^{-8}$.

\begin{figure}
 \centering
 \includegraphics[width=1.0\textwidth]{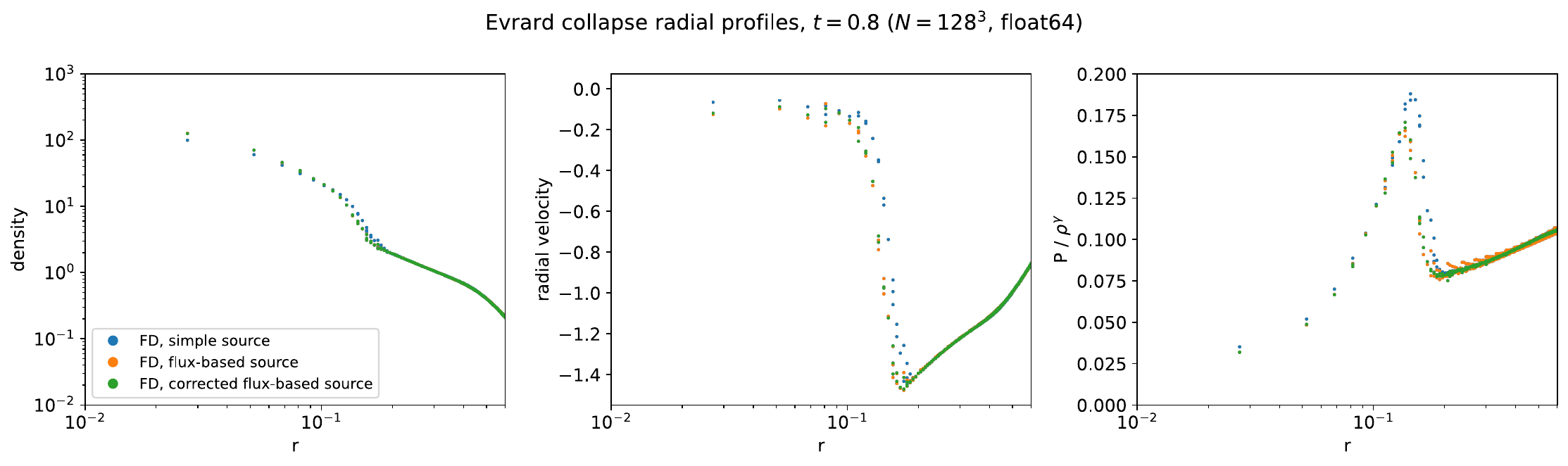}
 \caption{Radial profiles in Evrard's collapse at $t = 0.8$ for the different self-gravity coupling schemes ($128^3$ cells, double precision).}
 \label{fig:evrard_profiles}
\end{figure}

For reference, the radial profiles at $t = 0.8$ are given in Fig. \ref{fig:evrard_profiles}.

We observed crashes of the flux-based (conservative) source schemes at low
resolution (e.g.\ $32^3$) for the cold Evrard collapse. Enabling the
positivity-preserving (PP) flux limiter prevents them. Where the limiter is
active, it reduces the temporal convergence of the energy error to first order.
The limiter sizes the admissible antidiffusive flux from the low-order updated
state over the current step. As a result, the interface flux, and hence the
right-hand side of each Runge--Kutta stage, becomes an explicit function of the
time step, breaking high-order cancellation. When the limiter does not
engage, for example for a milder collapse, the full RK4 order is recovered. Even if it engages, the
resulting errors remain very small. For Evrard's collapse the final
relative energy error is $\Delta E / E_0 \lesssim 10^{-6}$, compared with
$\Delta E / E_0 \sim 7 \cdot 10^{-1}$ for the non-conservative simple-source
scheme, all at a resolution of $32^3$.

\begin{figure}
 \centering
 \includegraphics[width=1.0\textwidth]{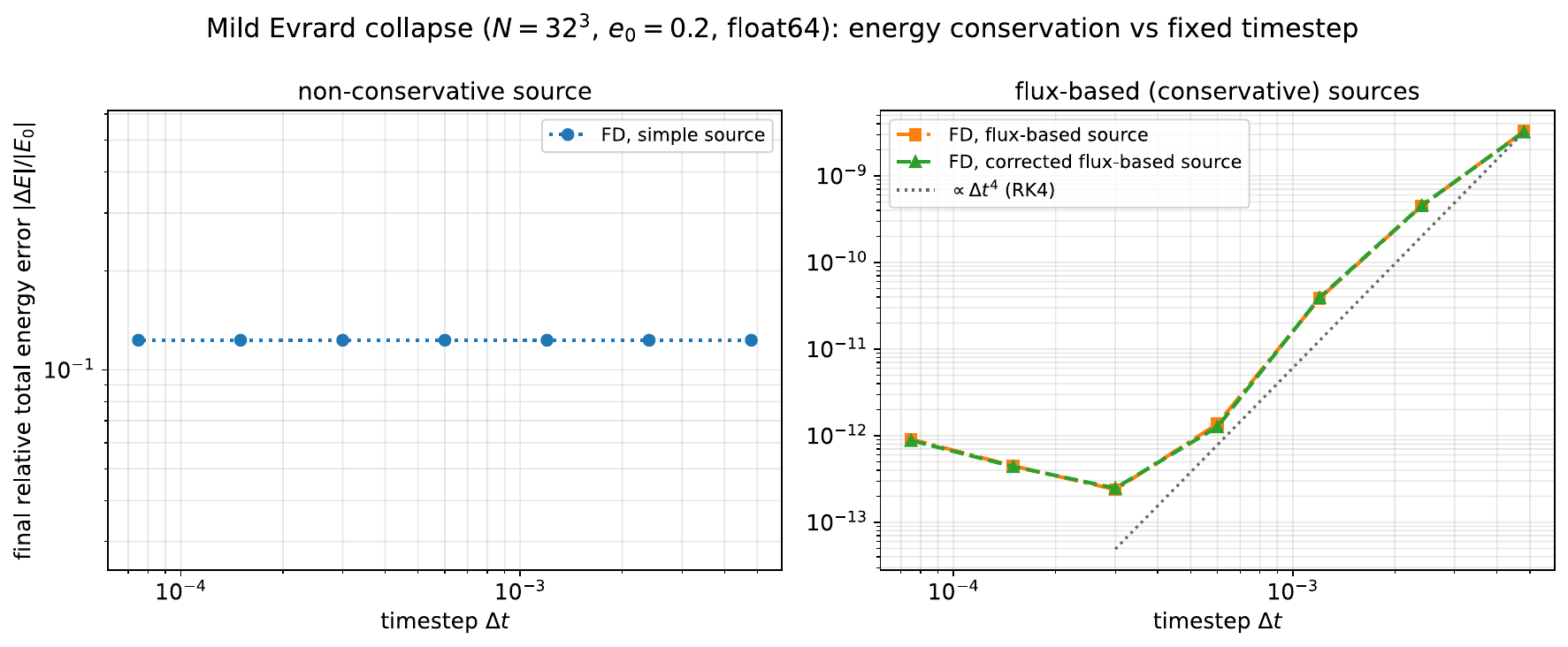}
 \caption{Convergence of the total energy error with the time step for the mild version of Evrard's collapse ($e_{th,0} = 0.2$).}
 \label{fig:mild_timestep_convergence}
\end{figure}

\begin{figure}
 \centering
 \includegraphics[width=1.0\textwidth]{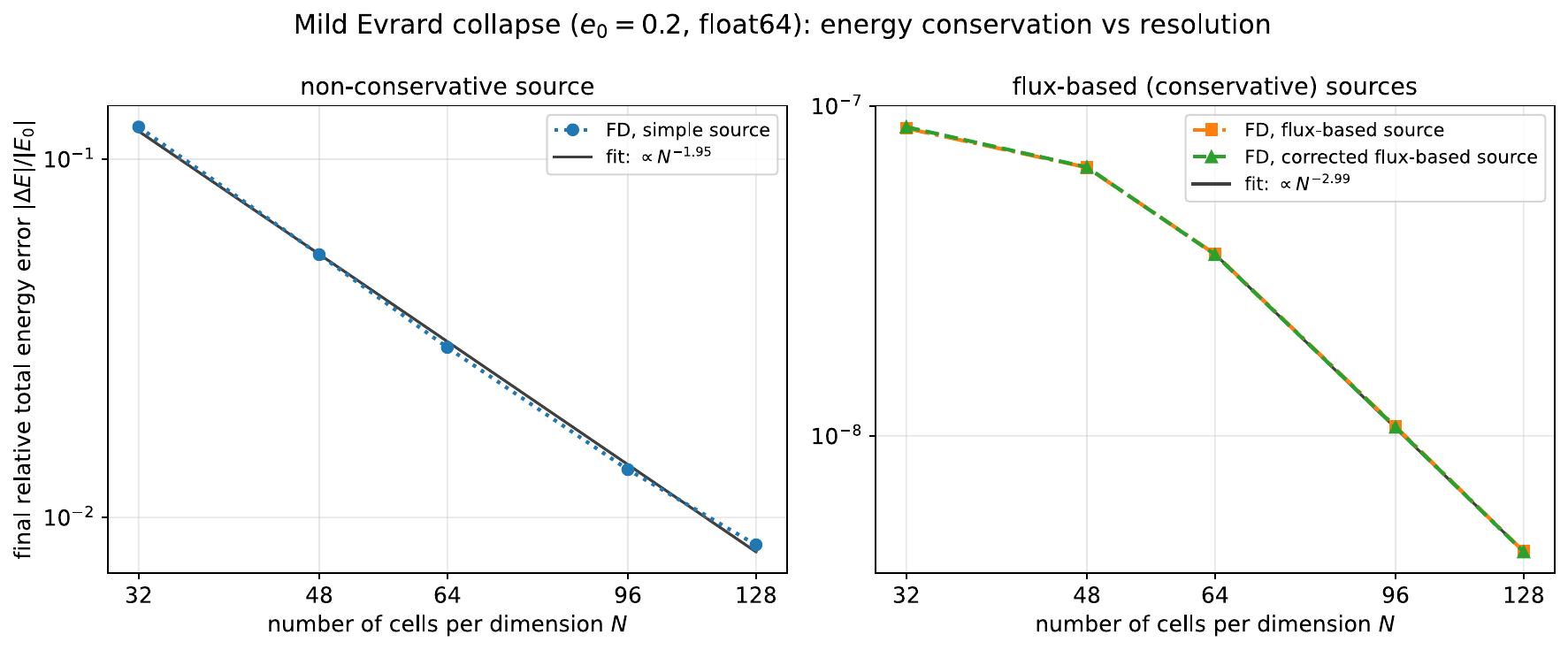}
 \caption{Convergence of the total energy error with spatial resolution (under a CFL time step) for the mild version of Evrard's collapse ($e_{th,0} = 0.2$).}
 \label{fig:mild_energy}
\end{figure}

For a milder collapse with a warmer massive sphere
with $e_{th,0} = 0.2$ we demonstrate the convergence of the energy error with the
time step in Fig. \ref{fig:mild_timestep_convergence}. As expected, the energy error
of the flux-based schemes converges with $\mathcal{O}(\Delta t^4)$ while it is constant
for the simple source term scheme. A convergence study over the spatial resolution
with CFL time step is given in Fig. \ref{fig:mild_energy}.

\section{Forward-performance measurements and scaling test}
\label{sec:performance}

In the following we present a single-GPU performance comparison
against AthenaPK (Sec. \ref{sec:single_performance}), strong scaling results to $8$ GPUs 
(Sec. \ref{sec:strong_scaling}) and weak scaling results to $4$ nodes with $4$ GPUs 
each (Sec. \ref{sec:weak_scaling}).

\subsection{Single-GPU runtimes and memory usage}
\label{sec:single_performance}

\begin{figure}
    \centering
    \includegraphics[width=0.6\linewidth]{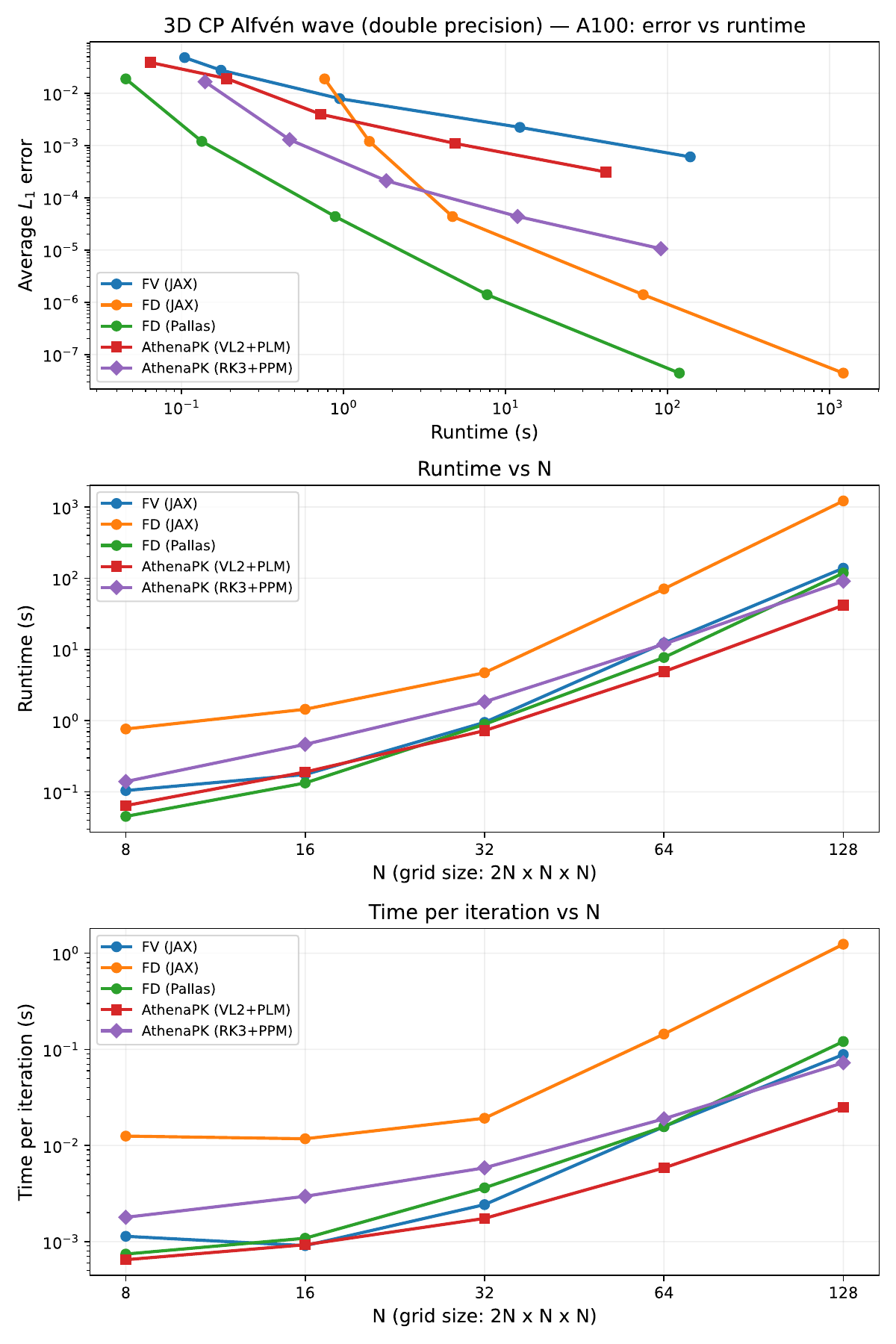}
    \caption{Single-GPU performance on the 3D circularly polarized
    Alfv\'en-wave test (double precision), comparing FV (JAX), FD (Pallas)
    and AthenaPK in a second-order (VL2+PLM) and third-order (RK3+PPM) setup,
    run on an NVIDIA A100 in double precision. \emph{Top:} average $L_1$ error
    versus wall-clock runtime; \emph{middle:} runtime versus resolution $N$
    (grid $2N\times N\times N$); \emph{bottom:} time per iteration versus $N$.
    At fixed runtime FD (Pallas) reaches orders-of-magnitude lower error,
    reflecting its higher convergence order.}
    \label{fig:sg_runtime}
\end{figure}

\begin{figure}
    \centering
    \includegraphics[width=\linewidth]{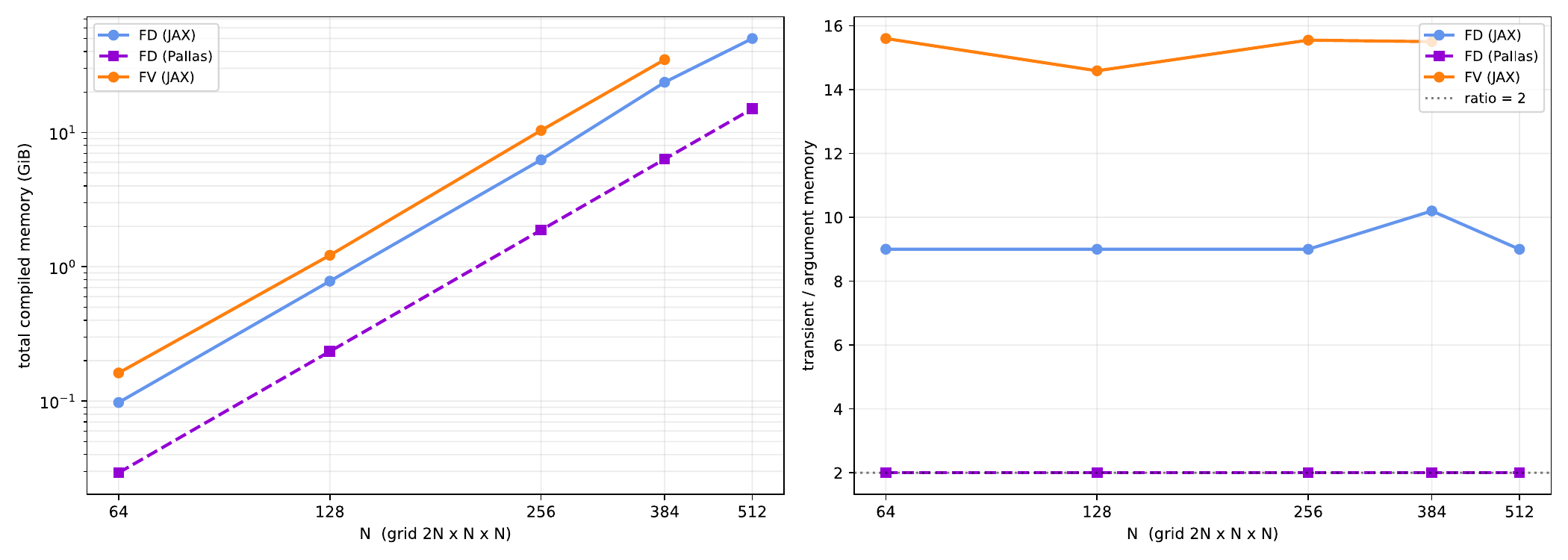}
    \caption{Single-GPU memory for the 3D sound-wave (hydro) test versus
    resolution $N$ (grid $2N\times N\times N$): total compiled memory
    (left) and transient-to-argument memory ratio (right). FD (Pallas) is
    leanest, with a transient/argument ratio close to~2.}
    \label{fig:sg_mem_hydro}
\end{figure}

\begin{figure}
    \centering
    \includegraphics[width=\linewidth]{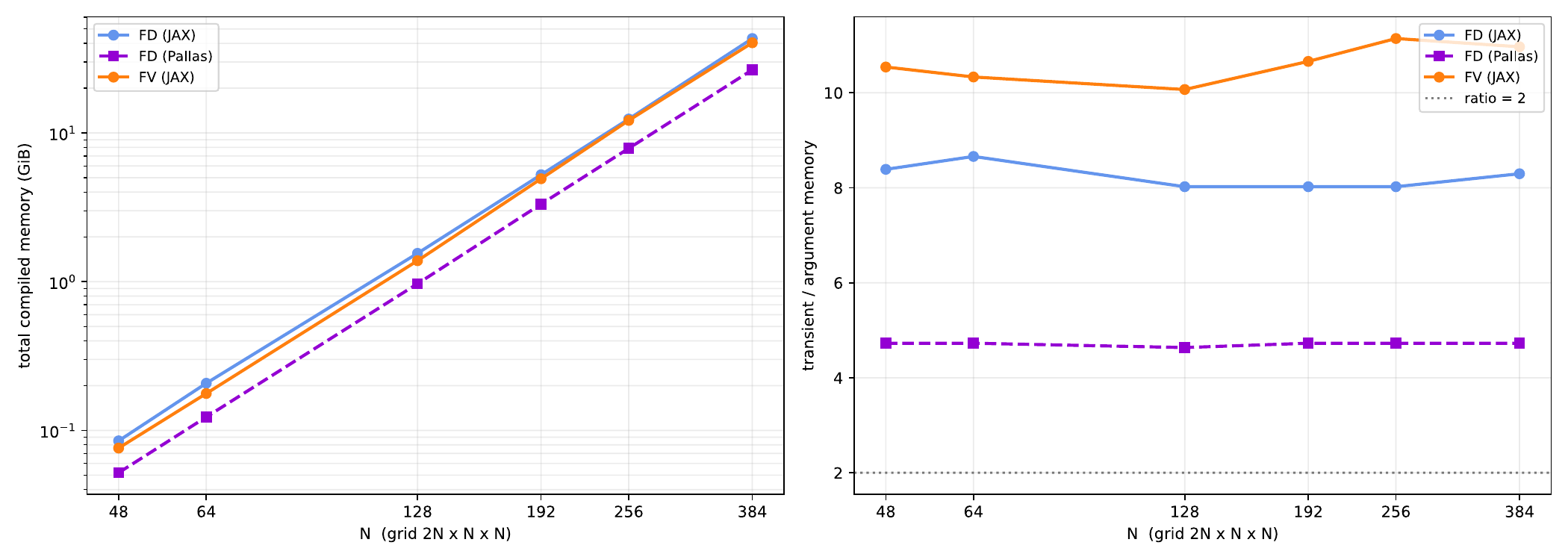}
    \caption{As Fig.~\ref{fig:sg_mem_hydro} but for the MHD (Alfv\'en-wave)
    test. The richer variable set raises every solver's footprint; FD
    (Pallas) still holds the lowest transient/argument ratio ($\sim\!4.7$).}
    \label{fig:sg_mem_mhd}
\end{figure}

Fig.~\ref{fig:sg_runtime} compares FV (JAX), FD (Pallas) and AthenaPK with
a second order (specifically a second order van Leer time integration with piecewise linear reconstruction, called VL2+PLM)
and third order (specifically a third order Runge--Kutta integrator with piecewise parabolic reconstruction, called RK3+PPM)
setup on the 3D circularly polarized Alfv\'en-wave test run on an NVIDIA A100 in double precision. Results on an
NVIDIA H100 are additionally provided in brackets.

FV (JAX) was run at $C_\mathrm{CFL} = 0.4$, AthenaPK VL2+PLM at $C_\mathrm{CFL} = 0.3$ and
RK3+PPM at $C_\mathrm{CFL} = 0.4$, and FD (Pallas) at $C_\mathrm{CFL} = 1.5$.

The top panel shows the $L^1$ error obtained after a given runtime, the center panel shows 
the runtime and the bottom panel the time per iteration as a function of the grid size. Per 
iteration, FD (Pallas) is slower than both AthenaPK setups at $N = 128$, 
by $\approx\!1.7\times$ [$\approx\!1.4\times$] relative to the third-order RK3+PPM 
and $\approx\!4.8\times$ [$\approx\!3.8\times$] relative to the second-order VL2+PLM. 
Such a difference is expected for a higher-order scheme compared to a lower-order scheme: 
the spatial stencil is larger and the accompanying higher-order Runge-Kutta integrator 
requires more right-hand side evaluations per time step. This gap is reduced in the total 
runtime comparison because of the larger $C_\mathrm{CFL}$ allowed by the Runge-Kutta 
integrator used for the FD scheme. At $N = 128$ the total FD (Pallas) runtime 
($\approx\!120$ s [$\approx\!64$ s]) is comparable to that of RK3+PPM 
($\approx\!91$ s [$\approx\!58$ s]). Due to its fifth-order convergence, 
the FD (Pallas) scheme reaches far lower errors in a given runtime on smooth 
problems compared to the second-order FV (JAX) and VL2+PLM schemes as well as 
the third-order RK3+PPM scheme: at a runtime of around $100$ [$60$] seconds its 
$L^1$ error is roughly four orders of magnitude below FV (JAX) and more than two 
orders of magnitude below RK3+PPM. This could for instance be interesting for 
subsonic turbulence simulations. Note that single precision runs, currently 
unsupported in \texttt{AthenaPK}, can be $3$ to $5$ times faster in 
\texttt{astronomix} than double precision runs. The Pallas
backend is between roughly five and nine times faster across the tested resolutions
compared to the native JAX backend.

The memory footprints for FV (JAX), FD (Pallas) and FD (JAX) are shown in 
Figs.~\ref{fig:sg_mem_hydro} and~\ref{fig:sg_mem_mhd}. Across resolutions FD (Pallas)
has the lowest transient-to-argument ratio and overall memory footprint.
The ratio is close to the ideal $2$ of the storage-saving Runge-Kutta 
scheme (Sec. \ref{app:rk:lsrk4}) for the hydrodynamical sound-wave test, 
much lower than the ratio of about $9$ of the native JAX backend,
and at $\sim\!4.7$ for the richer 
MHD variable set.

\subsection{Strong scaling}
\label{sec:strong_scaling}

\begin{figure}
    \centering
    \includegraphics[width=0.7\linewidth]{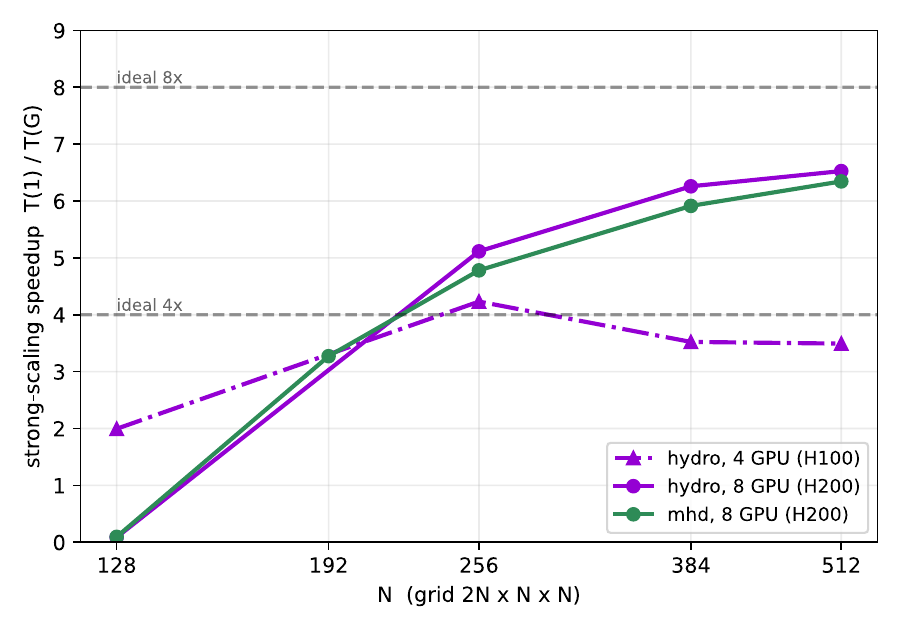}
    \caption{Strong-scaling speedup $T(1)/T(G)$ versus resolution for the
    FD (Pallas) solver: hydro on 4~GPUs (H100) and hydro/MHD on 8~GPUs
    (H200). Baselines exceeding single-GPU memory use a power-law
    extrapolation.}
    \label{fig:strong_speedup}
\end{figure}

Fig. \ref{fig:strong_speedup} shows the strong scaling speedup
on $4$ (H100) and $8$ (H200) GPUs. All simulations were run in 
single precision.

The strong scaling speedup improves with problem size as the halo-exchange overhead becomes 
subdominant: the 8$\times$H200 FD (Pallas) runs reach a speedup of $\sim\!6.5$ (hydro) and
$\sim\!6.3$ (MHD) at $N=512$, the 4$\times$H100 runs
$3.5$ (Fig.~\ref{fig:strong_speedup}). The smallest grids are dominated by
kernel-launch and warm-up overhead and fall below linear speedup.

\subsection{Weak scaling}
\label{sec:weak_scaling}

For the weak-scaling test the per-GPU sub-domain is held fixed at
$128\times2048\times2048$ ($5.37\cdot10^{8}$ cells) while the device count
grows, so the global grid is $128\,G\times2048\times2048$ and ideal scaling
is a flat wall-clock curve. The initial condition is already built in a 
globally sharded manner (each process materializes only its local shard), so the
full grid never resides on a single host. The largest resolution is reached at
$2048^{3}$ cells on 16~GPUs across 4~nodes. We quote per-GPU throughput as
Mcell/s/GPU: millions of cell updates (grid cells $\times$ time steps) per
second and per GPU. Again, these tests were run in single precision.

\begin{table}
    \centering
    \caption{Weak-scaling results for the 3D sound-wave (hydro) test with
    the FD (Pallas) solver on NVIDIA~H100 GPUs (fp32, LSRK4, Pallas block
    shape $(4,4,8)$, 10 fixed steps, one process per GPU, 1D $x$
    decomposition). The per-GPU sub-domain is fixed at
    $128\times2048\times2048$. Efficiency is $T(1)/T(G)$ (ideal $=100\%$);
    throughput is per-GPU cell updates per second.}
    \label{tab:weak_scaling}
    \begin{tabular}{rrlrrrr}
        \toprule
        GPUs & Nodes & Global grid & Cells & Runtime [s] & Eff.\ [\%] & Mcell/s/GPU \\
        \midrule
         1 & 1 & $128\times2048^2$  & $5.4\cdot10^{8}$ & 19.47 & 100.0 & 276 \\
         2 & 1 & $256\times2048^2$  & $1.1\cdot10^{9}$ & 23.76 &  81.9 & 226 \\
         4 & 1 & $512\times2048^2$  & $2.1\cdot10^{9}$ & 24.02 &  81.0 & 224 \\
         8 & 2 & $1024\times2048^2$ & $4.3\cdot10^{9}$ & 25.37 &  76.7 & 212 \\
        16 & 4 & $2048^{3}$         & $8.6\cdot10^{9}$ & 25.75 &  75.6 & 209 \\
        \bottomrule
    \end{tabular}
\end{table}

The one-time $\sim\!18\%$ drop from 1 to 2 GPUs marks the onset of
halo exchange; beyond that the curve is nearly flat, losing only a few
percent out to 16~GPUs and just $\sim\!1\%$ for the final inter-node
doubling ($8\to16$). At the largest size the code sustains
$\sim\!209$~Mcell/s/GPU (cell updates per second per GPU) on a $2048^{3}$
grid at $75.6\%$ weak-scaling
efficiency (Table~\ref{tab:weak_scaling}).

\section{Automatic differentiation I: Verification and performance}

Automatic differentiation allows for exact derivatives through
our discrete numerics, possibly with respect to hundreds of thousands
of parameters or more -- computationally infeasible with finite differencing.

While standard tests like linear waves 
or a shock tube exist for forward validation, no tests 
have been established to verify automatic differentiation through 
our simulator.

In the following, we introduce two such  tests.
The first is a validation against analytical functional derivatives
with respect to the full initial conditions. This test validates that 
differentiation through the simulator approaches the continuous
backward dynamics in the small perturbation limit. While the mathematical
tools used to derive the analytical derivatives are standard, we are not aware
of the use of such tests in the differentiable simulator literature. The second
test covers differentiation through nonlinear dynamics and 
shocks for a few parameters. We check if finite differencing converges
to the automatic differentiation results.

Finally, we also provide measurements of the performance of reverse-mode automatic
differentiation through adaptive time stepping and the optimal choice of the
number of checkpoints set by the user.

\subsection{Analytic functional derivatives of linearized equations}
\label{sec:analytical_gradients}

Consider the following initial conditions with small perturbations

\begin{equation}
    \begin{aligned}
        \rho &= \rho_B + \rho_P, \quad &\rho_P \ll \rho_B \\
        \vec{v} &= \vec{v}_P, \quad &\vec{v}_P^2 \ll e_{th}, \left \| \vec{v}_P \right \| \ll c_s \\
        P &= P_B + P_P, \quad &P_P \ll P_B,
    \end{aligned}
\end{equation}

where $\rho_B$ is a constant background density, $P_B$ a constant
background pressure, $c_s$ is a constant and the background velocity is zero. 
All background fields are constant throughout space and time.

We collect the \emph{perturbations} in the state vector $\vec{U} = (\rho_P, \vec{v}_P)$.

We initialize the pressure with

\begin{equation}
    P_B = \frac{c_s^2 \rho_B}{\gamma}, \quad P_{P,0} = c_s^2 \rho_{P,0}.
\end{equation}

The analytical forward solution in Fourier space is given by

\begin{equation}
    \begin{aligned}
        \hat{\rho}(\vec{k}, t) &= \hat{\rho}_0(\vec{k}) \cos(\omega t) - i \frac{\rho_B}{c_s} (\hat{\vec{v}}_0(\vec{k}) \cdot \hat{\vec{k}}) \sin(\omega t) \\
        \hat{\vec{v}}(\vec{k}, t) &= \hat{\vec{v}}_0(\vec{k}) - \hat{\vec{k}}(\hat{\vec{v}}_0(\vec{k}) \cdot \hat{\vec{k}}) \Big( 1 - \cos(\omega t) \Big) - i \frac{c_s}{\rho_B} \hat{\rho}_0(\vec{k}) \hat{\vec{k}} \sin(\omega t),
    \end{aligned}
\end{equation}

with the acoustic dispersion relation $\omega = c_s |\vec{k}|$, which is derived in Sec. \ref{app:analytical_forward}.

The analytical gradient (functional derivative) of the $L^2$-norm of
the final state with respect to the initial state

\begin{equation}
    J(\vec{U}_0) = \frac{1}{2} \langle S_t \vec{U}_0, S_t \vec{U}_0 \rangle,
\end{equation}

where $S_t$ is the time evolution operator, is given in Fourier space by (see Sec. \ref{app:analytical_gradients} for a derivation)

\begin{equation}
\begin{aligned}
\nabla_{\hat{\rho}_0} J &= \left[ \cos^2(\omega t) + \left(\frac{c_s}{\rho_B}\right)^2 \sin^2(\omega t) \right] \hat{\rho}_0 - i \sin(\omega t) \cos(\omega t) \left( \frac{\rho_B}{c_s} - \frac{c_s}{\rho_B} \right) \left(\hat{\vec{k}} \cdot \hat{\vec{v}}_0\right) \\
\nabla_{\hat{\vec{v}}_0} J &= i \sin(\omega t) \cos(\omega t) \left( \frac{\rho_B}{c_s} - \frac{c_s}{\rho_B} \right) \hat{\rho}_0 \hat{\vec{k}} + \hat{\vec{v}}_0 + \left( \left(\frac{\rho_B}{c_s}\right)^2 - 1 \right) \sin^2(\omega t) \left(\hat{\vec{k}} \cdot \hat{\vec{v}}_0\right) \hat{\vec{k}},
\end{aligned}
\end{equation}

which transforms to real space via

\begin{equation}
    \vec{\nabla} \vec{U} = \mathcal{F}^{-1}(\vec{\nabla} \hat{\vec{U}}).
\end{equation}

We refer to Sec. \ref{app:discretized_gradients} for the proper discretization of such
functional derivatives.

As a specific test we set $\rho_B = 1$, $c_s = 2$ and $\gamma = 5/3$, 
and seed a single density perturbation of amplitude $\varepsilon = 10^{-6}$
with vanishing initial velocity, so that nonlinear corrections are negligible. We use two initial
conditions: a smooth, axis-angled sinusoid $\rho_{P,0} = \varepsilon \sin(\vec{k}
\cdot \vec{x})$ with two wavelengths along each axis ($k_d = 4\pi/L$), and a
centered Gaussian blob $\rho_{P,0} = \varepsilon \exp(-|\vec{x} - \vec{x}_c|^2 /
2\sigma^2)$ with $\sigma = L/10$. 

In both cases the reverse-mode AD gradient
$\nabla_{U_0} J$ of the loss is compared to the previously 
derived analytical result.

\begin{figure}
 \centering
 \includegraphics[width=1.0\textwidth]{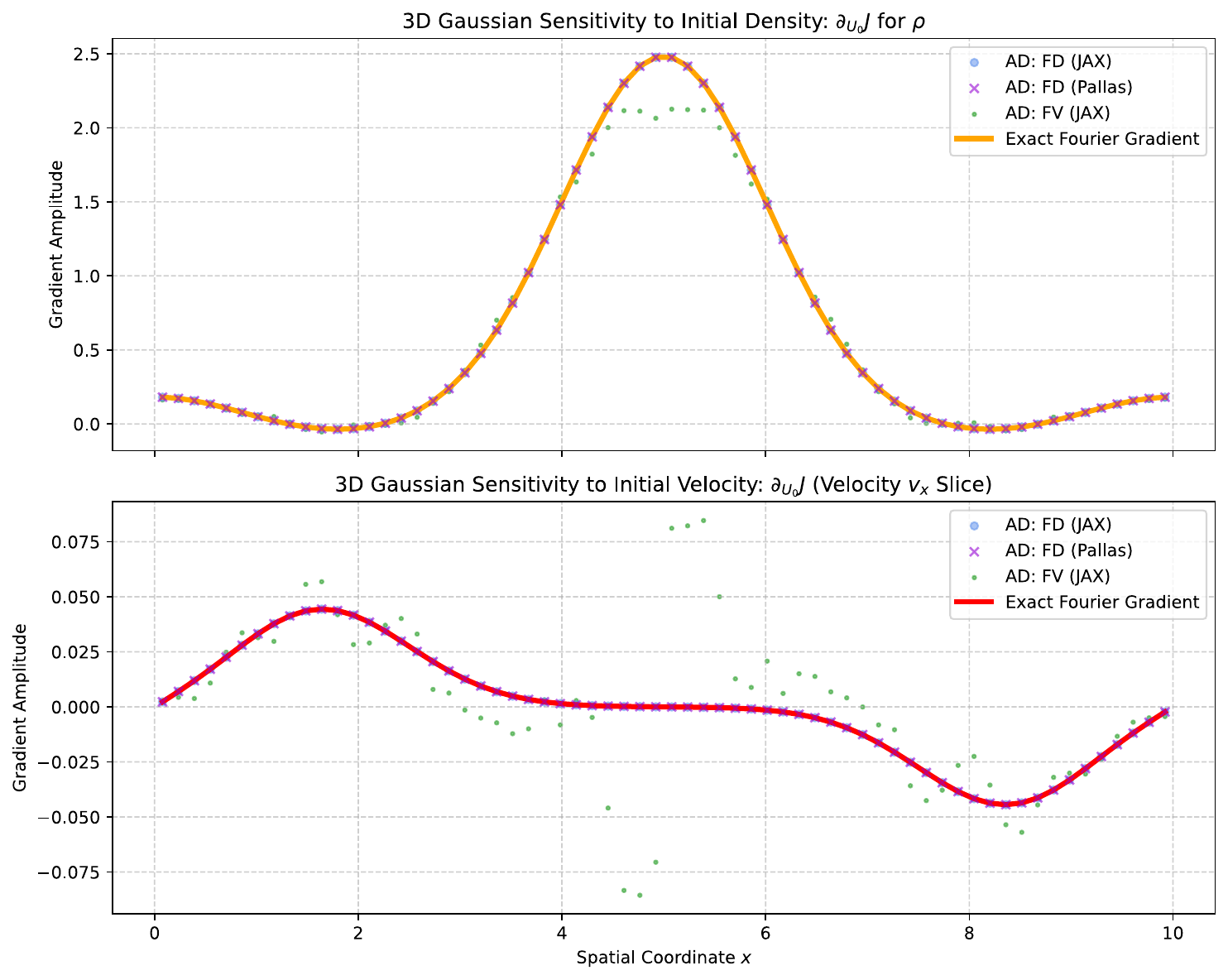}
 \caption{Reverse-mode AD gradient of the final-state $L^2$ cost with respect to
 the full initial density field, for a 3D Gaussian density perturbation
 ($N = 64$, evolved until it has separated into outgoing acoustic waves), compared
 against the exact analytic (Fourier-space) functional derivative.}
 \label{fig:gradient_3d_gaussian}
\end{figure}

Fig.~\ref{fig:gradient_3d_gaussian} shows the full-field 3D Gaussian case
($N = 64$, $L = 10$, evolved to $t = 1.5$, by which time the blob has separated
into outgoing acoustic waves). Based on its higher order, the results
of the finite difference scheme are already converged to the analytical solution
while the finite volume results deviate.

\begin{figure}
 \centering
 \includegraphics[width=1.0\textwidth]{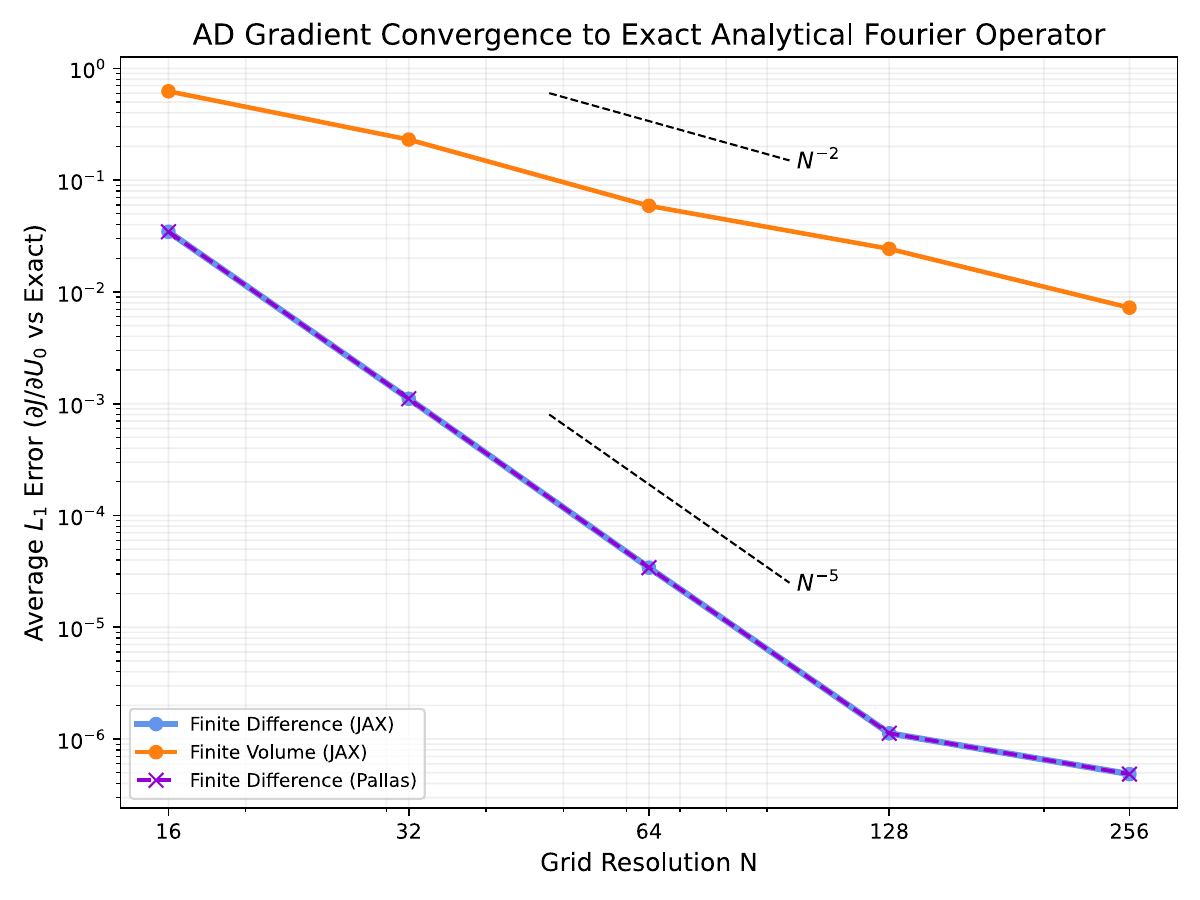}
 \caption{Convergence of the reverse-mode AD gradient to the exact analytic
 (Fourier-space) functional derivative of the linearized acoustic problem, as a
 function of grid resolution $N$. The finite-difference solver converges at its
 nominal fifth order ($N^{-5}$), the finite-volume solver at second order ($N^{-2}$); the JAX and
 Pallas finite-difference backends coincide.}
 \label{fig:gradient_convergence_test}
\end{figure}

To quantify the difference in convergence, we use the 1D sinusoidal test ($L = 1$, $t = 0.15$) by sweeping the resolution
$N \in \{16, 32, 64, 128, 256\}$ (Fig.~\ref{fig:gradient_convergence_test}). The
mean $L^1$ error of the AD gradient against the exact Fourier gradient drops at
the spatial order of the scheme: the fifth-order finite-difference solver
converges as $\mathcal{O}(N^{-5})$ ($3.4\cdot10^{-2}$ down to $1.1\cdot10^{-6}$ at $N = 16 \dots 128$), while the finite-volume solver converges
as $\mathcal{O}(N^{-2})$ ($6.2\cdot10^{-1}$ down to $7.2\cdot10^{-3}$).
The native-JAX and the Pallas backends of the finite-difference solver produce identical errors.

\subsection{Finite difference validation on a shock tube setting}
\label{sec:shock_gradients}

The linearized test above probes the smooth regime; differentiation through
fully nonlinear dynamics and shocks is verified on a standard Sod shock tube
with piecewise constant left and right state (split at $x = 0.5$), parameterized by
$\theta = (\rho_L, P_L, u_L, \rho_R, P_R, u_R) = (1, 1, 0, 0.125, 0.1, 0)$ with
$\gamma = 5/3$. The tube ($L = 1$, $N = 200$, open boundaries) is
evolved to $t = 0.2$ with $250$ fixed timesteps. Fixed timesteps are used so the 
finite-differencing baseline is not contaminated by changes in the 
step count as $\theta$ is varied. We differentiate the objective
$J(\theta) = \tfrac{1}{2}\int (\rho^2 + v^2 + P^2)\, \mathrm{d}x$ evaluated on the
final (post-shock) state.

\begin{figure}
 \centering
 \includegraphics[width=1.0\textwidth]{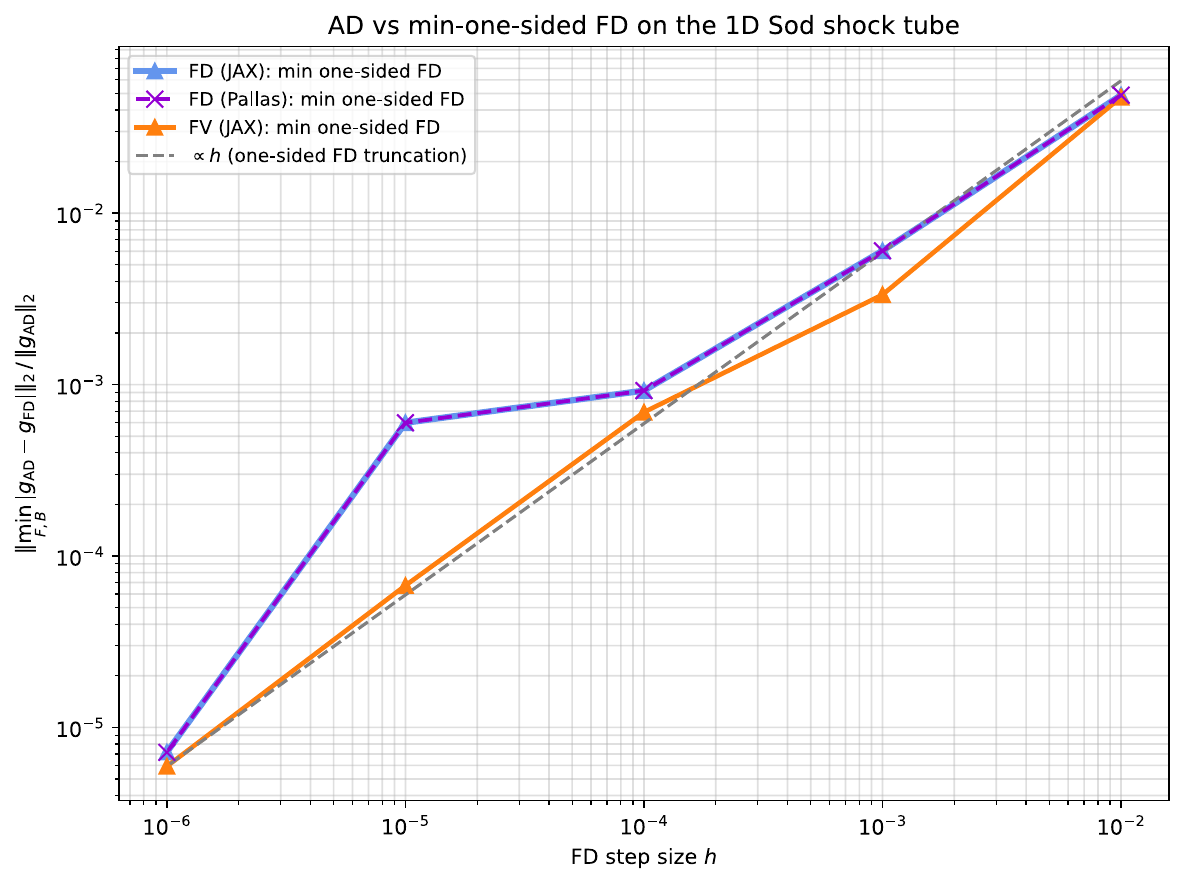}
 \caption{Shock-tube AD verification: the kink-immune min-one-sided
 finite-difference residual $\|\min_{F,B}|g_{\mathrm{AD}} - g_{\mathrm{FD}}|\|_2 /
 \|g_{\mathrm{AD}}\|_2$ as a function of the FD step $h$. It tracks the one-sided
 truncation $\propto h$ to $\sim\!7\cdot10^{-6}$; the JAX and Pallas
 finite-difference backends overlap.}
 \label{fig:shock_tube_sensitivity_convergence}
\end{figure}

For such a test with a low number of parameters, a convergence study against finite
differencing can be done, i.e. we check if finite differencing converges
to the automatic differentiation result. However, $\nabla J(\theta)$ is not continuous
and central differencing would average across jumps, whereas automatic differentiation
returns the derivative of one branch of the discrete map. We therefore certify AD against the
\emph{minimum} of the two one-sided finite differences per parameter,
$\min(|g_{\mathrm{AD}} - g_F|, |g_{\mathrm{AD}} - g_B|)$.
Note that this test certifies only that AD is consistent with the discrete
solver map, not the continuum problem.
Fig.~\ref{fig:shock_tube_sensitivity_convergence} shows this
residual against the finite-difference step $h$: it decreases at the expected
one-sided truncation rate $\mathcal{O}(h)$ down to $7.2\cdot10^{-6}$ at
$h = 10^{-6}$ for the finite-difference solver ($5.9\cdot10^{-6}$ for finite
volume), confirming that AD reproduces at least one one-sided slope to round-off.
As with the linearized test, the JAX and Pallas finite-difference backends are
indistinguishable.

\subsection{Performance of backpropagation through adaptive time-stepping}
\label{sec:checkpointing}

For the backward pass in reverse-mode automatic differentiation, the \textit{primals}
at each time-step must be known. To keep the memory cost of automatic differentiation under 
control, primals are recomputed from \textit{checkpoints}. For differentiation through
simulations with adaptive time-stepping, online checkpointing is used \citep{stumm10}.
The number of checkpoints controls the balance between memory and recomputation cost.

With $c$ checkpoints, the largest forward length that can be reversed with a
repetition number (maximum number of times any time step is performed, including the
initial forward pass) of at most
$r$ is the binomial capacity
\begin{equation}
    \beta(c, r) = \binom{c + r}{c},
\end{equation}
so for $N$ steps the repetition number is the smallest $r$ that reaches it,
\begin{equation}
    r(c, N) = \min\left\{\, r \in \mathbb{N}_0 \;:\; \binom{c+r}{c} \ge N \,\right\}.
\end{equation}
The minimal number of \emph{recomputations} (extra forward steps beyond the
original $N$) is then
\begin{equation}
    R(c, N) = \max\!\left(0,\; r\,N - \binom{c + r}{\,c+1\,} - N\right),
    \qquad r \equiv r(c, N).
\end{equation}

Note this assumes that in addition to the checkpoints and the current state
the previous state is also stored.

Two typical choices of the number of checkpoints $c$ are when the repetition number
falls under two or three, given by

\begin{equation}
    c_{r \le 2}(N) = \min\left\{\, c : \binom{c+2}{2} \ge N \,\right\},
    \qquad
    c_{r \le 3}(N) = \min\left\{\, c : \binom{c+3}{3} \ge N \,\right\},
\end{equation}

\begin{figure}
 \centering
 \includegraphics[width=1.0\textwidth]{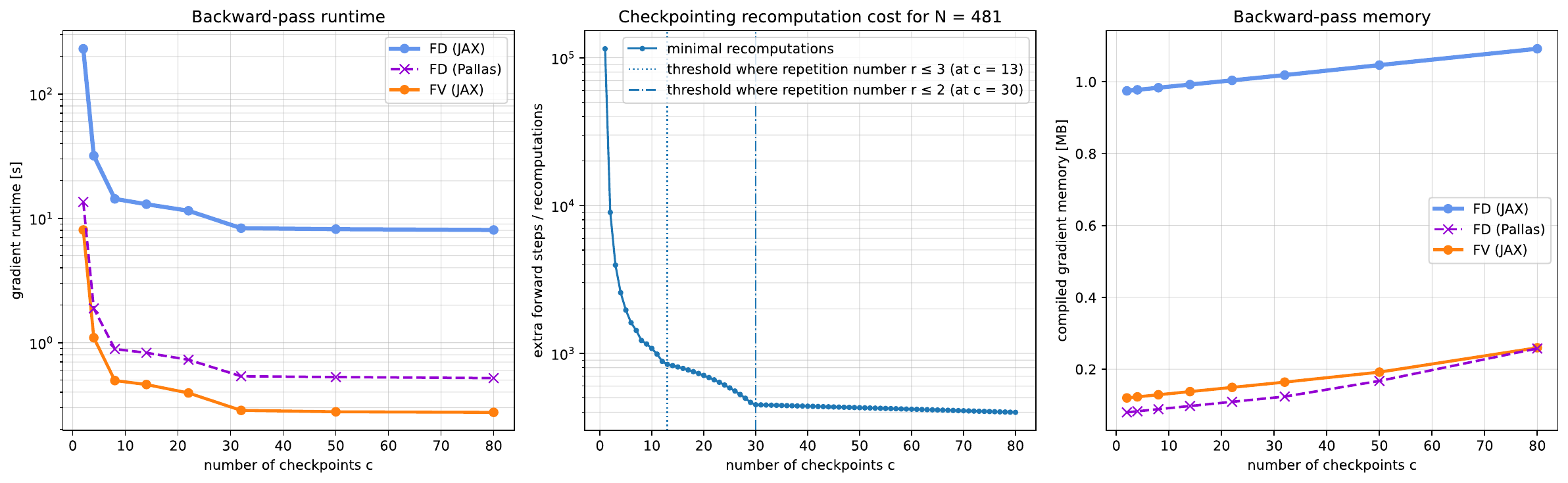}
 \caption{Reverse-mode performance versus the number of checkpoints $c$ for
 $N_{\rm steps} = 481$ adaptive steps: backward-pass runtime (left), theoretical
 minimal recomputation count (middle) and compiled backward-pass memory (right).
 The Pallas finite-difference adjoint backend based on custom adjoint kernels
 is $\sim\!15$--$17\times$ faster and requires $\sim\!10\times$ less memory than the native-JAX
 finite-difference adjoint backend.}
 \label{fig:checkpoint_scaling}
\end{figure}

A practical example on an NVIDIA A100 is given in Fig.~\ref{fig:checkpoint_scaling}.
In the left panel, the runtime of a backward pass through FV (JAX), FD (JAX) and FD (Pallas)
is plotted over the number of checkpoints. In the center panel, the expected number of recomputations
is plotted with $c_{r \le 2}(N)$ and $c_{r \le 3}(N)$ marked. The right panel shows memory usage. 
The runtimes closely reflect the number of recomputations;
the memory usage increases linearly with the number of checkpoints, as expected. The Pallas
FD backend is $\sim\!15$--$17\times$ faster and requires $\sim\!10\times$ less memory than the native-JAX
finite-difference backward.

\section{Automatic differentiation II: Applications}

With automatic differentiation we can efficiently calculate 
exact gradients through our discrete numerical simulator.
These gradients generally differ from ones obtained via
finite differencing, which are inexact, or analytical 
gradients of the continuum model which do not factor in 
the numerics of our solver.

Automatic differentiability enables applications in sensitivity analysis, 
PDE-constrained optimization, inverse modeling and optimization of parametric models
inside the simulator (so-called solver-in-the-loop techniques).
One might for instance infer full initial condition fields
\citep{Horowitz2025,Horowitz2025b}, train a machine learning model to correct coarse
simulator dynamics \citep{um21, wei25} or learn unresolved 
physical processes \citep{Kochkov2024}.

\begin{figure}
 \centering
 \includegraphics[width=1.0\textwidth]{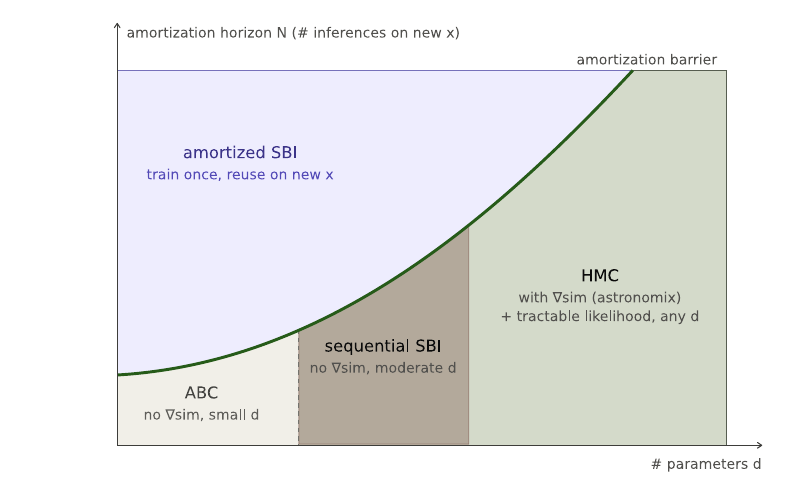}
 \caption{Inverse modeling landscape. The choice of inference method depends on the
 number of predicted parameters $d$ and the number of inferences on new observations $N$.
 The upfront training cost of amortized simulation-based inference (SBI) methods increases
 with parameter space complexity, moving the amortization barrier to higher $N$. Below 
 this amortization barrier, per-observation methods are preferred.
 When the likelihood cannot be evaluated, classical approximate Bayesian computation (ABC) or sequential SBI are the preferred options but these become costly in high-dimensional settings. Gradients
 through the simulator become indispensable to navigate high-dimensional parameter spaces. A typical gradient-based method is Hamiltonian Monte Carlo (HMC).}
 \label{fig:inverse_modeling_landscape}
\end{figure}

Differentiable simulators can be a valuable component in the inverse modeling
landscape. While simulation-based inference methods are successful in 
amortized settings with few parameters \citep{sbi} and the amortization
barrier might be lowered by sequential methods \citep{greenberg2019},
in settings with many parameters and insufficient amortization, gradient-based
sampling \citep{hmc} with a differentiable simulator is the preferred option.
This is illustrated in Fig. \ref{fig:inverse_modeling_landscape}.

In the following, we demonstrate using gradients through
\texttt{astronomix} for sensitivity analysis and eigen-initialization
of a Kelvin--Helmholtz instability, discuss gradient and optimization
pathologies with multiple shooting as a possible solution and present 
a specific example of optimizing initial conditions in 3D. 
For solver-in-the-loop applications we refer to our previous
publication \cite{storcks2025}.

\subsection{Sensitivity analysis and eigen-initialization}
\label{sec:khi_eigen}

Turbulent mixing layers and the Kelvin--Helmholtz instability,
a fluid instability resulting from velocity shear \citep{chandrasekhar1961}, 
are relevant in a range of astrophysical contexts, such 
as wind-blown bubbles \citep{lachlan24}, interstellar
clouds \citep{Vietri1997} or jets \citep{Bodo1989}.
An important field of study is how the Mach number \citep{mandelker16},
magnetic field strength \citep{chandrasekhar1961} or viscosity
\citep{Roediger2013} influence the instability. To initialize Kelvin--Helmholtz
simulations one typically wants to use eigenmodes to avoid 
transients in the growth behavior. While analytical eigenmodes for
the Euler equations have been derived \citep{mandelker16}, simplifying assumptions 
compared to the simulator setup are required (e.g. a sharp interface
instead of a smooth transition) and there might still be numerical transients.
With a differentiable simulator, we can find exact numerical eigenmodes 
and perform stability analysis at no additional implementation cost, in contrast
to linearizing the forward model by hand.

The following analysis is performed on two-dimensional setups, but the
generalization to three dimensions is straightforward.

Our differentiable solver defines a right-hand side
\begin{equation}
    \frac{\mathrm{d} q}{\mathrm{d} t} = \mathcal{R}(q), \qquad
    q \in \mathbb{R}^{n_\mathrm{var} \times N_x \times N_y},
\end{equation}
where $q$ is the conserved state ($n_\mathrm{var} = 4$: $\rho$, $\rho v_x$,
$\rho v_y$, $E$). Linearizing
about a base state $q_0$, automatic differentiation of the solver evaluates the
action of the Jacobian
\begin{equation}
    \mathcal{J} \;=\; \left.\frac{\partial \mathcal{R}}{\partial q}\right|_{q_0},
    \qquad \delta \dot{q} = \mathcal{J}\,\delta q,
\end{equation}
as a matrix-free Jacobian-vector product, evaluated from a single forward-mode pass
through the solver. Note that here we consider the right-hand side of the time integration,
not a finite-time integration.

This Jacobian has dimension $(n_\mathrm{var} N_x N_y) \times (n_\mathrm{var} N_x N_y)$
and is far too large to form or diagonalize densely. 
Typically, a few eigenpairs are extracted by a Krylov method matrix-free, which most
robustly returns the eigenvalues of largest magnitude, while finding the eigenpairs
with the largest real part, relevant for stability analysis, can be challenging. For this demonstration
we want to display the entire eigenspectrum.

This becomes tractable by exploiting the streamwise homogeneity of the base state and periodicity in $x$-direction. For a grid periodic in $x$ with $N_x$ cells and length $L_x$, any state $q$
can be written as
\begin{equation}
    q(x_i, y_j) = \sum_{m = 0}^{N_x - 1} \hat{q}_m(y_j)\,
    e^{\mathrm{i} k_{x,m} x_i},
    \qquad k_{x,m} = \frac{2\pi m}{L_x},
\end{equation}
where reality of $q$ implies $\hat{q}_{N_x - m} = \overline{\hat{q}_m}$, so
that modes pair up across $\pm k_x$ and it suffices to consider
$0 \le m \le N_x / 2$. Let us therefore consider the reduced right-hand side
\begin{equation}
    \hat{\mathcal{R}}_{k_x} :\;
    \mathbb{C}^{\,n_\mathrm{var} N_y} \to \mathbb{C}^{\,n_\mathrm{var} N_y},
    \qquad
    \hat{\mathcal{R}}_{k_x}(\hat{q})
    \;=\; P_{k_x}\, \mathcal{R}\!\left(q_0 + \hat{q}(y)\, e^{\mathrm{i} k_x x}\right),
\end{equation}
where $P_{k_x}$ extracts the $+k_x$ Fourier coefficient,
$(P_{k_x} F)(y_j) = N_x^{-1} \sum_i e^{-\mathrm{i} k_x x_i} F(x_i, y_j)$.
Let us also define the reduced Jacobian\footnote{Since the solver is
real-valued, the complex tangent action is assembled from two real
Jacobian-vector products,
$\mathcal{J}(a + \mathrm{i} b) = \mathcal{J} a + \mathrm{i}\, \mathcal{J} b$.}
\begin{equation}
    \hat{\mathcal{J}}(k_x)
    \;=\; \left.\frac{\partial \hat{\mathcal{R}}_{k_x}}{\partial \hat{q}}
          \right|_{\hat{q} = 0}
    \;\in\; \mathbb{C}^{\,n_\mathrm{var} N_y \times n_\mathrm{var} N_y}.
\end{equation}
which is tractable and can be diagonalized densely.

For a streamwise-homogeneous base state and periodic boundaries in $x$, the Jacobian $\mathcal{J}$ commutes with the periodic grid shift. Hence for every perturbation of the form
$\delta q = \hat{q}(y)\, e^{\mathrm{i} k_x x}$,
\begin{equation}
    \mathcal{J}\left(\hat{q}(y)\, e^{\mathrm{i} k_x x}\right)
    \;=\; \big(\hat{\mathcal{J}}(k_x)\, \hat{q}\big)(y)\;
          e^{\mathrm{i} k_x x}
    \label{eq:mode_action}
\end{equation}
--- the action of the full Jacobian on a single mode is represented exactly
by the reduced Jacobian acting on its transverse profile.
Eigenpairs of the reduced Jacobian therefore reconstruct eigenpairs of the
full operator: if $\hat{\mathcal{J}}(k_x)\, \hat{q} = \lambda\, \hat{q}$,
then by \eqref{eq:mode_action}
\begin{equation}
    \mathcal{J}\left(\hat{q}(y)\, e^{\mathrm{i} k_x x}\right)
    = \lambda\; \hat{q}(y)\, e^{\mathrm{i} k_x x},
\end{equation}
an exact eigenpair of $\mathcal{J}$, with physical perturbation
$\mathrm{Re}\!\left[\hat{q}(y)\, e^{\mathrm{i} k_x x}\, e^{\lambda t}\right]$,
growth rate $\mathrm{Re}(\lambda)$ and temporal frequency
$\mathrm{Im}(\lambda)$.

We apply this to a pure hydrodynamical simulation, without explicit viscosity, on a
$[0,1]^2$ domain (periodic in $x$, open in $y$) at a resolution of $300^2$ cells. The base state 
is built from a single hyperbolic-tangent transition across
$y_c = 1/2$,
\begin{equation}
    f(y) = f_- + (f_+ - f_-)\,\tfrac{1}{2}\!\left[1 + \tanh\!\frac{y - y_c}{\sigma}\right],
\end{equation}
applied to the density ($f_- = \chi\rho_b = 10$ below, $f_+ = \rho_b = 1$ above)
and the streamwise velocity ($f_\mp = \mp v_\mathrm{shear}/2$), while the pressure
is uniform ($P = 1$) and $v_y \equiv 0$. Here $v_\mathrm{shear} = M_b c_b$ with
$c_b = \sqrt{\gamma P / \rho_b} = \sqrt{5/3}$, $M_b = 0.5$, $\gamma = 5/3$, and the
smoothing length is $\sigma = \lambda_x/10$ with $\lambda_x = L_x/2$. The sharp
reference interface is the $\sigma \to 0$ limit of the same expression.
We calculate the eigenspectra of the following homogeneous-in-$x$ states: two
constant reference states (the dilute upper and dense lower fluids in isolation) and
the sharp and smoothed shear interfaces.

\begin{figure}
    \centering
    \includegraphics[width=\linewidth]{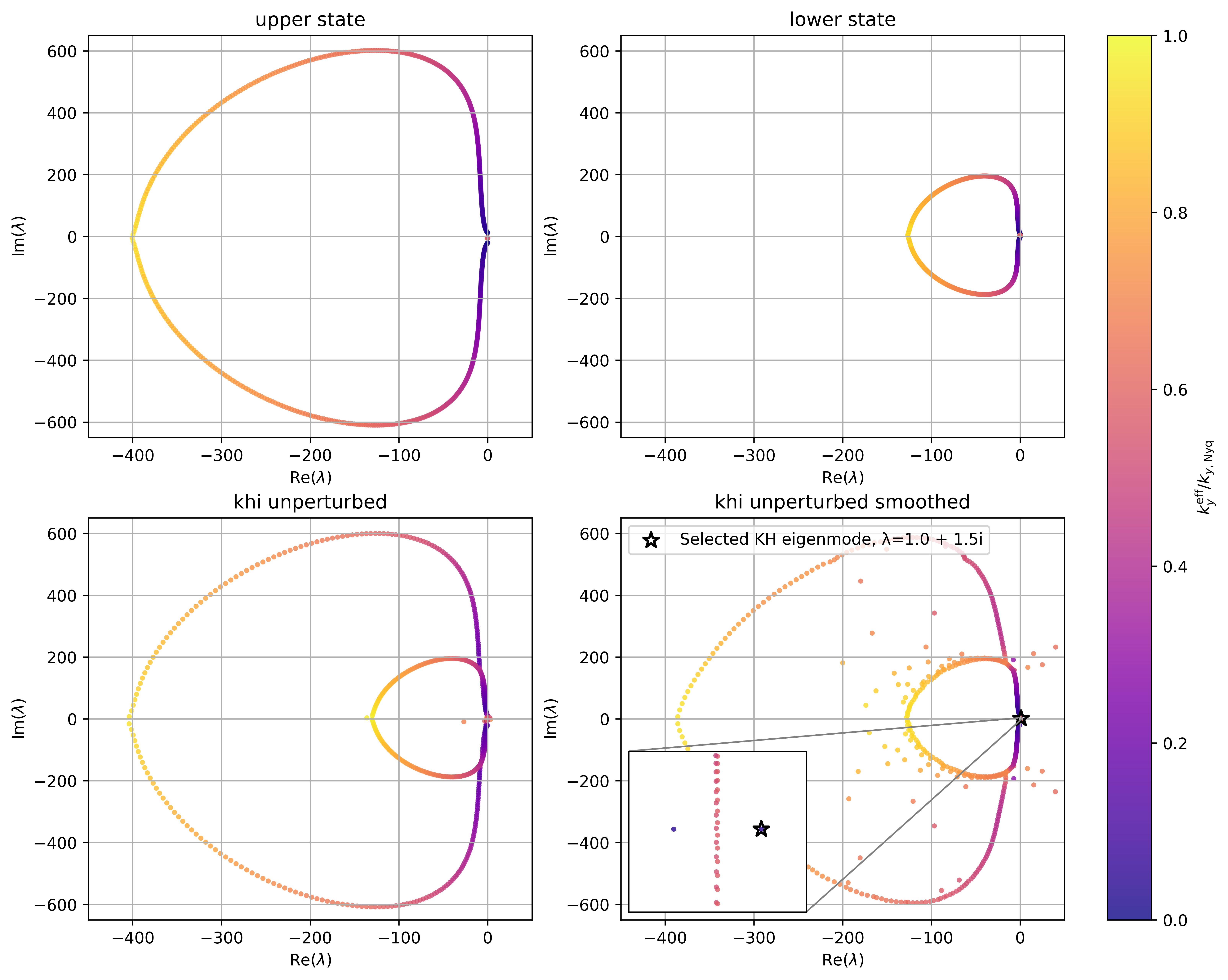}
    \caption{Eigenvalue spectra of the AD-linearized solver Jacobian for a single
    $x$-Fourier mode, for the homogeneous upper and lower states and for the sharp
    and smoothed shear interfaces. Points are colored by the effective transverse
    wavenumber $k_y^\mathrm{eff}/k_{y,\mathrm{Nyq}}$. The homogeneous states yield
    purely damped, oscillatory modes; the interface states additionally support the
    growing Kelvin--Helmholtz mode near the origin (star,
    $\lambda \approx 0.99 + 1.5\,i$), selected by the physical scoring described in
    the text. The inset zooms in on the unstable mode.}
    \label{fig:kh-spectrum}
\end{figure}

The resulting eigenspectra, with the effective transverse wavenumber of the eigenstates 
encoded in color, are plotted in Fig. \ref{fig:kh-spectrum}. The solver strongly damps high-frequency modes in 
time and space. The damped eigenvalues of the homogeneous states form loops with a size determined by
the sound speed. The eigenspectra of the Kelvin--Helmholtz interfaces also include unstable modes in addition
to the superposition of the loops of the homogeneous states.

To seed a clean Kelvin--Helmholtz instability, we select a single growing
eigenmode using two physical diagnostics evaluated on the primitive
transverse-velocity profile $\delta\hat{v}(y)$ (recovered from the conserved
eigenvector). The first is the interface localization: with the envelope
$w(y) = e^{-k_x|y - y_c|}$ normalized to unit maximum,
$L = \sum_y |\delta\hat{v}|^2 w / \sum_y |\delta\hat{v}|^2$. The second is the
effective transverse wavenumber $k_y^\mathrm{eff}/k_{y,\mathrm{Nyq}}$, the
spectral centroid
$k_y^\mathrm{eff} = \big(\sum_{k_y} k_y^2\,|\hat{q}(k_y)|^2 /
\sum_{k_y} |\hat{q}(k_y)|^2\big)^{1/2}$ of the eigenvector's transverse Fourier
spectrum normalized to the Nyquist wavenumber, the same quantity used to
color the spectra in Fig.~\ref{fig:kh-spectrum}. This number separates
the physical mode from the spurious modes of the discretization: the
fastest-growing eigenmodes of the operator are grid-scale oscillations in $y$
with $k_y^\mathrm{eff}/k_{y,\mathrm{Nyq}} \approx 0.6$ (the high-$k_y^\mathrm{eff}$
points in Fig.~\ref{fig:kh-spectrum}), whereas the physical Kelvin--Helmholtz
mode is smooth across the smoothed interface,
$k_y^\mathrm{eff}/k_{y,\mathrm{Nyq}} \approx 0.05$, since structure finer than
the interface is unresolved. We therefore accept a mode if it grows,
$\mathrm{Re}(\lambda) > 0$, is interface-localized, $L > 0.2$, and is transversely
smooth, $k_y^\mathrm{eff}/k_{y,\mathrm{Nyq}} < 0.25$, and among the accepted modes
take the fastest growing one. Note that these cutoffs are not finely tuned. The smooth
localized modes ($k_y^\mathrm{eff}/k_{y,\mathrm{Nyq}} \lesssim 0.05$) and the
grid-scale modes ($\gtrsim 0.6$) are separated by a wide gap, and the selection
is unchanged for any localization cutoff in $[0.1, 0.3]$ and wavenumber cutoff in
$[0.15, 0.4]$.

For comparison, we also initialize the instability with
a typical transverse-velocity perturbation \citep{Roediger2013},
\begin{equation}
    v_y(x, y) = A\,\sin(k_x x)\,\exp\!\left[-\left(\frac{y - y_c}{\sigma_v}\right)^{\!2}\right],
    \qquad \sigma_v = 0.2\,\lambda_x,
\end{equation}
which leaves $\rho$, $v_x$ and $P$ at their base-state
values. Both initializations use the same streamwise wavenumber
$k_x = 2\pi/\lambda_x$ with $\lambda_x = L_x/2$, and are scaled to the same peak
transverse velocity $A = M_b c_b/20$. This velocity perturbation is not a pure
eigenmode.

\begin{figure}
    \centering
    \includegraphics[width=\linewidth]{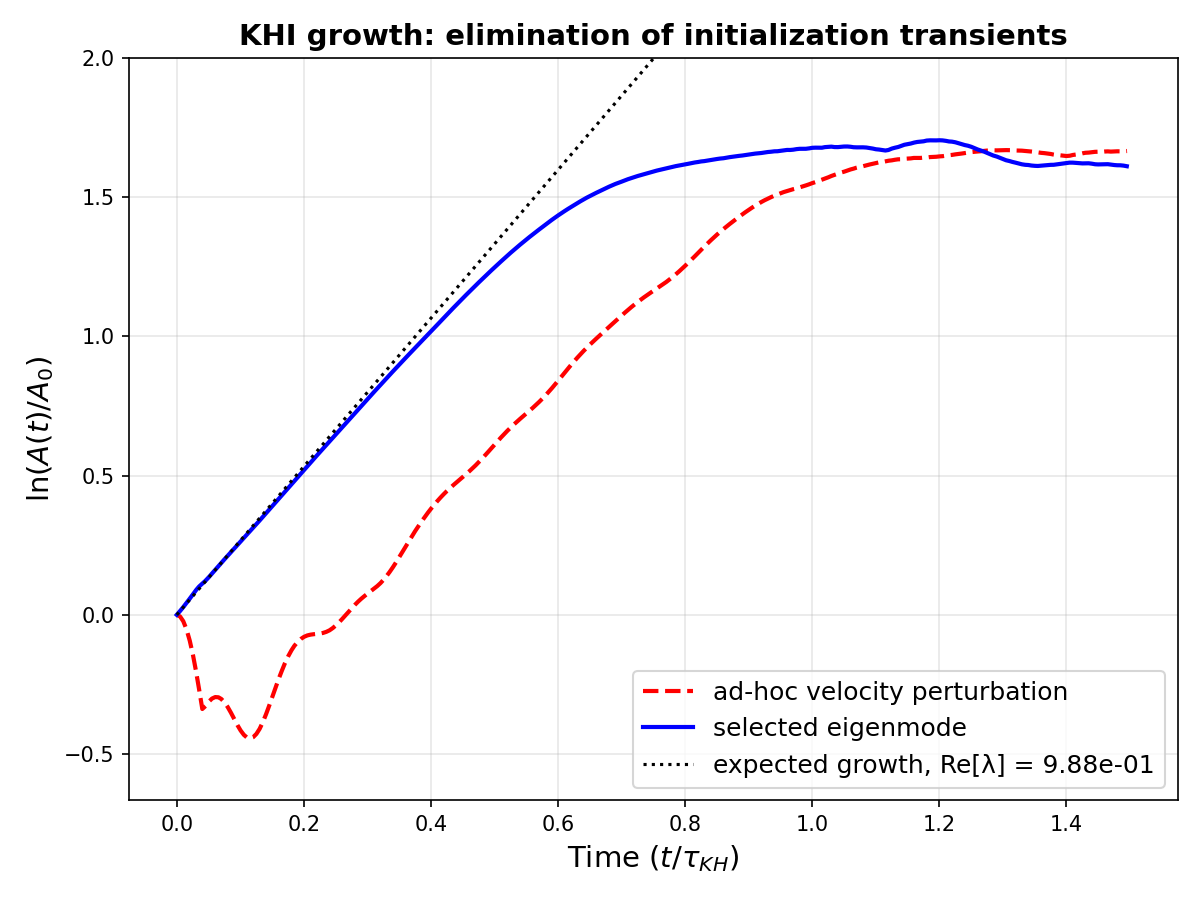}
    \caption{Growth of the mode amplitude $\ln(A(t)/A_0)$ versus $t/\tau_\mathrm{KH}$.
    The eigenmode initialization (blue) follows the predicted growth rate
    $\mathrm{Re}(\lambda)$ (dotted) from $t = 0$, while the pure velocity
    perturbation (red dashed) shows an initialization transient before joining the
    growing branch. Both saturate at comparable amplitude.}
    \label{fig:kh-transient}
\end{figure}

\begin{figure}
    \centering
    \includegraphics[width=\linewidth]{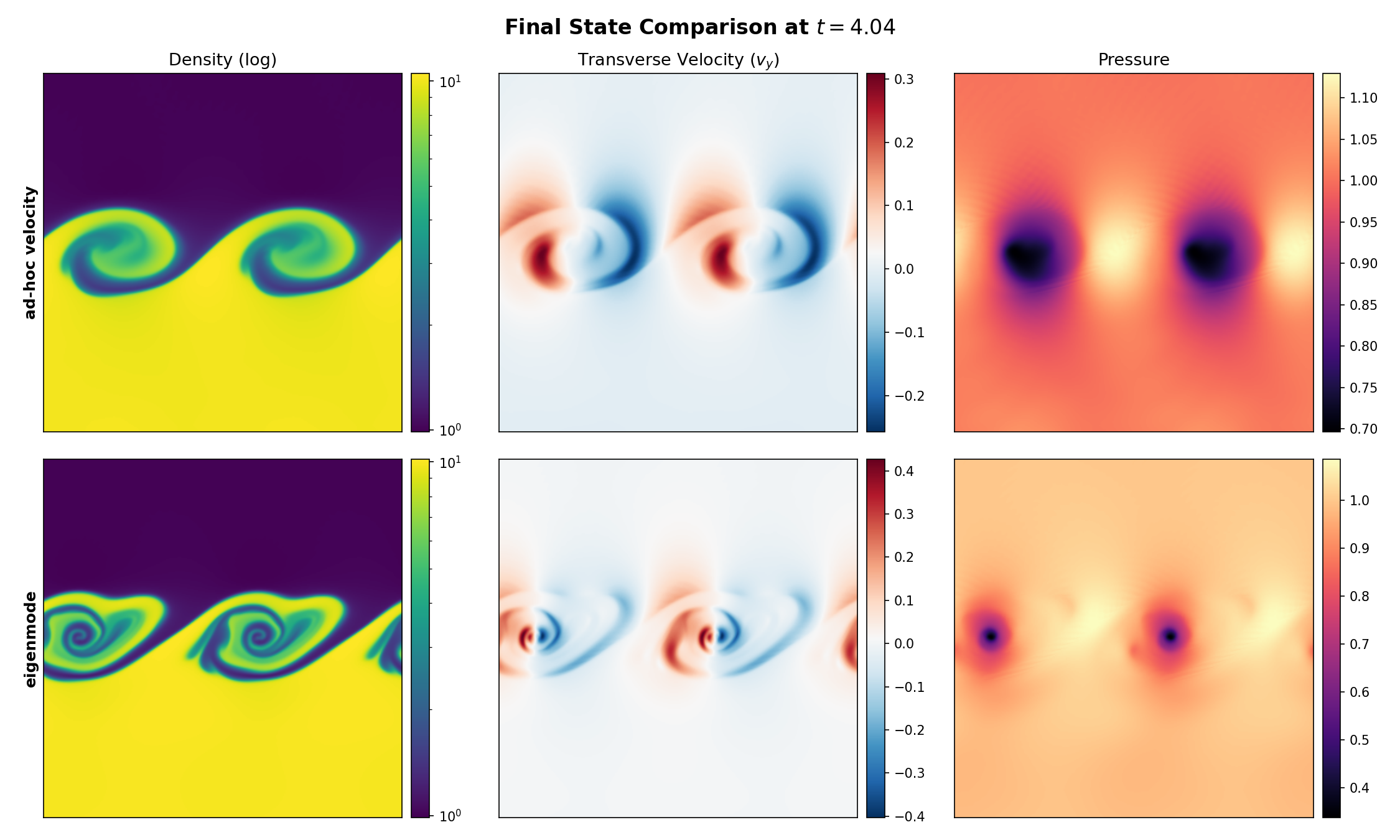}
    \caption{Final states at $t = 4.04 \simeq 1.5\,\tau_\mathrm{KH}$ for the pure
    velocity initialization (top) and the eigenmode initialization (bottom), showing
    log density, transverse velocity $v_y$, and pressure. The eigenmode case, free of
    the initial transient, reaches a more developed billow stage at the same time.}
    \label{fig:kh-final}
\end{figure}

The two initializations are compared in Figs.~\ref{fig:kh-transient}
and~\ref{fig:kh-final}. To quantify instability growth, we project the transverse 
velocity onto the streamwise
Fourier pair at $k_x$,
\begin{equation}
    a_s(y, t) = \frac{2}{N_x}\sum_{i} v_y(x_i, y, t)\,\sin(k_x x_i),
    \qquad
    a_c(y, t) = \frac{2}{N_x}\sum_{i} v_y(x_i, y, t)\,\cos(k_x x_i),
\end{equation}
and take the transverse peak of the resulting modulus,
\begin{equation}
    A(t) = \max_y \sqrt{a_s(y, t)^2 + a_c(y, t)^2}.
\end{equation}
Because we projected onto the Fourier-pair, $A(t)$ is insensitive
to the $x$-phase of the traveling wave. Plotted as $\ln[A(t)/A(0)]$, the eigenmode run
(Fig.~\ref{fig:kh-transient}, blue) grows at the predicted rate $\mathrm{Re}(\lambda)$
from $t = 0$, tracing the dotted reference slope with no initialization transient
until it saturates near $t \simeq \tau_\mathrm{KH}$. The velocity perturbation
(red) instead first \emph{loses} amplitude, as the off-eigenmode and acoustic content
it injected radiate away, and the field reorganizes onto the growing eigenmode.
Both saturate at comparable amplitude. The final states at $t = 4.04 \simeq 1.5\,\tau_\mathrm{KH}$
are compared in Fig.~\ref{fig:kh-final}. The eigenmode initialization has spent 
longer in the nonlinear roll-up phase and therefore shows more tightly
wound billows.

\subsection{Alleviating gradient and optimization pathologies with multiple shooting}
\label{sec:multiple_shooting}

\paragraph{How transitioning to a lifted space helps in theory.}
When we try to perform optimizations through long-time-horizon 
simulations, problems can arise. We will discuss this in the context
of optimizing an initial state through the simulator such that a final
state is reconstructed. However, the main insight of transitioning to a lifted
space while still using the simulator capabilities generalizes.

We can organize optimization approaches for initial state reconstruction along the 
reduced-space vs. full-space optimization discussion \citep{Biros2005}:

\begin{itemize}
    \item reduced-space: we evolve the PDE using a simulator and optimize the 
    initial state by taking derivatives through the simulator, the PDE constraint 
    is enforced by the simulator; in the ML community this ties to differentiable simulators
    \item full-space \citep{Biros2005}: we optimize the full (discretized) 
    trajectory through time, the PDE constraint is enforced by the optimizer 
    (the constraint might either be hard or soft); in the ML community this ties 
    to ODIL \citep{Karnakov2023} and PINNs \citep{Raissi2019}; as examples of full-space 
    methods also see direct collocation \citep{Kelly2017}
    \item multiple shooting \citep{Bock1984,hesse2008,Baake1992}: the trajectory in time 
    is divided into segments which are evolved in parallel and coupled via the optimization 
    problem, the PDE constraint on a segment is enforced by the simulator 
    and in-between segments (hard or soft) by the optimizer; multiple-shooting has been 
    applied in the neural ODE context \citep{iakovlev2023, Turan2022}; also see 
    the 4D-var literature \citep{Tremolet2006}
\end{itemize}

\begin{figure}
 \centering
 \includegraphics[width=1.0\textwidth]{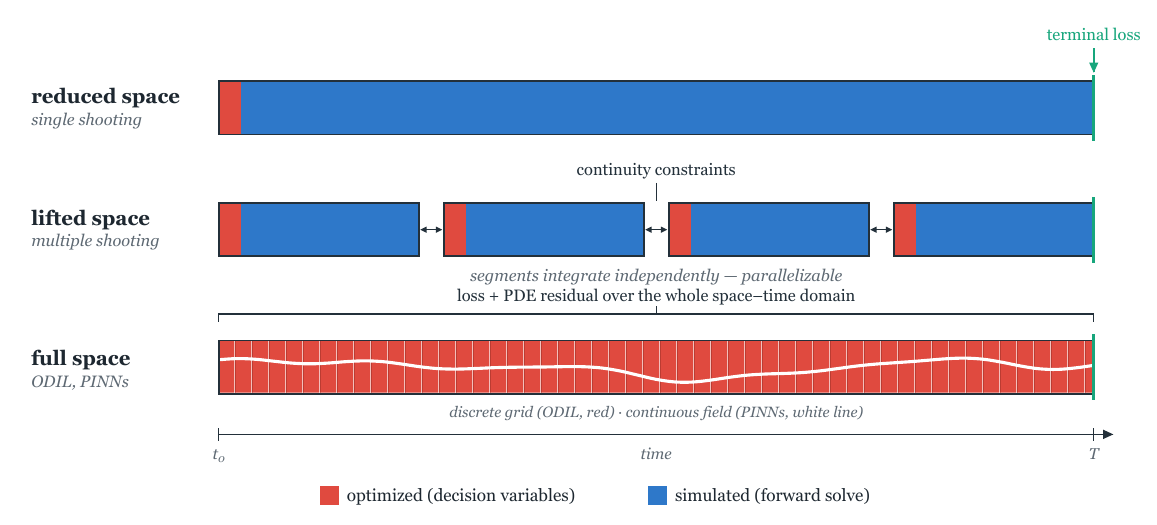}
 \caption{Illustration of PDE-constrained optimization paradigms 
 along the full-space vs reduced-space discussion. In single shooting,
 the PDE-constraint is fully enforced by the simulator while in full-space
 methods like PINNs and ODIL the PDE-constraint is also enforced by 
 the optimizer, multiple shooting sits in-between.}
 \label{fig:optimization_paradigms}
\end{figure}

These different approaches are illustrated in Fig. \ref{fig:optimization_paradigms}.
Also consider the connected shadowing literature for gradients of time-averaged quantities,
e.g. \cite{least_squares_shadowing}.

The broad types of failure in PDE-constrained optimization are

\begin{itemize}
    \item type 1 failure: the constraining power is there, the optimization setup could 
    work in theory but cannot (directly) be translated into a feasible numerical scheme 
    (e.g. gradient explosion to infinity, memory limitations, ...)
    \item type 2 failure: the constraining power is there but the optimization is plagued 
    by local minima, a problem of a multimodal loss landscape (e.g. because we can only 
    move through physical states in single shooting)
    \item type 3 failure: we lack the constraining power, nothing to be truly done about 
    this (except adding constraining power)
\end{itemize}

Type 1 can for instance appear when tangents or adjoints are naively formed over the full time horizon, e.g. in single shooting via gradient descent. Tangent and adjoint variables evolve under linear time-dependent equations and can grow exponentially even when the primal trajectory remains bounded, for instance in a chaotic system (for a short theoretical motivation see Sec.~\ref{app:exploding_gradients}). Even before the growing modes overflow in finite floating-point precision, this leads to a numerical loss of information in the subdominant modes, which cannot be resolved accurately against the largest growing mode. One way to control the numerical degeneracy caused by exponentially growing tangents is to reorthonormalize the propagated tangents on segments \citep{Benettin1980a, Benettin1980b}. While tangents and adjoints are only formed over segments in multiple shooting, a type 1 failure might still reappear when one condenses the problem.

Let us continue to type 2. In single shooting, the optimizer 
can only move through the space of physical trajectories, a difficult task on a 
chaotic flow map (see e.g. \cite{Pires1996}). The loss landscape becomes more
complex with the length of the time-window (see \cite{Ribeiro2020} and Appendix 
A in \cite{iakovlev2023}). Like gradient magnitude explosion this is also rooted 
in the Lyapunov exponent \citep[Theorem 1]{Ribeiro2020}. Also consider the adjacent 
machine learning literature, e.g. \cite{martinelli2026, Ribeiro2020B}.

The solution strategy here is to move to a lifted space where one can move through unphysical
states and the basins of attraction are enlarged. Even if we enforce the full PDE constraints
in full-space methods \citep{Biros2005,Kelly2017,Karnakov2023} or segment continuity in multiple
shooting via hard constraints, these only hold at convergence and the optimizer can move through
unphysical states in the lifted space. Multiple shooting can be seen as a sweet spot where
one does not overburden the optimizer with also enforcing the PDE everywhere (possibly a large,
ill-conditioned problem) compared to full-space methods but still retains advantages 
of a lifted space.

\paragraph{Single shooting vs. multiple shooting on the Kelvin--Helmholtz instability.}

We make the discussion on single vs.\ multiple shooting concrete with a twin
experiment on a two-dimensional Kelvin--Helmholtz instability (KH). 

We use a first-order
optimizer (Adam) for both single and multiple shooting, with identical step
budgets and identical initializations. Note that in the PDE-constrained optimization literature,
a second-order Gauss-Newton solver would be more typical and under hard constraints the multiple
shooting system might be condensed \citep{Bock1984}. However, we opted for the first-order optimizer
for simplicity and to strengthen the connection to the neural network optimization
discussion \citep{martinelli2026}.

\textbf{Problem setup.}
The problem domain is $[0,1]^2$ discretized by $256^2$ cells, periodic in the
streamwise direction $x$ and open in the transverse direction $y$.
The known base
flow is a $\tanh$ shear layer
$v_x^{\mathrm{base}}(y)=\tfrac{\Delta V}{2}\tanh\!\big((y-y_c)/\delta\big)$ with
$\Delta V=1$, $\delta=0.02$, $y_c=0.5$, uniform $\rho_0=1$, low Mach number
$\mathrm{Ma}=0.3$ (effectively incompressible), $\gamma=5/3$, and no explicit
viscosity. We specify time in units of the KH growth time $t_g=\delta/\Delta V$.

The velocity perturbation $v_y'(x,y)$ which we later want to reconstruct is given by
\begin{equation}
  u_0(p) = \mathrm{env}(y)\sum_{k=2}^{6}
    \big[p_k^{\cos}\cos(2\pi k x)+p_k^{\sin}\sin(2\pi k x)\big]\,\hat e_y,
  \qquad p\in\mathbb{R}^{5\times 2},
  \label{eq:khi_control}
\end{equation}
where $\mathrm{env}(y)=\exp[-((y-y_c)/w)^2]$ with $w=0.04$. We use a control
$p\in\mathbb{R}^{5\times 2}$ rather than e.g.\ the full $\mathbb{R}^{256\times 256}$
field to focus on the optimization difficulty in a simplified problem compared to
full field-level inference (an example of which is given in the following
Sec.~\ref{sec:inverse_example}). This reduction also removes the type-3 failure
of finding a full $\mathbb{R}^{256\times 256}$ field that matches the observation.

Let $S_t:\,u_0\mapsto u(t)$ denote the flow map of the differentiable solver. In the
forward simulation the perturbed layer rolls up into discrete Kelvin--Helmholtz
billows, giving the terminal state $u_T = S_T\big(u_0(p)\big)$. As our terminal
observation we use the velocity field
\begin{equation}
  y = H(u_T), \qquad u_T = S_T\big(u_0(p^{\mathrm{truth}})\big),
\end{equation}
where $H$ extracts the velocity from the full state.

We test reconstruction by optimization on a short time horizon
$T=20\,t_g$ (mild folding, $\sim 730$ time steps) and a long time horizon $T=60\,t_g$ (strong nonlinear
folding, of order $\times 40$ tangent growth, $\sim 2200$ time steps).

\textbf{Reduced-space optimization (single shooting).}
Single shooting minimizes the reduced objective
\begin{equation}
  J(p) = \tfrac12\big\| H\big(S_T(u_0(p))\big)-y \big\|_2^2 ,
  \label{eq:khi_single}
\end{equation}
by Adam with a cosine-decay learning rate. The gradient $\nabla_p J$ is formed
by the reverse-mode adjoint over the whole time span $T$.

\textbf{Lifted optimization with multiple shooting.}
We split $[0,T]$ into $M=8$ segments of length $h=T/M$ and add the segment-start
states $s_1,\dots,s_{M-1}$ to the optimization variables ($s_0=u_0(p)$ is tied to the control).
The optimization variable is
thus $\theta=(p,s_1,\dots,s_{M-1})$, and the reconstruction becomes the
equality-constrained least-squares problem
\begin{equation}
  \min_{\theta}\;\;\tfrac12\big\|H\big(S_h(s_{M-1})\big)-y\big\|_2^2
  \qquad\text{s.t.}\qquad
  r_c^j := S_h(s_j)-s_{j+1}=0,\;\; j=0,\dots,M-2 .
  \label{eq:khi_ms}
\end{equation}
The PDE constraint is enforced within a segment by the simulator and
between segments by the continuity constraints.

We can write the constrained optimization problem in terms of the augmented Lagrangian
\begin{equation}
  \mathcal{L}(\theta;\lambda,\rho)=
  \tfrac12\big\|H\big(S_h(s_{M-1})\big)-y\big\|_2^2
  +\sum_{j=0}^{M-2}\Big(\langle\lambda_j,\,r_c^j\rangle+\tfrac{\rho}{2}\|r_c^j\|_2^2\Big),
  \label{eq:khi_al}
\end{equation}
which we optimize alternating inner Adam sweeps over $\theta$ at fixed $(\lambda,\rho)$ with outer
Lagrange multiplier updates $\lambda_j\!\leftarrow\!\lambda_j+\rho\,r_c^j$ and 
penalty weight increases $\rho\!\leftarrow\!1.6\,\rho$ 
whenever the total defect fails to decrease.

We initialize the segments with a globally consistent trajectory (the cold 
control propagated forward, so the defects start at zero).

\begin{figure}
 \centering
 \includegraphics[width=1.0\textwidth]{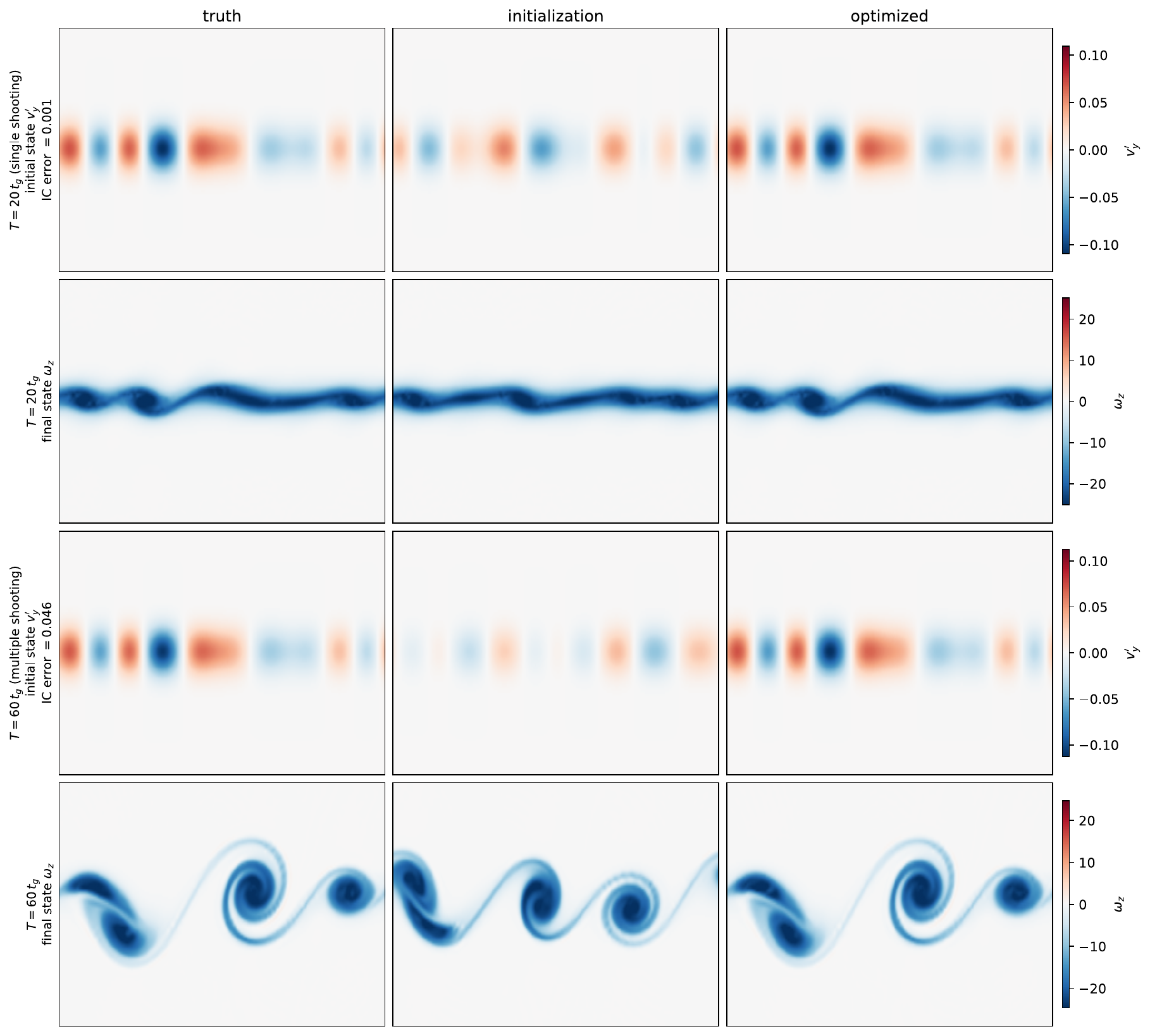}
 \caption{Reconstruction for one successful initialization per horizon
 (recovered $\mathrm{ic\_err}<0.1$): single shooting at $T=20\,t_g$ (top two rows,
 $\mathrm{ic\_err}=0.001$) and multiple shooting at $T=60\,t_g$ (bottom two rows,
 $\mathrm{ic\_err}=0.046$). Columns show the truth, the cold-start initialization,
 and the optimized solution; for each horizon the upper row is the initial
 transverse-velocity perturbation $v_y'$ and the lower row the corresponding
 final-state vorticity $\omega_z$ obtained by forward-integrating that initial state.
 The cold start evolves into billows at the wrong locations, whereas the optimized
 initial state both matches the true seed and reproduces the observed final state.
 Panels span the box $[0,1]^2$ zoomed onto the shear layer $y\in[0.28,0.72]$.}
 \label{fig:reconstruction_example}
\end{figure}

\begin{figure}
 \centering
 \includegraphics[width=1.0\textwidth]{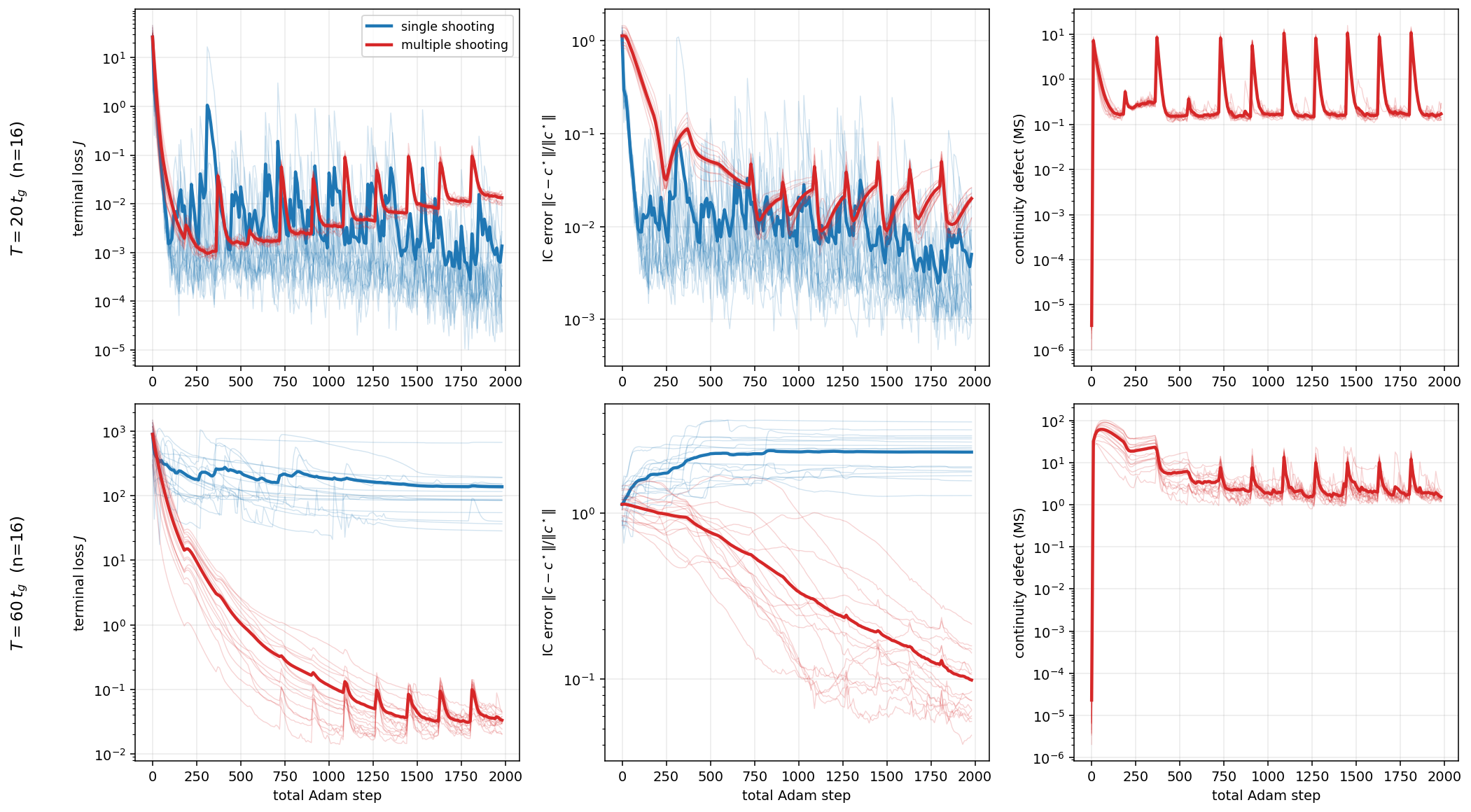}
 \caption{Single vs.\ multiple shooting on the KH instability with an Adam
 optimizer, $N=256$, $M=8$, $16$ cold initializations per configuration. Rows: short horizon
 $T=20\,t_g$ (top, single shooting wins) and long horizon $T=60\,t_g$ (bottom,
 multiple shooting wins). Columns: terminal data loss $J$, relative initial-condition
 error $\mathrm{ic\_err}=\|u_0(p)-u_0^{\mathrm{truth}}\|/\|u_0^{\mathrm{truth}}\|$,
 and the multiple-shooting continuity defect $\sum_j\|r_c^j\|^2$, all versus the total
 number of Adam steps. Thin lines are individual initializations, thick lines the mean
 (single shooting blue, multiple shooting red).}
 \label{fig:khi_shooting}
\end{figure}

\textbf{Results.}
To obtain robust results, we run single- and multiple-shooting from 
multiple ($16$) initial conditions. The terminal loss, IC error and
continuity defect over the number of optimizer iterations are shown 
in Fig. \ref{fig:khi_shooting} for the short (upper row) and long (lower row)
time horizon, an example reconstruction in Fig. \ref{fig:reconstruction_example}.
For the shorter time horizon, multiple shooting is not advantageous.
For the longer time horizon, multiple shooting obtains on average three orders of magnitude
better terminal losses and one order of magnitude lower IC errors, while the optimizer
stalls in the multimodal loss landscape of single shooting.
Note that the multiple-shooting terminal loss is evaluated on the
lifted trajectory, whose segments are not (yet) continuous, so
the three-orders-of-magnitude figure is measured on an infeasible trajectory. The
feasible, directly comparable metric is instead the IC error.
Fig.~\ref{fig:reconstruction_example} additionally confirms the recovered initial state by
forward-integrating it to the observation.

\subsection{Inverse modeling 3D example}
\label{sec:inverse_example}

\begin{figure}
 \centering
 \includegraphics[width=1.0\textwidth]{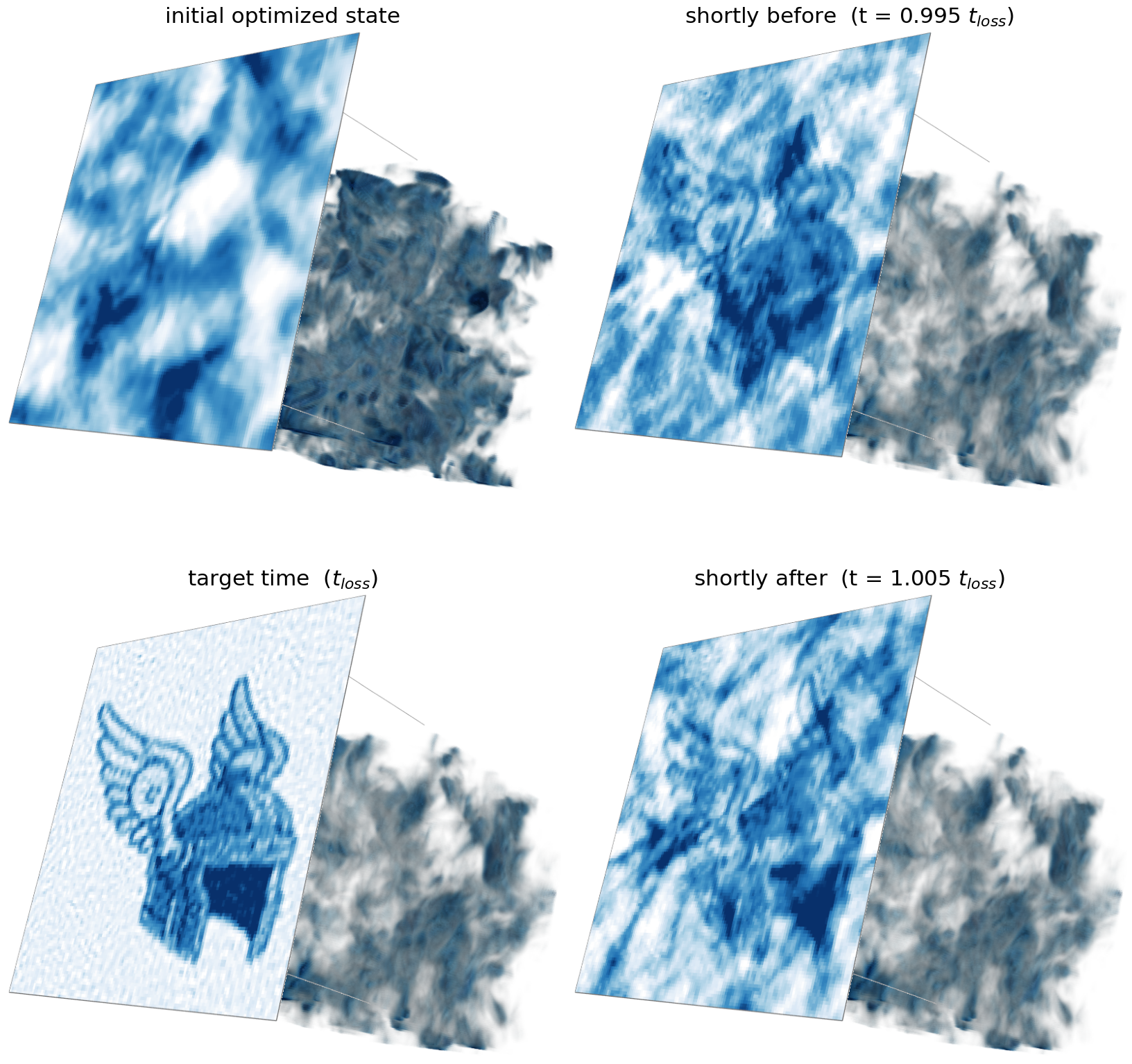}
 \caption{Field-level inference with the differentiable \texttt{astronomix} MHD
  solver. The initial velocity field of a driven $128^3$ turbulence box is
  optimized so that its $z$-axis density projection forms the
  \texttt{astronomix} logo after $\sim$ one turbulent crossing time. Panels show
  volume renderings of the density cube with the column-density projection cast
  onto the side screen, at (left to right) the optimized initial state, shortly
  before, at, and shortly after the target time.}
 \label{fig:field_level_inference}
\end{figure}

In Fig.~\ref{fig:field_level_inference} an example of field-level inference 
through an MHD simulation with the finite difference solver is shown. 
We also use this example to highlight that \texttt{astronomix} makes it easy to
work with physical units via \texttt{astropy}, here employing code units
$\ell_0 = 3\,\mathrm{pc}$, $v_0 = 100\,\mathrm{km\,s^{-1}}$ and $m_0 =
100\,M_\odot$. We choose an ISM-like medium with $n_\mathrm{H} = 2\,\mathrm{cm^{-3}}$,
$p_0/k_B = 3\cdot10^{4}\,\mathrm{K\,cm^{-3}}$,
$B_0 = 13.5\,\mu\mathrm{G}$ and $\gamma = 5/3$.

We first generate a saturated turbulent state by stochastic turbulent
driving at an injection rate $\dot{E} = 4.3\cdot10^{34}\,
\mathrm{erg\,s^{-1}}$ for $\sim 3.5$ crossing times, reaching an rms velocity
$v_\mathrm{rms} \approx 0.42\,v_0 \approx 42\,\mathrm{km\,s^{-1}}$.
The driving is then switched off and the full
$3\times128^3$ initial velocity field is taken as the control variable
(the turbulent density, pressure and magnetic field providing the fixed
background), initialized from the driven velocity. It is optimized so that, after
freely evolving for $t_\mathrm{end} = \tfrac{1}{2}\,\ell_0/v_\mathrm{rms}$, roughly one turbulent crossing time, the density projected
along the $z$-axis reproduces the
\texttt{astronomix} logo. The objective is the mean-squared error between the projected
and target column densities. Gradients of this objective with respect to the
$\sim\!6.3\cdot10^{6}$ velocity degrees of freedom are obtained by reverse-mode
automatic differentiation through the (checkpointed) MHD time integration. We use
Adam as the optimizer. Note that in a physical initial condition recovery setting one would typically want
to apply regularization (a form of prior), e.g. total variation or $H^1$ regularization, which we
do not discuss here.

Optimization on high-resolution simulations is costly
in terms of memory and compute: The number of cells scales
with $N^3$ and, due to the CFL criterion, the number of time steps
typically with $N$, such that the computation cost scales with $N^4$.
Regarding backpropagation a further factor can play a role: At higher
resolution we can afford fewer checkpoints in memory, necessitating
more recomputation. We will explore ideas of reducing field level inference costs in future work (see also Sec. \ref{sec:conclusion}).

Despite optimizing through about one turbulent crossing time covered by $\sim 1100$ time steps, we successfully obtained a final projection visibly resembling the \texttt{astronomix} logo at a mean-squared error of $\sim 5.6 \cdot 10^{-6}$.

\section{Summary and outlook}
\label{sec:conclusion}

We have introduced \texttt{astronomix}, a differentiable astrophysical 
fluid simulator written in \texttt{JAX}. Building upon the fifth-order finite difference WENO
scheme with a constrained transport right-hand side evolving interface
magnetic fields by \cite{how_mhd}, we have introduced a fourth-order
semi-discretely energy-conserving self-gravity scheme and improved
stability mechanisms. The forward simulator was tested on standard problems 
as well as highly magnetized and high-Mach turbulence and a challenging
gravitational collapse.

We showed that custom GPU kernels written by coding agents
in \texttt{Pallas}, the JAX kernel language, based on our \texttt{JAX}
code as a numerical specification language, could vastly improve forward
(in our 3D Alfvén wave test a factor of $\sim 5$ to $9$), backward performance
(up to a factor of $17$) and memory efficiency (by a factor of $5$ to $10$).
On a single GPU, \texttt{astronomix} displayed
runtimes comparable to those of AthenaPK but at vastly lower compute time
to a given error on smooth problems owing to the higher order. Strong scaling
to $8$ NVIDIA H200 GPUs with a speedup of up to $\sim 6.5$ and weak scaling
to $16$ H100 GPUs on four nodes with a weak-scaling efficiency of $76\%$ were
demonstrated.

\texttt{astronomix} is differentiable: functions of the outcomes of astrophysical
simulations can be differentiated with respect to the simulation parameters including
the initial conditions. We introduced standard tests to verify such gradients: analytic
functional derivatives for smooth problems and comparisons against finite differencing
for a shock tube setup. We demonstrated how automatic differentiation can be used for seeding eigenmodes
of a Kelvin--Helmholtz instability and for optimizing initial conditions on a target loss.
We showcased how issues in optimization through long-time-horizon simulations can
be alleviated via multiple shooting, a technique from the PDE-constrained optimization
literature.

This work lays the foundations for a number of future contributions, ranging from expanding the numerical capabilities of \texttt{astronomix} through foundational method development to forward and gradient-based applications. Regarding the numerical capabilities, a multigrid method could be employed to speed up the Poisson solve in simulations with self-gravity \citep{tomida23}. Additionally, the turbulent driving module of \citet{turbgen} might be implemented, expanding on our current formulation in terms of driving modes and amplitude control. With respect to stability mechanisms, a dual-energy formalism could be implemented for simulations with hypersonic bulk flows \citep[Sec. 2.4]{Bryan1995}. Further physics modules to be included are N-body dynamics \citep{viterbo2025}, radiative transfer \citep{branca2025} and chemistry \citep{gijs2026}, the latter two possibly in the form of neural surrogates \citep{Janssen2026, rost2025}. Regarding foundational method development, we are, for instance, interested in reducing the cost of optimizing initial conditions through the simulator. On this problem, one might progress from low- to high-$k$ modes in Fourier space during the optimization, with corresponding changes in simulation resolution. In terms of forward applications, high-resolution simulations with driven magnetohydrodynamical turbulence and self-gravity would be of interest. Regarding gradient-based applications, we are targeting field-level inference problems based on astronomical observations, such as inferring the initial conditions of the interstellar medium from observations of supernova remnants. First examples of training machine learning models inside \texttt{astronomix} can be found in \cite{storcks2025}. We are working on generalizing these models both in terms of problem setups and resolutions.  An application using \texttt{astronomix} as an efficient forward simulator for training data generation can be found in \cite{niklas2026}.

As \texttt{astronomix} is public and open source, we are excited to see what kind
of forward simulation, inverse modeling and machine learning applications
it will enable for the broader community.

\printcredits

\section*{Acknowledgements}
We would like to thank Philip Mocz, Benjamin Horowitz and Ulrich Steinwandel for helpful discussions on differentiable hydrodynamics. We would like to thank Philipp Grete for discussions on turbulence and performance. We would like to thank Max Gronke and Lachlan Lancaster for general discussions.
This project was made possible by funding from the Carl-Zeiss-Stiftung.
This work was supported by the Deutsche Forschungsgemeinschaft (DFG, German Research Foundation) under Germany’s Excellence Strategy EXC 2181/1 - 390900948 (the Heidelberg STRUCTURES Excellence Cluster). We acknowledge the usage of the AI-clusters Tom and Jerry funded by the Field of Focus 2 of Heidelberg University. AI-tools such as Claude were used as sparring partners in discussions, for proof-reading and adaptations to the originally fully human-written codebase.

\printbibliography

\appendix

\section{Explicit Runge--Kutta integrators available in \texttt{astronomix}}
\label{app:rk_integrators}

Time integration is performed with explicit Runge--Kutta methods. The implementation
of these methods in \texttt{astronomix} is very general: all operations act componentwise
on a general pytree $U$ whose time derivative is given by the right-hand side $L(U)$,
i.e.\ $\mathrm{d}U/\mathrm{d}t = L(U)$.
Additionally, at the start of each stage a hook $\mathcal{P}$ can be applied, e.g.\
the boundary handling. One can also pass a finalization operation $\mathcal{F}$,
e.g.\ the calculation of the cell-centered magnetic fields from the staggered
interface fields. In the stage formulae below, each right-hand-side evaluation acts
on the hooked state ($L(U^{(i)})$ stands for $L(\mathcal{P}(U^{(i)}))$)
and $\mathcal{F}$ is applied once to the returned state $U_{n+1}$.

\subsection{SSPRK4: five-stage, fourth-order SSP}
\label{app:rk:ssprk4}

The default time integrator for the finite-difference method
is the optimal five-stage, fourth-order
strong-stability-preserving Runge--Kutta method of
\citet{SpiteriRuuth2002}, written here in Shu--Osher convex-combination
form:
\begin{align}
  U^{(1)} &= U_n + 0.391752226571890\,\Delta t\,L(U_n), \notag\\
  U^{(2)} &= 0.444370493651235\,U_n + 0.555629506348765\,U^{(1)} \notag\\
          &\quad + 0.368410593050371\,\Delta t\,L(U^{(1)}), \notag\\
  U^{(3)} &= 0.620101851488403\,U_n + 0.379898148511597\,U^{(2)} \notag\\
          &\quad + 0.251891774271694\,\Delta t\,L(U^{(2)}), \notag\\
  U^{(4)} &= 0.178079954393132\,U_n + 0.821920045606868\,U^{(3)} \notag\\
          &\quad + 0.544974750228521\,\Delta t\,L(U^{(3)}), \notag\\
  U_{n+1} &= 0.517231671970585\,U^{(2)} + 0.096059710526147\,U^{(3)} \notag\\
          &\quad + 0.063692468666290\,\Delta t\,L(U^{(3)}) \notag\\
          &\quad + 0.386708617503269\,U^{(4)}
                 + 0.226007483236906\,\Delta t\,L(U^{(4)}).
  \label{eq:ssprk4}
\end{align}
Every stage is a convex combination of the current iterates advanced by
forward-Euler substeps with non-negative weights, so the method inherits any
convex functional bound (monotonicity, positivity of density and pressure,
TVD) that holds for the forward-Euler update $U+\Delta t\,L(U)$, now valid up
to the rescaled step $\mathcal{C}\,\Delta t_{\mathrm{FE}}$
\citep{GottliebShuTadmor2001,GKS2011}. Its SSP (CFL) coefficient is
$\mathcal{C}\approx 1.508$, i.e.\ an effective coefficient
$\mathcal{C}/s\approx 0.302$ over $s=5$ stages. Five stages is the practical
minimum here: no four-stage, fourth-order explicit SSP Runge--Kutta method
with a positive SSP coefficient exists, and no explicit SSP Runge--Kutta
method with a positive SSP coefficient can exceed fourth order
\citep{GKS2011}.

\subsection{LSRK4: five-stage, fourth-order, low-storage}
\label{app:rk:lsrk4}

When memory is the binding constraint we use the five-stage, fourth-order
$2N$-storage Runge--Kutta scheme of
\citet{CarpenterKennedy1994}. It carries only two registers, the state $U$
and an increment accumulator $\Delta U$, and is given by
\begin{equation}
  \Delta U^{(i)} = A_i\,\Delta U^{(i-1)} + \Delta t\,L\!\bigl(U^{(i-1)}\bigr),
  \qquad
  U^{(i)} = U^{(i-1)} + B_i\,\Delta U^{(i)},
  \label{eq:lsrk4}
\end{equation}
for $i=1,\dots,5$, initialized with $U^{(0)}=U_n$ and $\Delta U^{(0)}=0$, and
returning $U_{n+1}=U^{(5)}$. Since $A_1=0$, the zero initialization makes the
result independent of the (otherwise arbitrary) starting register and renders
the first stage a plain forward-Euler micro-step. The coefficients are listed
in Table~\ref{tab:lsrk4}.
\begin{table}[t]
  \centering
  \caption{Carpenter--Kennedy $2N$-storage LSRK4 coefficients used in
  \eqref{eq:lsrk4}.}
  \label{tab:lsrk4}
  \begin{tabular}{c r r}
    \toprule
    $i$ & $A_i$ & $B_i$ \\
    \midrule
    $1$ & $0$
        & $\tfrac{1432997174477}{9575080441755}$ \\[2pt]
    $2$ & $-\tfrac{567301805773}{1357537059087}$
        & $\tfrac{5161836677717}{13612068292357}$ \\[2pt]
    $3$ & $-\tfrac{2404267990393}{2016746695238}$
        & $\tfrac{1720146321549}{2090206949498}$ \\[2pt]
    $4$ & $-\tfrac{3550918686646}{2091501179385}$
        & $\tfrac{3134564353537}{4481467310338}$ \\[2pt]
    $5$ & $-\tfrac{1275806237668}{842570457699}$
        & $\tfrac{2277821191437}{14882151754819}$ \\
    \bottomrule
  \end{tabular}
\end{table}
This scheme is not strong-stability-preserving, but combines fourth-order
accuracy with a large linear stability region and a storage footprint of only
two state arrays.

\subsection{RK2-SSP: two-stage, second-order SSP}
\label{app:rk:rk2}

The two-stage, second-order SSP Runge--Kutta method \citep{ShuOsher1988}, i.e.\
Heun's method / the explicit trapezoidal rule, is given by
\begin{align}
  U^{(1)} &= U_n + \Delta t\,L(U_n), \notag\\
  U_{n+1} &= \tfrac{1}{2}\,U_n
             + \tfrac{1}{2}\Bigl(U^{(1)} + \Delta t\,L\!\bigl(U^{(1)}\bigr)\Bigr)
           \;=\; U_n + \tfrac{\Delta t}{2}\Bigl(L(U_n) + L\!\bigl(U^{(1)}\bigr)\Bigr).
  \label{eq:rk2ssp}
\end{align}
It has SSP coefficient $\mathcal{C}=1$ (effective coefficient
$\mathcal{C}/s=0.5$). Both stages are convex combinations of the iterates
advanced by forward-Euler substeps, so the method preserves the same convex
bounds as the forward-Euler update at the full CFL number.

\section{(Magneto)hydrodynamical equations and their eigen-structure}
\label{app:equations}

For every system below the governing equations are written in conservative form
\begin{equation}
  \partial_t \vec{U} + \partial_x \vec{F}_x + \partial_y \vec{F}_y + \partial_z \vec{F}_z = \vec{0},
\end{equation}
with conserved state vector $\vec{U}$ and physical fluxes $\vec{F}_{x,y,z}$. Only the
$x$-flux $\vec{F}_x$ and the eigen-structure of the associated flux Jacobian
$\mathsf{A}_x \equiv \partial \vec{F}_x / \partial \vec{U}$ are listed; the fluxes and
eigen-structure in the $y$- and $z$-directions follow by cyclic relabelling of the
velocity (and, for MHD, magnetic-field) components. The right eigenvectors
$\vec{r}_i$ and left eigenvectors $\vec{\ell}_i$ satisfy
$\mathsf{A}_x \vec{r}_i = \lambda_i \vec{r}_i$, $\vec{\ell}_i \mathsf{A}_x = \lambda_i \vec{\ell}_i$,
and are normalized to the bi-orthonormal pair $\vec{\ell}_i \cdot \vec{r}_j = \delta_{ij}$.

Throughout we abbreviate $v^2 = v_x^2+v_y^2+v_z^2$ and (for MHD) $B^2 = B_x^2+B_y^2+B_z^2$,
write $\bar\gamma \equiv \gamma-1$, and denote by $a$ the sound speed,
\begin{equation}
  a = \sqrt{\frac{\gamma P}{\rho}} \quad\text{(adiabatic)},
  \qquad
  a = c_s = \text{const.} \quad\text{(isothermal, } P = c_s^2\rho\text{)} .
\end{equation}
The eigenvectors are written in terms of the local primitive variables. In the
implementation they are evaluated at cell interfaces from momentum-averaged
(Roe-type) interface states, and the square roots and reciprocals are regularized
with small floors and a differentiable square root; these are numerical safeguards
that do not alter the algebraic structure given here.

\subsection{Adiabatic hydrodynamics}
\label{app:hydro_adiabatic}

The conserved state and $x$-flux are
\begin{equation}
  \vec{U} =
  \begin{pmatrix} \rho \\ \rho v_x \\ \rho v_y \\ \rho v_z \\ E \end{pmatrix},
  \qquad
  \vec{F}_x =
  \begin{pmatrix}
    \rho v_x \\
    \rho v_x^2 + P \\
    \rho v_x v_y \\
    \rho v_x v_z \\
    (E + P)\, v_x
  \end{pmatrix},
  \qquad
  E = \frac{P}{\bar\gamma} + \tfrac12 \rho v^2 .
\end{equation}
With the specific enthalpy $H = (E+P)/\rho$, the eigenvalues are
\begin{equation}
  \lambda = \bigl(\, v_x - a,\; v_x,\; v_x,\; v_x,\; v_x + a \,\bigr),
\end{equation}
i.e.\ a left-/right-going acoustic wave ($v_x\mp a$) and a triply degenerate
$v_x$ wave (entropy plus two transverse shear waves). The right eigenvectors,
ordered $(\,v_x-a,\ \text{entropy},\ \text{shear-}y,\ \text{shear-}z,\ v_x+a\,)$, form the matrix
\begin{equation}
  \mathsf{R} =
  \begin{pmatrix}
    1            & 1              & 0   & 0   & 1            \\
    v_x - a      & v_x            & 0   & 0   & v_x + a      \\
    v_y          & v_y            & 1   & 0   & v_y          \\
    v_z          & v_z            & 0   & 1   & v_z          \\
    H - v_x a    & \tfrac12 v^2   & v_y & v_z & H + v_x a
  \end{pmatrix}.
\end{equation}
The corresponding left eigenvectors (rows) are
\begin{align}
  \vec{\ell}_{v_x-a}     &= \frac{1}{2a^2}\Bigl(\, \tfrac12\bar\gamma v^2 + v_x a,\;\; -(\bar\gamma v_x + a),\;\; -\bar\gamma v_y,\;\; -\bar\gamma v_z,\;\; \bar\gamma \,\Bigr), \\
  \vec{\ell}_{\text{ent}} &= \frac{1}{a^2}\Bigl(\, a^2 - \tfrac12\bar\gamma v^2,\;\; \bar\gamma v_x,\;\; \bar\gamma v_y,\;\; \bar\gamma v_z,\;\; -\bar\gamma \,\Bigr), \\
  \vec{\ell}_{\text{shear-}y} &= \bigl(\, -v_y,\;\; 0,\;\; 1,\;\; 0,\;\; 0 \,\bigr), \\
  \vec{\ell}_{\text{shear-}z} &= \bigl(\, -v_z,\;\; 0,\;\; 0,\;\; 1,\;\; 0 \,\bigr), \\
  \vec{\ell}_{v_x+a}     &= \frac{1}{2a^2}\Bigl(\, \tfrac12\bar\gamma v^2 - v_x a,\;\; -(\bar\gamma v_x - a),\;\; -\bar\gamma v_y,\;\; -\bar\gamma v_z,\;\; \bar\gamma \,\Bigr).
\end{align}
The one- and two-dimensional systems are obtained by dropping the shear-$z$
(and, in 1D, the shear-$y$) waves together with the associated momentum
components.

\subsection{Isothermal hydrodynamics}
\label{app:hydro_isothermal}

There is no energy equation; the pressure is fixed by $P = a^2\rho$ with constant
sound speed $a = c_s$. The conserved state and $x$-flux are
\begin{equation}
  \vec{U} =
  \begin{pmatrix} \rho \\ \rho v_x \\ \rho v_y \\ \rho v_z \end{pmatrix},
  \qquad
  \vec{F}_x =
  \begin{pmatrix}
    \rho v_x \\
    \rho v_x^2 + a^2 \rho \\
    \rho v_x v_y \\
    \rho v_x v_z
  \end{pmatrix}.
\end{equation}
The eigenvalues are
\begin{equation}
  \lambda = \bigl(\, v_x - a,\; v_x,\; v_x,\; v_x + a \,\bigr),
\end{equation}
with no entropy wave (only the two acoustic waves $v_x\mp a$ and the two
transverse shear waves at $v_x$). The right eigenvectors, ordered
$(\,v_x-a,\ \text{shear-}y,\ \text{shear-}z,\ v_x+a\,)$, are
\begin{equation}
  \mathsf{R} =
  \begin{pmatrix}
    1        & 0   & 0   & 1        \\
    v_x - a  & 0   & 0   & v_x + a  \\
    v_y      & 1   & 0   & v_y      \\
    v_z      & 0   & 1   & v_z
  \end{pmatrix},
\end{equation}
and the left eigenvectors are
\begin{align}
  \vec{\ell}_{v_x-a}          &= \frac{1}{2a^2}\bigl(\, a^2 + v_x a,\;\; -a,\;\; 0,\;\; 0 \,\bigr)
                                = \tfrac12\bigl(\, 1 + v_x/a,\;\; -1/a,\;\; 0,\;\; 0 \,\bigr), \\
  \vec{\ell}_{\text{shear-}y} &= \bigl(\, -v_y,\;\; 0,\;\; 1,\;\; 0 \,\bigr), \\
  \vec{\ell}_{\text{shear-}z} &= \bigl(\, -v_z,\;\; 0,\;\; 0,\;\; 1 \,\bigr), \\
  \vec{\ell}_{v_x+a}          &= \frac{1}{2a^2}\bigl(\, a^2 - v_x a,\;\; a,\;\; 0,\;\; 0 \,\bigr)
                                = \tfrac12\bigl(\, 1 - v_x/a,\;\; 1/a,\;\; 0,\;\; 0 \,\bigr).
\end{align}

\subsection{Adiabatic magnetohydrodynamics}
\label{app:mhd_adiabatic}

The conserved state and $x$-flux are
\begin{equation}
  \vec{U} =
  \begin{pmatrix} \rho \\ \rho v_x \\ \rho v_y \\ \rho v_z \\ B_x \\ B_y \\ B_z \\ E \end{pmatrix},
  \qquad
  \vec{F}_x =
  \begin{pmatrix}
    \rho v_x \\
    \rho v_x^2 + P_T - B_x^2 \\
    \rho v_x v_y - B_x B_y \\
    \rho v_x v_z - B_x B_z \\
    0 \\
    B_y v_x - B_x v_y \\
    B_z v_x - B_x v_z \\
    (E + P_T)\, v_x - (\vec{v}\cdot\vec{B})\, B_x
  \end{pmatrix},
\end{equation}
with total pressure $P_T = P + \tfrac12 B^2$ and total energy
$E = P/\bar\gamma + \tfrac12\rho v^2 + \tfrac12 B^2$. The $B_x$ flux vanishes
($B_x$ is constant along $x$), so the flux Jacobian acts on the reduced seven-component
state $(\rho,\ \rho v_x,\ \rho v_y,\ \rho v_z,\ B_y,\ B_z,\ E)$, and the
eigenvectors below carry no $B_x$ entry.

Defining the squared sound, Alfv\'en, fast- and slow-magnetosonic speeds
\begin{align}
  a^2 &= \frac{\gamma P}{\rho}, \qquad C_A^2 = \frac{B_x^2}{\rho}, \\
  C_{f,s}^2 &= \frac12\!\left[\, \frac{B^2}{\rho} + a^2
    \;\pm\; \sqrt{\left(\frac{B^2}{\rho} + a^2\right)^2 - 4\,\frac{B_x^2}{\rho}\,a^2}\, \right]
    \quad (+ \to f,\; - \to s),
\end{align}
the seven eigenvalues are
\begin{equation}
  \lambda = \bigl(\, v_x - C_f,\; v_x - C_A,\; v_x - C_s,\; v_x,\; v_x + C_s,\; v_x + C_A,\; v_x + C_f \,\bigr).
\end{equation}

The eigenvectors use the renormalization quantities of \cite{Roe1996},
which remove the indeterminacy of the MHD eigenvectors at the degeneracies. With
\begin{equation}
  \alpha_f^2 = \frac{a^2 - C_s^2}{C_f^2 - C_s^2}, \qquad
  \alpha_s^2 = \frac{C_f^2 - a^2}{C_f^2 - C_s^2}, \qquad
  \alpha_f^2 + \alpha_s^2 = 1,
\end{equation}
the normalized tangential field
\begin{equation}
  \beta_y = \frac{B_y}{\sqrt{B_y^2 + B_z^2}}, \qquad
  \beta_z = \frac{B_z}{\sqrt{B_y^2 + B_z^2}}
  \qquad\bigl(\beta_y=\beta_z=\tfrac{1}{\sqrt2}\ \text{when}\ B_\perp\to 0\bigr),
\end{equation}
the sign $S = \operatorname{sgn}(B_x)$, the abbreviation
$\Gamma_2 \equiv \dfrac{\gamma-2}{\gamma-1}$, and the auxiliary sign
\begin{equation}
  \sigma =
  \begin{cases}
    \operatorname{sgn}(B_y), & B_y \neq 0,\\
    \operatorname{sgn}(B_z), & B_y = 0,
  \end{cases}
\end{equation}
the fast-family eigenvectors are multiplied by $\sigma$ when $a < C_A$ and the
slow-family eigenvectors are multiplied by $\sigma$ when $a \geq C_A$ (the
$\times\,\sigma$ tags below); this single sign keeps the fast/slow branches
continuous through the $a = C_A$ crossing.

\paragraph{Right eigenvectors.}
The components are listed in the order $(\rho,\ \rho v_x,\ \rho v_y,\ \rho v_z,\ B_y,\ B_z,\ E)$.
\begin{align}
\vec{r}_{v_x-C_f} &=
  \begin{pmatrix}
    \alpha_f \\
    \alpha_f (v_x - C_f) \\
    \alpha_f v_y + \alpha_s C_s \beta_y S \\
    \alpha_f v_z + \alpha_s C_s \beta_z S \\
    a\,\alpha_s \beta_y/\sqrt{\rho} \\
    a\,\alpha_s \beta_z/\sqrt{\rho} \\
    \alpha_f\!\left(C_f^2 - C_f v_x + \tfrac12 v^2 - \Gamma_2 a^2\right) + \alpha_s C_s (\beta_y v_y + \beta_z v_z) S
  \end{pmatrix}
  \;(\times\,\sigma\ \text{if } a<C_A),
  \\[4pt]
\vec{r}_{v_x-C_A} &=
  \begin{pmatrix}
    0 \\ 0 \\ -\beta_z \\ \beta_y \\ -\beta_z S/\sqrt{\rho} \\ \beta_y S/\sqrt{\rho} \\ \beta_y v_z - \beta_z v_y
  \end{pmatrix},
  \\[4pt]
\vec{r}_{v_x-C_s} &=
  \begin{pmatrix}
    \alpha_s \\
    \alpha_s (v_x - C_s) \\
    \alpha_s v_y - \alpha_f C_f \beta_y S \\
    \alpha_s v_z - \alpha_f C_f \beta_z S \\
    -a\,\alpha_f \beta_y/\sqrt{\rho} \\
    -a\,\alpha_f \beta_z/\sqrt{\rho} \\
    \alpha_s\!\left(C_s^2 - C_s v_x + \tfrac12 v^2 - \Gamma_2 a^2\right) - \alpha_f C_f (\beta_y v_y + \beta_z v_z) S
  \end{pmatrix}
  \;(\times\,\sigma\ \text{if } a\geq C_A),
  \\[4pt]
\vec{r}_{v_x} &=
  \begin{pmatrix}
    1 \\ v_x \\ v_y \\ v_z \\ 0 \\ 0 \\ \tfrac12 v^2
  \end{pmatrix},
  \\[4pt]
\vec{r}_{v_x+C_s} &=
  \begin{pmatrix}
    \alpha_s \\
    \alpha_s (v_x + C_s) \\
    \alpha_s v_y + \alpha_f C_f \beta_y S \\
    \alpha_s v_z + \alpha_f C_f \beta_z S \\
    -a\,\alpha_f \beta_y/\sqrt{\rho} \\
    -a\,\alpha_f \beta_z/\sqrt{\rho} \\
    \alpha_s\!\left(C_s^2 + C_s v_x + \tfrac12 v^2 - \Gamma_2 a^2\right) + \alpha_f C_f (\beta_y v_y + \beta_z v_z) S
  \end{pmatrix}
  \;(\times\,\sigma\ \text{if } a\geq C_A),
  \\[4pt]
\vec{r}_{v_x+C_A} &=
  \begin{pmatrix}
    0 \\ 0 \\ -\beta_z \\ \beta_y \\ \beta_z S/\sqrt{\rho} \\ -\beta_y S/\sqrt{\rho} \\ \beta_y v_z - \beta_z v_y
  \end{pmatrix},
  \\[4pt]
\vec{r}_{v_x+C_f} &=
  \begin{pmatrix}
    \alpha_f \\
    \alpha_f (v_x + C_f) \\
    \alpha_f v_y - \alpha_s C_s \beta_y S \\
    \alpha_f v_z - \alpha_s C_s \beta_z S \\
    a\,\alpha_s \beta_y/\sqrt{\rho} \\
    a\,\alpha_s \beta_z/\sqrt{\rho} \\
    \alpha_f\!\left(C_f^2 + C_f v_x + \tfrac12 v^2 - \Gamma_2 a^2\right) - \alpha_s C_s (\beta_y v_y + \beta_z v_z) S
  \end{pmatrix}
  \;(\times\,\sigma\ \text{if } a<C_A).
\end{align}

\paragraph{Left eigenvectors.}
Same component ordering $(\rho,\ \rho v_x,\ \rho v_y,\ \rho v_z,\ B_y,\ B_z,\ E)$.
The fast and slow rows carry an overall factor $1/(2a^2)$ and the same
$\times\,\sigma$ tag as their right-eigenvector partners; the Alfv\'en rows carry
a factor $1/2$.
\begin{align}
\vec{\ell}_{v_x-C_f} &= \frac{1}{2a^2}\Bigl(
    \alpha_f\!\bigl(\tfrac12\bar\gamma v^2 + C_f v_x\bigr) - \alpha_s C_s (\beta_y v_y + \beta_z v_z) S,\;
    -\alpha_f(\bar\gamma v_x + C_f), \notag\\
  &\hphantom{={}\frac{1}{2a^2}\Bigl(}\;
    -\bar\gamma\alpha_f v_y + \alpha_s C_s \beta_y S,\;
    -\bar\gamma\alpha_f v_z + \alpha_s C_s \beta_z S, \notag\\
  &\hphantom{={}\frac{1}{2a^2}\Bigl(}\;
    -\bar\gamma\alpha_f B_y + a\,\alpha_s \beta_y \sqrt{\rho},\;
    -\bar\gamma\alpha_f B_z + a\,\alpha_s \beta_z \sqrt{\rho},\;
    \bar\gamma\alpha_f \Bigr)\;(\times\,\sigma\ \text{if } a<C_A),
  \\[4pt]
\vec{\ell}_{v_x-C_A} &= \frac{1}{2}\Bigl(
    \beta_z v_y - \beta_y v_z,\; 0,\; -\beta_z,\; \beta_y,\;
    -\beta_z S \sqrt{\rho},\; \beta_y S \sqrt{\rho},\; 0 \Bigr),
  \\[4pt]
\vec{\ell}_{v_x-C_s} &= \frac{1}{2a^2}\Bigl(
    \alpha_s\!\bigl(\tfrac12\bar\gamma v^2 + C_s v_x\bigr) + \alpha_f C_f (\beta_y v_y + \beta_z v_z) S,\;
    -\bar\gamma\alpha_s v_x - \alpha_s C_s, \notag\\
  &\hphantom{={}\frac{1}{2a^2}\Bigl(}\;
    -\bar\gamma\alpha_s v_y - \alpha_f C_f \beta_y S,\;
    -\bar\gamma\alpha_s v_z - \alpha_f C_f \beta_z S, \notag\\
  &\hphantom{={}\frac{1}{2a^2}\Bigl(}\;
    -\bar\gamma\alpha_s B_y - a\,\alpha_f \beta_y \sqrt{\rho},\;
    -\bar\gamma\alpha_s B_z - a\,\alpha_f \beta_z \sqrt{\rho},\;
    \bar\gamma\alpha_s \Bigr)\;(\times\,\sigma\ \text{if } a\geq C_A),
  \\[4pt]
\vec{\ell}_{v_x} &= \frac{\bar\gamma}{a^2}\Bigl(
    \frac{a^2}{\bar\gamma} - \tfrac12 v^2,\; v_x,\; v_y,\; v_z,\; B_y,\; B_z,\; -1 \Bigr),
  \\[4pt]
\vec{\ell}_{v_x+C_s} &= \frac{1}{2a^2}\Bigl(
    \alpha_s\!\bigl(\tfrac12\bar\gamma v^2 - C_s v_x\bigr) - \alpha_f C_f (\beta_y v_y + \beta_z v_z) S,\;
    -\bar\gamma\alpha_s v_x + \alpha_s C_s, \notag\\
  &\hphantom{={}\frac{1}{2a^2}\Bigl(}\;
    -\bar\gamma\alpha_s v_y + \alpha_f C_f \beta_y S,\;
    -\bar\gamma\alpha_s v_z + \alpha_f C_f \beta_z S, \notag\\
  &\hphantom{={}\frac{1}{2a^2}\Bigl(}\;
    -\bar\gamma\alpha_s B_y - a\,\alpha_f \beta_y \sqrt{\rho},\;
    -\bar\gamma\alpha_s B_z - a\,\alpha_f \beta_z \sqrt{\rho},\;
    \bar\gamma\alpha_s \Bigr)\;(\times\,\sigma\ \text{if } a\geq C_A),
  \\[4pt]
\vec{\ell}_{v_x+C_A} &= \frac{1}{2}\Bigl(
    \beta_z v_y - \beta_y v_z,\; 0,\; -\beta_z,\; \beta_y,\;
    \beta_z S \sqrt{\rho},\; -\beta_y S \sqrt{\rho},\; 0 \Bigr),
  \\[4pt]
\vec{\ell}_{v_x+C_f} &= \frac{1}{2a^2}\Bigl(
    \alpha_f\!\bigl(\tfrac12\bar\gamma v^2 - C_f v_x\bigr) + \alpha_s C_s (\beta_y v_y + \beta_z v_z) S,\;
    -\bar\gamma\alpha_f v_x + \alpha_f C_f, \notag\\
  &\hphantom{={}\frac{1}{2a^2}\Bigl(}\;
    -\bar\gamma\alpha_f v_y - \alpha_s C_s \beta_y S,\;
    -\bar\gamma\alpha_f v_z - \alpha_s C_s \beta_z S, \notag\\
  &\hphantom{={}\frac{1}{2a^2}\Bigl(}\;
    -\bar\gamma\alpha_f B_y + a\,\alpha_s \beta_y \sqrt{\rho},\;
    -\bar\gamma\alpha_f B_z + a\,\alpha_s \beta_z \sqrt{\rho},\;
    \bar\gamma\alpha_f \Bigr)\;(\times\,\sigma\ \text{if } a<C_A).
\end{align}

\subsection{Isothermal magnetohydrodynamics}
\label{app:mhd_isothermal}

With the isothermal closure $P = a^2\rho$ ($a = c_s$ constant) there is no energy
equation. The conserved state and $x$-flux are
\begin{equation}
  \vec{U} =
  \begin{pmatrix} \rho \\ \rho v_x \\ \rho v_y \\ \rho v_z \\ B_x \\ B_y \\ B_z \end{pmatrix},
  \qquad
  \vec{F}_x =
  \begin{pmatrix}
    \rho v_x \\
    \rho v_x^2 + P_T - B_x^2 \\
    \rho v_x v_y - B_x B_y \\
    \rho v_x v_z - B_x B_z \\
    0 \\
    B_y v_x - B_x v_y \\
    B_z v_x - B_x v_z
  \end{pmatrix},
  \qquad P_T = a^2\rho + \tfrac12 B^2 .
\end{equation}
Again $B_x$ is constant, so the Jacobian acts on the reduced six-component state
$(\rho,\ \rho v_x,\ \rho v_y,\ \rho v_z,\ B_y,\ B_z)$. The Alfv\'en, fast- and
slow-magnetosonic speeds are as before with the constant $a^2 = c_s^2$ in place of
$\gamma P/\rho$,
\begin{equation}
  C_A^2 = \frac{B_x^2}{\rho}, \qquad
  C_{f,s}^2 = \frac12\!\left[\, \frac{B^2}{\rho} + a^2
    \;\pm\; \sqrt{\left(\frac{B^2}{\rho} + a^2\right)^2 - 4\,\frac{B_x^2}{\rho}\,a^2}\, \right],
\end{equation}
and the six eigenvalues are
\begin{equation}
  \lambda = \bigl(\, v_x - C_f,\; v_x - C_A,\; v_x - C_s,\; v_x + C_s,\; v_x + C_A,\; v_x + C_f \,\bigr).
\end{equation}
There is no entropy wave. The auxiliaries $\alpha_f,\alpha_s,\beta_y,\beta_z,S,\sigma$
are defined exactly as in Sec.~\ref{app:mhd_adiabatic} (with the constant $a$).

\paragraph{Right eigenvectors.}
Components in the order $(\rho,\ \rho v_x,\ \rho v_y,\ \rho v_z,\ B_y,\ B_z)$.
\begin{align}
\vec{r}_{v_x-C_f} &=
  \begin{pmatrix}
    \alpha_f \\
    \alpha_f (v_x - C_f) \\
    \alpha_f v_y + \alpha_s C_s \beta_y S \\
    \alpha_f v_z + \alpha_s C_s \beta_z S \\
    a\,\alpha_s \beta_y/\sqrt{\rho} \\
    a\,\alpha_s \beta_z/\sqrt{\rho}
  \end{pmatrix}\!(\times\,\sigma\ \text{if } a<C_A),
&
\vec{r}_{v_x-C_A} &=
  \begin{pmatrix}
    0 \\ 0 \\ -\beta_z \\ \beta_y \\ -\beta_z S/\sqrt{\rho} \\ \beta_y S/\sqrt{\rho}
  \end{pmatrix},
  \\[4pt]
\vec{r}_{v_x-C_s} &=
  \begin{pmatrix}
    \alpha_s \\
    \alpha_s (v_x - C_s) \\
    \alpha_s v_y - \alpha_f C_f \beta_y S \\
    \alpha_s v_z - \alpha_f C_f \beta_z S \\
    -a\,\alpha_f \beta_y/\sqrt{\rho} \\
    -a\,\alpha_f \beta_z/\sqrt{\rho}
  \end{pmatrix}\!(\times\,\sigma\ \text{if } a\geq C_A),
&
\vec{r}_{v_x+C_s} &=
  \begin{pmatrix}
    \alpha_s \\
    \alpha_s (v_x + C_s) \\
    \alpha_s v_y + \alpha_f C_f \beta_y S \\
    \alpha_s v_z + \alpha_f C_f \beta_z S \\
    -a\,\alpha_f \beta_y/\sqrt{\rho} \\
    -a\,\alpha_f \beta_z/\sqrt{\rho}
  \end{pmatrix}\!(\times\,\sigma\ \text{if } a\geq C_A),
  \\[4pt]
\vec{r}_{v_x+C_A} &=
  \begin{pmatrix}
    0 \\ 0 \\ -\beta_z \\ \beta_y \\ \beta_z S/\sqrt{\rho} \\ -\beta_y S/\sqrt{\rho}
  \end{pmatrix},
&
\vec{r}_{v_x+C_f} &=
  \begin{pmatrix}
    \alpha_f \\
    \alpha_f (v_x + C_f) \\
    \alpha_f v_y - \alpha_s C_s \beta_y S \\
    \alpha_f v_z - \alpha_s C_s \beta_z S \\
    a\,\alpha_s \beta_y/\sqrt{\rho} \\
    a\,\alpha_s \beta_z/\sqrt{\rho}
  \end{pmatrix}\!(\times\,\sigma\ \text{if } a<C_A).
\end{align}

\paragraph{Left eigenvectors.}
Same component ordering. The fast/slow rows carry $1/(2a^2)$ and the same
$\times\,\sigma$ tag as their right partners; the Alfv\'en rows carry $1/2$.
\begin{align}
\vec{\ell}_{v_x-C_f} &= \frac{1}{2a^2}\Bigl(
    \alpha_f(a^2 + C_f v_x) - \alpha_s C_s (\beta_y v_y + \beta_z v_z) S,\;
    -\alpha_f C_f,\;
    \alpha_s C_s \beta_y S, \notag\\
  &\hphantom{={}\frac{1}{2a^2}\Bigl(}\;
    \alpha_s C_s \beta_z S,\;
    a\,\alpha_s \beta_y \sqrt{\rho},\;
    a\,\alpha_s \beta_z \sqrt{\rho} \Bigr)\;(\times\,\sigma\ \text{if } a<C_A),
  \\[4pt]
\vec{\ell}_{v_x-C_A} &= \frac{1}{2}\Bigl(
    \beta_z v_y - \beta_y v_z,\; 0,\; -\beta_z,\; \beta_y,\;
    -\beta_z S \sqrt{\rho},\; \beta_y S \sqrt{\rho} \Bigr),
  \\[4pt]
\vec{\ell}_{v_x-C_s} &= \frac{1}{2a^2}\Bigl(
    \alpha_s(a^2 + C_s v_x) + \alpha_f C_f (\beta_y v_y + \beta_z v_z) S,\;
    -\alpha_s C_s,\;
    -\alpha_f C_f \beta_y S, \notag\\
  &\hphantom{={}\frac{1}{2a^2}\Bigl(}\;
    -\alpha_f C_f \beta_z S,\;
    -a\,\alpha_f \beta_y \sqrt{\rho},\;
    -a\,\alpha_f \beta_z \sqrt{\rho} \Bigr)\;(\times\,\sigma\ \text{if } a\geq C_A),
  \\[4pt]
\vec{\ell}_{v_x+C_s} &= \frac{1}{2a^2}\Bigl(
    \alpha_s(a^2 - C_s v_x) - \alpha_f C_f (\beta_y v_y + \beta_z v_z) S,\;
    \alpha_s C_s,\;
    \alpha_f C_f \beta_y S, \notag\\
  &\hphantom{={}\frac{1}{2a^2}\Bigl(}\;
    \alpha_f C_f \beta_z S,\;
    -a\,\alpha_f \beta_y \sqrt{\rho},\;
    -a\,\alpha_f \beta_z \sqrt{\rho} \Bigr)\;(\times\,\sigma\ \text{if } a\geq C_A),
  \\[4pt]
\vec{\ell}_{v_x+C_A} &= \frac{1}{2}\Bigl(
    \beta_z v_y - \beta_y v_z,\; 0,\; -\beta_z,\; \beta_y,\;
    \beta_z S \sqrt{\rho},\; -\beta_y S \sqrt{\rho} \Bigr),
  \\[4pt]
\vec{\ell}_{v_x+C_f} &= \frac{1}{2a^2}\Bigl(
    \alpha_f(a^2 - C_f v_x) + \alpha_s C_s (\beta_y v_y + \beta_z v_z) S,\;
    \alpha_f C_f,\;
    -\alpha_s C_s \beta_y S, \notag\\
  &\hphantom{={}\frac{1}{2a^2}\Bigl(}\;
    -\alpha_s C_s \beta_z S,\;
    a\,\alpha_s \beta_y \sqrt{\rho},\;
    a\,\alpha_s \beta_z \sqrt{\rho} \Bigr)\;(\times\,\sigma\ \text{if } a<C_A).
\end{align}

\section{Finite volume scheme}
\label{app:finite_volume}

Magnetohydrodynamics for the finite volume scheme is handled in an operator-split fashion,
such that we can first independently discuss the pure hydrodynamical update step.

To obtain the fluxes $\vec{F}^*_{i\pm\frac{1}{2}}$ we perform

\begin{enumerate}
    \item a slope-limited linear interpolation of the primitive
    state to the interfaces of each cell to obtain $\vec{W}_{j-\frac{1}{2}R}$ and  $\vec{W}_{j+\frac{1}{2}L}$
    (and the corresponding conservative forms $\vec{U}_{j-\frac{1}{2}R},\vec{U}_{j+\frac{1}{2}L}$)
    \item and a Riemann solve, $\vec{F}^*_{i+\frac{1}{2}}(\vec{U}_{j+\frac{1}{2}L},\vec{U}_{j+\frac{1}{2}R})$
\end{enumerate}

\subsection{Interpolation to the interfaces}

Linear numerical schemes for solving partial differential equations (PDEs), 
having the property of not generating new extrema (monotone scheme), 
can be at most first-order accurate (\textbf{Godunov's theorem}, 
\cite{godunov1959}).

For second-order accuracy in space we linearly extrapolate
from the average fluid state assumed to hold at the cell center
to the boundaries.

However, independently extrapolating the conserved quantities
\begin{equation}
    \vec{U} = \begin{pmatrix}
        \rho \\
        \rho v\\
        \rho e
    \end{pmatrix}
\end{equation}
with $e = e_{th} + \frac{\vec{v}^2}{2}$ risks corrupting the
thermal energy $e_{th}$. If we extrapolate such that $\rho e < \frac{\rho\vec{v}^2}{2}$,
we can even obtain a negative pressure \citep{vanLeer2003}. We therefore extrapolate the 
primitive variables
\begin{equation}
    \begin{gathered}
        \vec{W} = \begin{pmatrix}
        \rho \\
        v \\
        P
    \end{pmatrix} = \begin{pmatrix}
        U_1 \\
        U_2 / U_1 \\
        (\gamma - 1) (U_3 - 0.5 U_2^2 / U_1)
    \end{pmatrix}, \quad
    \vec{U} = \begin{pmatrix}
        \rho \\
        \rho v\\
        \rho e
    \end{pmatrix} = \begin{pmatrix}
        W_1 \\
        W_1 W_2\\
        W_3/(\gamma - 1) + 0.5 W_1 W_2^2
    \end{pmatrix}, \\
    \mat{F} = \begin{pmatrix} W_1 W_2 \\ W_1 W_2^2 + W_3 \\ W_3 W_2 \gamma /(\gamma - 1) + 0.5 W_1 W_2^3 \end{pmatrix}
    \end{gathered}
    \end{equation}

Consider the 1D Euler equations as a function of $\vec{W}$

\begin{equation}
    \partial_t \vec{U}(\vec{W}) + \partial_x \vec{F}(\vec{U}(\vec{W})) = \underbrace{\partial_{\vec{W}} \vec{U}}_{\mat{J}_{\vec{U}}(\vec{W})} \partial_t \vec{W} + \underbrace{\partial_{\vec{W}}\vec{F}}_{\mat{J}_{\vec{F}}(\vec{W})} \partial_x \vec{W} = 0 \rightarrow \partial_t \vec{W} + \underbrace{\mat{J}^{-1}_{\vec{U}}(\vec{W}) \mat{J}_{\vec{F}}(\vec{W})}_{=:\mat{A}} \partial_x \vec{W} = 0,
\end{equation}

with (using the inverse function theorem, so $\mat{J}^{-1}_{\vec{U}}(\vec{W}) = \mat{J}_{\vec{W}}(\vec{U})$)

\begin{equation}
    \begin{gathered}
    \mat{A} = \begin{pmatrix}
        1 & 0 & 0 \\
        -\frac{W_2}{W_1} & \frac{1}{W_1} & 0 \\
        (\gamma - 1)\frac{W_2^2}{2} & -(\gamma - 1)W_2 & (\gamma - 1)
    \end{pmatrix}
    \begin{pmatrix}
        W_2 & W_1 & 0 \\
        W_2^2 & 2 W_1 W_2 & 1 \\
        \frac{W_2^3}{2} & W_3 \frac{\gamma}{\gamma - 1} + \frac{3}{2} W_1 W_2^2 & W_2 \frac{\gamma}{\gamma - 1}
    \end{pmatrix} \\ = 
    \begin{pmatrix}
        W_2 & W_1 & 0 \\
        0 & W_2 & \frac{1}{W_1} \\
        0 & \gamma W_3 & W_2
    \end{pmatrix} = 
    \begin{pmatrix}
        v & \rho & 0 \\
        0 & v & \frac{1}{\rho} \\
        0 & \gamma P & v
    \end{pmatrix}
    \end{gathered}
\end{equation}

where we might use $c_s^2 = \gamma \frac{P}{\rho}$ to write $\gamma P = \rho c_s^2$. With this
we have derived Eq. 12 from \cite{vanLeer2003}, given there without derivation.

With $\mat{A}$ in $\partial_t \vec{W} + \mat{A} \partial_x \vec{W} = 0$ known and approximating
limited spatial gradients with

\begin{equation}
    \delta_x W_j = \text{ave}\left( \frac{W_j - W_{j-1}}{\Delta x}, \frac{W_{j + 1} - W_{j}}{\Delta x}\right),
\end{equation}

with e.g.

\begin{equation}
    \begin{aligned}
        \text{ave}(a,b) &= \operatorname{minmod}(a,b) = 0.5 \cdot (\operatorname{sgn}(a) + \operatorname{sgn}(b)) \cdot \min{(|a|,|b|)} 
        &&\text{(minmod)} \\
        \text{ave}(a,b) &= \begin{cases}\operatorname{minmod}\left(\frac{a+b}{2}, 2 a, 2 b\right), & a b>0, \\ 0, & a b \leq 0 ;\end{cases}
        &&\text{(double minmod)}\\
        \text{ave}(a,b) &= 
        \begin{cases}
            \text{minmod}(\text{maxmod}(a,b),\text{minmod}(2a,2b)), & ab>0 \\
            0, & ab \leq 0
        \end{cases}
        &&\text{(superbee)}
    \end{aligned}
\end{equation}

predict the boundary values as

\begin{equation}
    \begin{aligned}
        \vec{W}_{j-\frac{1}{2}R} &= \vec{W}_j - \frac{1}{2} \Delta x \delta_x W_j \\
        \vec{W}_{j+\frac{1}{2}L} &=\vec{W}_j + \frac{1}{2} \Delta x \delta_x W_j,
    \end{aligned}
\end{equation}

where \textit{R} stands for right of the boundary (between cell $j-1$ and cell $j$)
and \textit{L} for left of the boundary (between cell $j$ and $j+1$). $\vec{U}_{j-\frac{1}{2}R}$
and $\vec{U}_{j+\frac{1}{2}L}$ are the corresponding states in terms of the conserved quantities. In a MUSCL scheme
we would also make half a step in time via $\vec{\tilde{W}}_j = \vec{W}_j - \frac{\Delta t}{2} \mat{A} \delta_x W_j$.

Higher-order schemes have been developed (see e.g. \cite{shu99}) but the second-order
Godunov scheme can still be considered a solid, well-tested choice for many
applications \citep{weno_vs_godunov}.

\subsection{Riemann solvers}

The Riemann problem is to determine the subsequent evolution of two
piecewise-constant states meeting at a plane.

We calculate $\vec{F}^*_{i+\frac{1}{2}}(\vec{{U}}_{j+\frac{1}{2}L},\vec{{U}}_{j+\frac{1}{2}R})$
via approximate Riemann solvers.

These approximate Riemann solvers build upon estimates 
of the maximum wave speed to the left and right 
at the interface, e.g.

\begin{equation}
    \begin{aligned}
        S_R &= \max{(u_L + c_L, u_R + c_R)} \\
        S_L &= \min{(u_L - c_L, u_R - c_R)},
    \end{aligned}
\end{equation}
or the Einfeldt estimate \citep{einfeldt88}
\begin{equation}
    \begin{aligned}
        S_R &= \max{(u_R + c_R, \hat{u} + \hat{c})} \\
        S_L &= \min{(u_L - c_L, \hat{u} - \hat{c})},
    \end{aligned}
\end{equation}
with Roe averages of the sound speed
\begin{equation}
    \hat{c}^2=\frac{c_L^2 \cdot \sqrt{\rho_L}+c_R^2 \cdot \sqrt{\rho_R}}{\sqrt{\rho_L}+\sqrt{\rho_R}}+\frac{\gamma - 1}{2} \frac{\sqrt{\rho_L} \sqrt{\rho_R}}{\left(\sqrt{\rho_L}+\sqrt{\rho_R}\right)^2}\left(u_R-u_L\right)^2,
\end{equation}
and the velocity
\begin{equation}
    \hat{u} = \frac{\sqrt{\rho_L}u_L + \sqrt{\rho_R}u_R}{\sqrt{\rho_L} + \sqrt{\rho_R}}.
\end{equation}

\paragraph{HLL solver.} The Harten-Lax-van Leer (HLL) Riemann solver
is given by

\begin{equation}
    \begin{gathered}
        F^{HLL} = \frac{S_R^+F_L - S_L^- F_R + S_L^- S_R^+ (U_R - U_L)}{S_R^+ - S_L^-} \\
        S_R^+ = \max{(S_R,0)}, \quad S_L^- = \min{(S_L, 0)}.
    \end{gathered}
\end{equation}

We can only consider all intercell interfaces independent if with the maximum absolute
wave speed, i.e. $S_{max} = \max{(|S_R|,|S_L|)}$, one cannot cross a cell, so
$\Delta t = C_{CFL} \min_{\text{all cells}}\Delta x / S_{max} < \min_{\text{all cells}}\Delta x / S_{max}$ (CFL criterion,
e.g. with a CFL-number $C_{CFL} = 0.8$).

\paragraph{HLLC solver.} In the HLLC scheme, an additional middle wave is considered.
Following the central formulation from \cite{hllc_lm},
the HLLC scheme is given by:

\begin{itemize}
    \item calculate the contact wave speed as \citep{batten}
    \begin{equation}
        S_*=\frac{p_R-p_L+\rho_L u_L\left(S_L-u_L\right)-\rho_R u_R\left(S_R-u_R\right)}{\rho_L\left(S_L-u_L\right)-\rho_R\left(S_R-u_R\right)}.
    \end{equation}
    \item calculate intermediate states as (here for the 1D case)
    \begin{equation}
        \vec{U}_{* K} = \frac{S_K-u_K}{S_K-S_*} \begin{pmatrix}
            \rho_K \\
            \rho_K S_* \\
            E_K + (S_* - u_K)(\rho_K S_* + \frac{P_K}{S_K - u_K})
        \end{pmatrix}
    \end{equation}
    \item calculate the HLLC flux as
    \begin{equation}
        \vec{F}^{H L L C}= \begin{cases}\vec{F}_L & \text { if } S_L \geq 0 \\ \vec{F}_R & \text { if } S_R \leq 0 \\ \vec{F}_* & S_L < 0 \, \text{and} \, S_R > 0 \end{cases}
    \end{equation}
    with
    \begin{equation}
        \vec{F}_*=\frac{1}{2}\left(\vec{F}_L+\vec{F}_R\right) + \frac{1}{2}\left[S_L\left(\vec{U}_{* L}-\vec{U}_L\right)+\left|S_*\right|\left(\vec{U}_{* L}-\vec{U}_{* R}\right)+S_R\left(\vec{U}_{* R}-\vec{U}_R\right)\right]
        \label{eq:hllc_flux}
    \end{equation}
\end{itemize}

\paragraph{HLLC-AM scheme.} The HLLC scheme shows a numerical instability
on strong shocks in multidimensional settings \citep{baumgart24}, such
that improved variants have been developed \citep{hllc_am,hllc_lm}. One of them,
the AM-HLLC scheme, is given by blending a low-Mach corrected HLLC flux with
the standard HLLC flux inside numerical shock layers and additionally damping
the momentum dissipation in the low-Mach regime. It combines three ingredients.

First, following \cite{hllc_lm}, the nonlinear signal speeds entering the
dissipation term of \eqref{eq:hllc_flux} are reduced in the low-Mach regime to
rebalance the vanishing advection dissipation against the dominant acoustic
dissipation,
\begin{equation}
    S_L^{LM} = \phi\, S_L, \qquad S_R^{LM} = \phi\, S_R, \qquad
    \phi = \sin\!\left(\min\!\left(1, \frac{Ma_\text{local}}{Ma_\text{limit}}\right)\frac{\pi}{2}\right),
\end{equation}
with $Ma_\text{local} = \max\!\left(\left|u_L/c_L\right|, \left|u_R/c_R\right|\right)$ and
$Ma_\text{limit} = 0.1$. The reduced speeds replace $S_{L}, S_{R}$ only in the
final dissipation term, while the contact speed $S_*$ and the intermediate
states $\vec{U}_{*L/R}$ keep the original $S_{L}, S_{R}$. This yields the HLLC-LM
flux $\vec{F}^{HLLC\text{-}LM}$, which has the same form as \eqref{eq:hllc_flux}.

Second, because reducing the signal speeds everywhere spoils the resolution of
contact and rarefaction waves, the HLLC-LM flux is only applied inside numerical
shock layers and blended with the standard HLLC flux via a velocity-divergence
shock detector \citep{hllc_am},
\begin{equation}
    \vec{F}^{Hybrid} = g\, \vec{F}^{HLLC\text{-}LM} + (1-g)\, \vec{F}^{HLLC}, \qquad
    g = \begin{cases}
        1 & \text{if } \nabla\cdot\vec{u} < -C_{th}\, c / d, \\
        0 & \text{otherwise,}
    \end{cases}
\end{equation}
where $C_{th} = 0.05$, $c$ and $\nabla\cdot\vec{u}$ are the sound speed and the
(central-difference) velocity divergence evaluated at the cell center, and
$d = \max(\Delta x, \Delta y, \Delta z)$ is the mesh size. The detector marks
strongly compressive cells ($g=1$), so that the speed reduction acts only where
it is needed to stabilize the shock.

Third, to recover the correct incompressible limit, the excess dissipation in
the momentum equations is suppressed at low Mach number. Writing the dissipation
term of \eqref{eq:hllc_flux} as $\vec{D} = \vec{F}_* - \tfrac{1}{2}(\vec{F}_L + \vec{F}_R)$,
its mass and energy components are kept unchanged while each momentum component
is scaled by
\begin{equation}
    f = \min(1, \widetilde{Ma}), \qquad
    \widetilde{Ma} = \max\!\left(\frac{\lVert\vec{u}_L\rVert}{c_L}, \frac{\lVert\vec{u}_R\rVert}{c_R}\right),
    \quad \lVert\vec{u}_K\rVert = \sqrt{u_K^2 + v_K^2 + w_K^2}.
\end{equation}
Note that this fix uses the full velocity magnitude $\lVert\vec{u}\rVert$, whereas
the signal-speed reduction above uses only the interface-normal velocity $u$.

The AM-HLLC flux is obtained by applying this low-Mach dissipation fix to both
$\vec{F}^{HLLC}$ and $\vec{F}^{HLLC\text{-}LM}$ and inserting the result into the
hybrid blend, giving a single flux that is shock-stable at high Mach number and
accurate down to the incompressible limit \citep{hllc_am}.

\subsection{A simple operator-splitting scheme for MHD}

We draw our implementation from \cite{pang24}, who

\begin{itemize}
  \item split the MHD system into a magnetic and hydrodynamic subsystem (operator splitting)
  \item solve the magnetic part using a constrained transport fixed-point iteration ensuring $\vec{\nabla} \cdot \vec{B} = 0$
  \item introduce a provably positivity-preserving hydrodynamics scheme
\end{itemize}

The provably positivity-preserving part of \cite{pang24} is also implemented
in \texttt{astronomix} but relies on a Lax--Friedrichs Riemann solver which is
too diffusive for many practical applications, such that we do not present the rather technical
details of this method here.

The first step in the \cite{pang24} scheme is to split the MHD system into a magnetic and
hydrodynamic system. Different from their approach, here we motivate the split based
on the primitive form of the MHD equations (their reasoning is generally more formal).

Let us start with the MHD equations in primitive form (see Equation Set 2. in \cite{clarke15})

\begin{equation}
  \begin{aligned}
    \frac{\partial \rho}{\partial t} & =-\nabla \cdot(\rho \vec{v}) \\
    \frac{\partial \vec{v}}{\partial t} & = -(\vec{v} \cdot \nabla) \vec{v} - \frac{1}{\rho} \nabla p +\frac{1}{\rho}(\nabla \times \vec{B}) \times \vec{B} \\
    \frac{\partial p}{\partial t} & =- \vec{v} \cdot \nabla p -\gamma p \nabla \cdot \vec{v} \\
    \frac{\partial \vec{B}}{\partial t} & = \vec{\nabla} \times (\underbrace{\vec{v} \times \vec{B}}_{=-\vec{E}}).
  \end{aligned}
\end{equation}

Here $\vec{E}$ denotes the electric field (see \cite[Sec. 5]{clarke15} or \cite{antonsen19} for the derivation of this relation
for an ideal MHD fluid with zero viscosity and resistivity). In this primitive form, the magnetic
field only directly couples to the velocity as we would expect from the Lorentz force.

Let us now split this system into a hydrodynamic part (which we can turn back into the conservative form)

\begin{equation}
  \text{System A} \quad \begin{cases}
    \frac{\partial \rho}{\partial t} & =-\nabla \cdot(\rho \vec{v}) \\
    \frac{\partial \vec{v}}{\partial t} & = -(\vec{v} \cdot \nabla) \vec{v} - \frac{1}{\rho} \nabla p\\
    \frac{\partial p}{\partial t} & =- \vec{v} \cdot \nabla p -\gamma p \nabla \cdot \vec{v} \\
    \frac{\partial \vec{B}}{\partial t} & = 0
  \end{cases} \quad \Longleftrightarrow \quad \begin{cases}
    \partial_t \rho + \vec{\nabla} \cdot (\rho \vec{v}) = 0 \\
    \partial_t (\rho \vec{v}) + \vec{\nabla} \cdot (\rho \vec{v}\vec{v}^T + P\mat{1}) = 0 \\
    \partial_t (\rho e) + \vec{\nabla} \cdot \left[(\rho e + P)\vec{v} \right] = 0 \\
    \frac{\partial \vec{B}}{\partial t} = 0
  \end{cases} 
\end{equation}

and a magnetic part

\begin{equation}
  \text{System B} \quad \begin{cases}
    \frac{\partial \rho}{\partial t} & = 0 \\
    \frac{\partial \vec{v}}{\partial t} & = \frac{1}{\rho}(\nabla \times \vec{B}) \times \vec{B} \\
    \frac{\partial p}{\partial t} & = 0 \\
    \frac{\partial \vec{B}}{\partial t} & = \vec{\nabla} \times (\vec{v} \times \vec{B})
  \end{cases}
\end{equation}

Where for an MHD time-step we apply a second-order operator splitting
approach

\begin{equation}
  \vec{U}^{n+1} = S_A^{\frac{\Delta t}{2}} \circ S_B^{\Delta t} \circ S_A^{\frac{\Delta t}{2}} \vec{U}^n.
\end{equation}

We solve the hydrodynamical system A as before and for system B we use finite differencing and a fixed-point iteration.

On a grid in three dimensions, the curl of a field $\vec{A} \in \mathbb{R}^3$ is
discretized as

\begin{equation}
\vec{\nabla} \times \vec{A}_{i,j,k}
:=
\begin{pmatrix}
\frac{A^z_{i,j+1,k} - A^z_{i,j-1,k}}{2 \Delta y} - \frac{A^y_{i,j,k+1} - A^y_{i,j,k-1}}{2 \Delta z} \\
\frac{A^x_{i,j,k+1} - A^x_{i,j,k-1}}{2 \Delta z} - \frac{A^z_{i+1,j,k} - A^z_{i-1,j,k}}{2 \Delta x} \\
\frac{A^y_{i+1,j,k} - A^y_{i-1,j,k}}{2 \Delta x} - \frac{A^x_{i,j+1,k} - A^x_{i,j-1,k}}{2 \Delta y}
\end{pmatrix}
\end{equation}

and on a grid in two dimensions as (the field is still $\vec{A} \in \mathbb{R}^3$)

\begin{equation}
\vec{\nabla} \times \vec{A}_{i,j}
:=
\begin{pmatrix}
    \frac{A^z_{i,j+1} - A^z_{i,j-1}}{2 \Delta y} \\
  -\frac{A^z_{i+1,j} - A^z_{i-1,j}}{2 \Delta x} \\
  \frac{A^y_{i+1,j} - A^y_{i-1,j}}{2 \Delta x} - \frac{A^x_{i,j+1} - A^x_{i,j-1}}{2 \Delta y}
\end{pmatrix}
\end{equation}

The central difference divergence in 3D is given as

\begin{equation}
\vec{\nabla} \cdot \vec{A}_{i,j,k}
:=
\frac{A^x_{i+1,j,k} - A^x_{i-1,j,k}}{2 \Delta x} + \frac{A^y_{i,j+1,k} - A^y_{i,j-1,k}}{2 \Delta y} + \frac{A^z_{i,j,k+1} - A^z_{i,j,k-1}}{2 \Delta z},
\end{equation}

and in 2D as

\begin{equation}
\vec{\nabla} \cdot \vec{A}_{i,j}
:=
\frac{A^x_{i+1,j} - A^x_{i-1,j}}{2 \Delta x} + \frac{A^y_{i,j+1} - A^y_{i,j-1}}{2 \Delta y}.
\end{equation}

\bluebox{
Note that the discrete divergence of a discrete curl as defined above still vanishes. For the 3D field on the 2D grid,
this follows from
\begin{equation}
  \begin{gathered}
    \vec{\nabla} \cdot (\vec{\nabla} \times \vec{A}_{i,j}) = \frac{1}{2 \Delta x}\left( \frac{A^z_{i+1,j+1} - A^z_{i+1,j-1}}{2 \Delta y} - \frac{A^z_{i-1,j+1} - A^z_{i-1,j-1}}{2 \Delta y} \right) \\ + \frac{1}{2 \Delta y}\left( -\frac{A^z_{i+1,j+1} - A^z_{i-1,j+1}}{2 \Delta x} + \frac{A^z_{i+1,j-1} - A^z_{i-1,j-1}}{2 \Delta x} \right) \\
    = \frac{1}{4 \Delta y \Delta x}\big(A^z_{i+1,j+1} - A^z_{i+1,j+1} + A^z_{i+1,j-1} - A^z_{i+1,j-1} \\ + A^z_{i-1,j+1} - A^z_{i-1,j+1} + A^z_{i-1,j-1} - A^z_{i-1,j-1} \big) \\
    = 0.
  \end{gathered}
\end{equation}
}

We perform a time-step of System B given by

\begin{equation}
  \partial_t \begin{pmatrix}
    \vec{v} \\
    \vec{B}
  \end{pmatrix} = \begin{pmatrix}
    \frac{1}{\rho}(\nabla \times \vec{B}) \times \vec{B} \\
    \vec{\nabla} \times (\vec{v} \times \vec{B})
  \end{pmatrix} =: \vec{\Psi}\left(\begin{pmatrix}
    \vec{v} \\
    \vec{B}
  \end{pmatrix}\right)
\end{equation}

using an implicit midpoint method

\begin{equation}
  \vec{R}^{(n+1)} = \vec{R}^{(n)} + \Delta t \, \vec{\Psi}\left(\frac{1}{2}\left( \vec{R}^{(n)} + \vec{R}^{(n + 1)} \right) \right),
\end{equation}

with

\begin{equation}
  \vec{R} = \begin{pmatrix}
    \vec{v} \\
    \vec{B}
  \end{pmatrix}
\end{equation}

We solve this implicit equation using a fixed-point iteration

\begin{equation}
  \begin{gathered}
  \vec{R}^{(0)} = \vec{R}^{(n)}, \\
  \vec{R}^{(k+1)} = \vec{R}^{(n)} + \Delta t\, \vec{\Psi}\left( \frac{\vec{R}^{(n)} + \vec{R}^{(k)}}{2} \right), \quad k = 0, 1, \dots, \\
  \text{until } \max\left\{ \left\| \vec{B}^{(k+1)} - \vec{B}^{(k)} \right\|_{\infty}, \left\| \vec{v}^{(k+1)} - \vec{v}^{(k)} \right\|_{\infty} \right\} < \varepsilon_{\mathrm{tol}},
  \end{gathered}
\end{equation}
where the tolerance is chosen in accordance with the working precision:
$\varepsilon_{\mathrm{tol}} = 10^{-10}$ in double precision (as assumed here) and
$10^{-5}$ in single precision, where $10^{-10}$ would be below the round-off floor and
thus unreachable. The iteration is in any case capped at a maximum number of sweeps.

This iteration usually converges in a few ($5$ to $9$) iterations \citep{pang24} and thus is relatively inexpensive. Here we
leave extensive performance profiling for future work. Note that all our updates to the magnetic field are numerical curls of vector fields;
we therefore never add divergence to the magnetic field and retain $\vec{\nabla} \cdot \vec{B} = 0$ by design.

\subsection{Standard settings for the finite volume scheme}

The standard settings for the finite volume scheme are the HLL Riemann solver,
Minmod limiter and the RK2 time integrator from Sec. \ref{app:rk_integrators}.

\section{Self-gravity}

\subsection{Time derivative of the gravitational energy}
\label{app:gravitational_energy}

Given the density $\rho$, the gravitational potential is given by

\begin{equation}
    \Phi(\vec{x}) = \int K(\vec{x}, \vec{x}')\, \rho(\vec{x}')\, \mathrm{d}V',
\end{equation}

where the kernel is symmetric, only depending on the separation, 
$K(\vec{x}, \vec{x}') = K\!\left(|\vec{x} - \vec{x}'|\right) = K(\vec{x}', \vec{x})$.\footnote{
The explicit kernels are
\begin{equation}
    K_{1D}(r) = 2 \pi G r, \quad K_{2D}(r) = 2G \log r, \quad K_{3D}(r) = -\frac{G}{r}.
\end{equation}
}

The gravitational energy is given by

\begin{equation}
    E_\mathrm{grav} = \frac{1}{2} \int \rho \Phi \, \dif V = \frac{1}{2} \int \int K(\vec{x}, \vec{x}')\, \rho(\vec{x}) \, \rho(\vec{x}')\, \mathrm{d}V' \, \dif V,
\end{equation}

and its time derivative by

\begin{equation}
    \begin{aligned}
        \partial_t E_\mathrm{grav} &= \frac{1}{2} \int \int K(\vec{x}, \vec{x}')\, \dot{\rho}(\vec{x}) \, \rho(\vec{x}')\, \mathrm{d}V' \, \dif V + \frac{1}{2} \int \int K(\vec{x}, \vec{x}')\, \rho(\vec{x}) \, \dot{\rho}(\vec{x}')\, \mathrm{d}V' \, \dif V \\
                                   &= \frac{1}{2} \int \int K(\vec{x}, \vec{x}')\, \dot{\rho}(\vec{x}) \, \rho(\vec{x}')\, \mathrm{d}V' \, \dif V + \frac{1}{2} \int \int K(\vec{x}', \vec{x})\, \rho(\vec{x}') \, \dot{\rho}(\vec{x}) \, \mathrm{d}V' \, \dif V \\
                                   &= \int \int K(\vec{x}, \vec{x}')\, \dot{\rho}(\vec{x}) \, \rho(\vec{x}')\, \mathrm{d}V' \, \dif V \\
                                   &= \int \dot{\rho} \Phi \, \dif V,
    \end{aligned}
\end{equation}

where we used the symmetry of $K$. The same holds discretely, where $\Phi_i = \sum_j V_j K_{ij} \rho_j$ with $K_{ij} = K_{ji}$.

\subsection{Jeans Linear Waves}
\label{app:jeans}

We would like to find a self-gravity test problem with an analytical solution.

\subsubsection{Linearization of the Euler equations with self-gravity}

The idea is to consider an initial state with small perturbations in the density 
and velocity fields such that the Euler equations linearize (i.e. we obtain a linear PDE)
and simplify. We also assume that these perturbations remain sufficiently small on the
timescales we consider.

\greenbox{\textbf{Linearization of PDEs:} To linearize a PDE we split the state
into a background $B$ and a perturbation $P$ and only keep terms in the 
equations up to linear order in the perturbations.}

Consider

\begin{equation}
    \begin{aligned}
        \rho &= \rho_B + \rho_P, \quad &\rho_P \ll \rho_B \\
        \vec{v} &= \vec{v}_P, \quad &\vec{v}_P^2 \ll e_{th}, ||\vec{v}_P|| \ll c_s \\
        P &= P_B + P_P, \quad &P_P \ll P_B,
    \end{aligned}
\end{equation}

where $\rho_B$ is a constant background density, $P_B$ a constant 
background pressure and $c_s$ is a constant. All background fields
are constant throughout space and time.

We initialize the pressure with

\begin{equation}
    P_B = \frac{c_s^2 \rho_B}{\gamma}, \quad P_{P,0} = c_s^2 \rho_{P,0}.
\end{equation}

Let us now simplify the Euler equations with self-gravity. Let us start with the
potential equation

\begin{equation}
    \Phi = \Phi_B + \Phi_P,
\end{equation}

such that (with $\bar{\rho} \approx \rho_B$)

\begin{equation}
    \nabla^2 \Phi_P = 4 \pi G \rho_P, \quad \vec{g} = \vec{g}_P = -\vec{\nabla} \Phi_P,
\end{equation}

where there is no background acceleration due to the Jeans swindle, $\vec{g}_B = 0$, such
that $\vec{g} = \vec{g}_P$ is a first-order perturbation. The Jeans swindle is applicable here,
as the background medium can be assumed static on the
timescales of the perturbations we consider.

Next, consider the continuity equation

\begin{equation}
\begin{aligned}
    0 &= \partial_t \rho + \vec{\nabla} \cdot (\rho \vec{v}) \\
      &= \partial_t \rho_P + \vec{v}_P \cdot \vec{\nabla} \rho_P + \left(\rho_B + \rho_P \right) \vec{\nabla} \cdot \vec{v}_P \\
      &= \partial_t \rho_P + \rho_B \vec{\nabla} \cdot \vec{v}_P,
\end{aligned}
\end{equation}

where we ignored all terms higher than first order in the perturbations (e.g. $\vec{v}_P \cdot \vec{\nabla} \rho_P$).

We continue with the momentum equation and make simplifications in the same spirit as before.
By $\vec{v}_P^2 \ll e_{th}$ we can ignore the nonlinear term.

\begin{equation}
    \rho_B \partial_t \vec{v}_P + \vec{\nabla} P_P = \rho_B \vec{g}_P.
\end{equation}

Finally, we consider the energy equation. By $\vec{v}_P^2 \ll e_{th}$

\begin{equation}
    e = e_{th} + \frac{1}{2} \vec{v}^2_P = e_{th},
\end{equation}

and therefore

\begin{equation}
    \rho e = \frac{P}{\gamma - 1}, \quad \rho e + P = \frac{\gamma}{\gamma - 1} P.
\end{equation}

Additionally, we neglect $\vec{v}_P \cdot \vec{g}_P$ as it is a second-order term.

We obtain (applying simplifications of the same kind as before)

\begin{equation}
    \begin{aligned}
        0 &= \partial_t P_P + \gamma P_B \vec{\nabla} \cdot \vec{v}_P \\
          &= \partial_t P_P + c_s^2 \rho_{B} \vec{\nabla} \cdot \vec{v}_P \\
          &= \partial_t P_P - c_s^2 \partial_t \rho_P,
    \end{aligned}
\end{equation}

where in the last step we substituted in $\rho_B \vec{\nabla} \cdot \vec{v}_P = -\partial_t \rho_P$ from the
continuity equation. Integration with initial condition $P_{P,0} = c_s^2 \rho_{P,0}$ yields

\begin{equation}
    P_P = c_s^2 \rho_P.
\end{equation}

We therefore obtain the linearized Euler equations

\begin{equation}
    \begin{gathered}
        \partial_t \rho_P + \rho_B \vec{\nabla} \cdot \vec{v}_P = 0  \\
        \rho_B \partial_t \vec{v}_P + c_s^2 \vec{\nabla} \rho_P = \rho_B \vec{g}_P \\ \\
        \nabla^2 \Phi_P = 4 \pi G \rho_P, \quad \vec{g}_P = -\vec{\nabla} \Phi_P.
    \end{gathered}
    \label{eq:euler_grav_jeans}
\end{equation}

\subsubsection{Solving the linearized Euler equations with self-gravity}

Note that in the following we are only interested in one particular solution
we can use to test our simulator.

Start by eliminating the velocity from the linearized equations

\begin{equation}
    \begin{aligned}
        0 &= \partial_t^2 \rho_P + \rho_B \vec{\nabla} \cdot \partial_t \vec{v}_P \\
          &= \partial_t^2 \rho_P + \vec{\nabla} \cdot \left( \rho_B \vec{g}_P -  c_s^2 \vec{\nabla} \rho_P \right) \\
          &= \partial_t^2 \rho_P + \rho_B \vec{\nabla} \cdot \vec{g}_P - c_s^2 \nabla^2 \rho_P \\
          &= \partial_t^2 \rho_P - 4 \pi G \rho_B \rho_P  - c_s^2 \nabla^2 \rho_P.
    \end{aligned}
\end{equation}

We now make a plane wave Ansatz

\begin{equation}
    \rho_P = A_\rho \exp{\left( i\vec{k} \cdot \vec{x} - i \omega t\right)}
\end{equation}

(with $A_\rho \in \mathbb{R}$) which yields the dispersion relation

\begin{equation}
    \omega^2 = k^2 c_s^2 - 4 \pi G \rho_B,
\end{equation}

and we can construct a solution using the positive root

\begin{equation}
    \omega = \sqrt{k^2 c_s^2 - 4 \pi G \rho_B}.
\end{equation}

(we assume $k^2 c_s^2 > 4 \pi G \rho_B$ otherwise we are in an unstable regime).

Taking the same plane wave Ansatz for the velocity is consistent

\begin{equation}
    \vec{v}_P = \vec{A}_v \exp{\left( i\vec{k} \cdot \vec{x} - i \omega t\right)},
\end{equation}

such that

\begin{equation}
    \omega A_\rho - \rho_B \vec{k} \cdot \vec{A}_v = 0,
\end{equation}

which is solved by (assuming longitudinal waves, $\vec{A}_v \parallel \vec{k}$)

\begin{equation}
    \vec{A}_v = \frac{\omega A_\rho}{\rho_B k^2} \vec{k}.
\end{equation}

Finally, we use $P_P = c_s^2 \rho_P$ to obtain the pressure.

We choose $A_\rho = \epsilon \rho_B$ with $\epsilon \ll 1$.

A possible solution to the linearized Euler equations with self-gravity reads
(taking the imaginary part of the Ansatz)

\begin{equation}
    \begin{aligned}
        \rho &= \rho_B + \rho_B \epsilon \sin{\left( \vec{k} \cdot \vec{x} - \omega t\right)} \\
        \vec{v} &= \frac{\epsilon \omega \vec{k}}{k^2} \sin{\left(\vec{k} \cdot \vec{x} - \omega t\right)} \\
        P &= \frac{c_s^2 \rho_B}{\gamma} + c_s^2 \rho_B \epsilon \sin{\left( \vec{k} \cdot \vec{x} - \omega t\right)},
    \end{aligned}
\end{equation}

where for $\rho_B = 1, c_s^2 = 1, \gamma = \frac{5}{3}, 4\pi G = 1, \epsilon = 10^{-6}$ and
$\vec{k} = (2,4,4)^T$ we obtain the setup of \citet{hanawa25}. For the box dimensions
we have to be careful to ensure periodicity of our waves. The wavelength is given by

\begin{equation}
    \lambda = \frac{2\pi}{k},
\end{equation}

and in all dimensions the box size should be a clean multiple of the wavelength.

\section{Handling of boundary conditions}
\label{app:boundaries}

Boundary conditions can be handled using \textit{ghost cells}, extra cell layers with
which we pad the computational representation of the physical domain. For periodic boundaries
it is also very natural to use periodic array operations
in the form of \texttt{jnp.roll} throughout the code which will
automatically result in periodic boundaries.

If we are using ghost cells, the number of ghost cells necessary on each side
of the computational domain depends on the stencil size for the update of
an individual cell. For instance, in a second-order Godunov
scheme the gradient reconstruction depends on a cell's left and right neighbor
and the flux to a cell depends on the direct neighbors' fluid states and reconstructed
gradients - therefore two neighbors to the left and right influence a cell's update, and
we need a layer of two ghost cells around the computational domain.

Before each time step, the ghost cells have to be set depending on the boundary
conditions. Consider a one-dimensional domain with regular cells $1,...,N$ and 
fluid states $\vec{U}_1,...,\vec{U}_N$.

We will illustrate how the ghost cells are set for two ghost cells at each side of the domain.

\begin{itemize}
    \item For \textbf{periodic boundary conditions} and two ghost cells, the array of fluid states
    from left to right reads
    \begin{equation}
        \underbrace{\vec{U}_{N-1}, \vec{U}_{N}}_{\text{ghost cells}}, \vec{U}_1,...,\vec{U}_N, \underbrace{\vec{U}_1, \vec{U}_2}_{\text{ghost cells}}.
    \end{equation}
    \item For \textbf{open boundaries} and two ghost cells, the array of fluid states
    from left to right reads
    \begin{equation}
        \underbrace{\vec{U}_{1}, \vec{U}_{1}}_{\text{ghost cells}}, \vec{U}_1,...,\vec{U}_N, \underbrace{\vec{U}_N, \vec{U}_N}_{\text{ghost cells}}.
    \end{equation}
    \item For \textbf{reflective boundaries} and two ghost cells, the array of fluid states
    from left to right reads
    \begin{equation}
        \underbrace{\vec{V}_{-}(\vec{U}_{2}), \vec{V}_{-}(\vec{U}_{1})}_{\text{ghost cells}}, \vec{U}_1,...,\vec{U}_N, \underbrace{\vec{V}_{-}(\vec{U}_N), \vec{V}_{-}(\vec{U}_{N-1})}_{\text{ghost cells}},
    \end{equation}
    where the $V_{-}$ operator inverts the sign of the velocity
\end{itemize}

The generalization to more dimensions and more ghost cells is straightforward.

\section{Gradients}

Before we get to gradients we need to refresh some basics on linear spaces
and differentiability, here following \cite{blondel22}. We limit ourselves
to spaces over the real numbers.

\subsection{Linear spaces}

\begin{definition}[Linear space]
    A linear space $\mathcal{E}$ is a space with operations $+,\cdot$
    such that for any $\vec{u},\vec{v} \in \mathcal{E}$ and $a \in \mathbb{R}$,
    we have $\vec{u} + \vec{v} \in \mathcal{E}$ and $a \cdot \vec{u} \in \mathcal{E}$,
    as well as a number of other vector axioms, see 
    \url{https://en.wikipedia.org/wiki/Vector_space}.
\end{definition}

\begin{definition}[Functional]
    A function from a vector space to its base field, here $\mathcal{E} \to \mathbb{R}$ is called a functional.
\end{definition}

\begin{definition}[Dual space]
    For a vector space $\mathcal{E}$, the dual space $\mathcal{E}^*$ is the 
    vector space of linear functionals on $\mathcal{E}$ with addition and scaling
    defined pointwise.
\end{definition}

\begin{definition}[Inner product]
    An inner product on a linear space $\mathcal{E}$ is a function 
    $\langle \cdot, \cdot \rangle: \mathcal{E} \times \mathcal{E} \to \mathbb{R}$ that is
    \begin{itemize}
        \item bilinear: $\vec{v} \mapsto \langle \vec{v}, \vec{w} \rangle$ and $\vec{w} \mapsto \langle \vec{v}, \vec{w} \rangle$ 
        are linear for any $\vec{v}, \vec{w} \in \mathcal{E}$
        \item symmetric: $\langle \vec{v}, \vec{w} \rangle = \langle \vec{w}, \vec{v} \rangle$ for any 
        $\vec{w},\vec{v} \in \mathcal{E}$
        \item positive definite: $\langle \vec{w}, \vec{w} \rangle \geq 0$ for
        any $\vec{w} \in \mathcal{E}$ and $\langle \vec{w}, \vec{w} \rangle = 0$
        if and only if $\vec{w} = 0$.
    \end{itemize}
    An inner product defines a norm $\left \| \vec{w} \right \| := \sqrt{\langle \vec{w}, \vec{w} \rangle}$.
\end{definition}

\begin{definition}[Euclidean space]
    A linear space equipped with an inner product and a basis $\vec{e}_1,...,\vec{e}_p \in \mathcal{E}$
    such that any element $\vec{v} \in \mathcal{E}$ can be written as 
    $\vec{v} = \sum_{i=1}^{p} v_i \vec{e}_i$ with unique scalars $v_1,...,v_p \in \mathbb{R}$
    is called a \textbf{Euclidean space}.
\end{definition}

\begin{definition}[Hilbert space]
    A Hilbert space is a linear space equipped with an inner product complete with
    respect to the norm induced by the inner product.
\end{definition}

Examples of Hilbert spaces are

\begin{itemize}
    \item the Euclidean space $\mathbb{R}^P$ with $\langle \vec{w}, \vec{v} \rangle = \sum_{i=1}^{P} w_i v_i$
    \item the space of real square integrable functions $L^2(\mathbb{R})$ with
    \begin{equation}
        \langle f, g \rangle = \int_{-\infty}^\infty f(x) g(x) \, dx.
    \end{equation}
\end{itemize}

To define gradients later on, we also need \textbf{Riesz representation theorem}.

\begin{theorem}[Riesz representation theorem]
    Consider a Hilbert space $H$. For every continuous linear functional 
    $\phi \in H^*$ there exists a unique vector $f_\phi \in H$,
    called Riesz representation of $\phi$, such that
    \begin{equation}
        \phi(x) = \langle x, f_\phi \rangle \quad \forall x \in H.
    \end{equation}
\end{theorem}

For a proof and details see \url{https://en.wikipedia.org/wiki/Riesz_representation_theorem}.

\subsection{Fréchet differentiability and definition of the gradient}

Consider normed vector spaces $V$ and $W$. Let $U \subseteq V$ be an open subset of
$V$. A function $f: U \to W$ is called Fréchet differentiable at $w \in U$ if 
there exists a bounded linear operator $\partial f(w): V \to W$, called Fréchet derivative,
such that

\begin{equation}
  \lim _{\|v\|_V \rightarrow 0} \frac{\|f(w+v)-f(w)-\partial f(w)[v]\|_W}{\|v\|_V}=0.
\end{equation}

In Landau notation

\begin{equation}
    f(w+v) = f(w) + \partial f(w)[v] + o(\left \| v \right \|).
\end{equation}

If $V$ is a Hilbert space and $f: V \to \mathbb{R}$,
then $\partial f(w) \in V^*$ admits a Riesz representation $\nabla f(w) \in V$
called the \textbf{gradient}, such that

\begin{equation}
    \partial f(w)[v] = \langle \nabla f(w), v \rangle_V.
\end{equation}

\subsubsection{Example: Gradient of a quadratic form}
\label{sec:quad_form_grad}

Consider a bounded linear operator $S: H \to H$ on a Hilbert space $H$ and

\begin{equation}
    J(w) = \frac{1}{2} \langle Sw, Sw \rangle.
\end{equation}

Let us expand for any $v \in H$

\begin{equation}
    \begin{aligned}
        J(w+v) &= \frac{1}{2} \langle Sw + Sv, Sw + Sv \rangle \\
               &= \frac{1}{2} \langle Sw, Sw \rangle + \langle Sw, Sv \rangle + \frac{1}{2} \langle Sv, Sv \rangle \\
               &= J(w) + \langle Sw, Sv \rangle + O(\left \| v \right \|^2),
    \end{aligned}
\end{equation}

such that we can identify

\begin{equation}
    \partial J(w)[v] = \langle Sw, Sv \rangle = \langle S^* Sw, v \rangle,
\end{equation}

and based on the Riesz representation

\begin{equation}
    \partial J(w)[v] = \langle \nabla J(w), v \rangle,
\end{equation}

we can identify

\begin{equation}
    \nabla J(w) = S^* Sw.
\end{equation}

\subsubsection{Gradients of functionals in Fourier space}
\label{sec:grad_fourier}

Consider a functional $f(w)$ where $w$ is a square-integrable
function $w \in V = L^2$. Let $\hat{w}$ be the Fourier transform
of $w$

\begin{equation}
    \hat{w} = \mathcal{F}(w),
\end{equation}

and we define a functional $g$ such that

\begin{equation}
    g(\hat{w}) = f(w).
\end{equation}

Defining $\hat{v} = \mathcal{F}(v)$, we can equivalently write the
directional derivative as

\begin{equation}
    \partial f(w) [v] = \partial g(\hat{w}) [\hat{v}],
\end{equation}

i.e. in terms of the Riesz representation

\begin{equation}
    \langle \nabla_{\hat{w}} g(\hat{w}), \hat{v} \rangle_{\hat{V}} = \langle \nabla f(w), v \rangle_V.
\end{equation}

Assuming a unitary Fourier transform, Plancherel's theorem states

\begin{equation}
    \langle u, v \rangle_{V} = \langle \mathcal{F}(u), \mathcal{F}(v) \rangle_{\hat{V}},
\end{equation}

therefore

\begin{equation}
    \langle \nabla_{\hat{w}} g(\hat{w}), \hat{v} \rangle_{\hat{V}} = \langle \mathcal{F}(\nabla f(w)), \mathcal{F}(v) \rangle_{\hat{V}} = \langle \mathcal{F}(\nabla f(w)), \hat{v} \rangle_{\hat{V}},
\end{equation}

and $\nabla_{\hat{w}} g(\hat{w}) = \mathcal{F}(\nabla f(w))$ and

\begin{equation}
    \boxed{
        \nabla f(w) = \mathcal{F}^{-1}(\nabla_{\hat{w}} g(\hat{w}))
    }
\end{equation}

i.e. we can calculate a functional derivative by calculating the corresponding functional derivative in Fourier space
and applying the inverse Fourier transform.

\subsection{Analytical gradients for the linearized Euler equations}

\subsubsection{The Euler equations}

The Euler equations are given by
\begin{equation}
    \begin{gathered}
        \partial_t \rho + \vec{\nabla} \cdot (\rho \vec{v}) = 0  \\
        \partial_t (\rho \vec{v}) + \vec{\nabla} \cdot (\rho \vec{v}\vec{v}^\top + P\mat{1}) = 0 \\
        \partial_t (\rho e) + \vec{\nabla} \cdot \left[(\rho e + P)\vec{v} \right] = 0,
    \end{gathered}
    \label{eq:euler}
\end{equation}

with $e = e_{th} + \frac{1}{2} \vec{v}^2$ with closure $P = (\gamma - 1) \rho e_{th}$.

\subsubsection{Linearization of the Euler equations}

The idea is to consider an initial state with small perturbations in the density 
and velocity fields such that the Euler equations linearize (i.e. we obtain a linear PDE)
and simplify. We also assume that these perturbations remain sufficiently small on the
timescales we consider.

\greenbox{\textbf{Linearization of PDEs:} To linearize a PDE we split the state
into a background $B$ and a perturbation $P$ and only keep terms in the 
equations up to linear order in the perturbations. This is a standard
perturbation expansion.}

Consider

\begin{equation}
    \begin{aligned}
        \rho &= \rho_B + \rho_P, \quad &\rho_P \ll \rho_B \\
        \vec{v} &= \vec{v}_P, \quad &\vec{v}_P^2 \ll e_{th}, \left \| \vec{v}_P \right \| \ll c_s \\
        P &= P_B + P_P, \quad &P_P \ll P_B,
    \end{aligned}
\end{equation}

where $\rho_B$ is a constant background density, $P_B$ a constant 
background pressure, $c_s$ is a constant and the background velocity is zero. 
All background fields are constant throughout space and time.

We initialize the pressure with

\begin{equation}
    P_B = \frac{c_s^2 \rho_B}{\gamma}, \quad P_{P,0} = c_s^2 \rho_{P,0}.
\end{equation}

Let us start with the continuity equation

\begin{equation}
\begin{aligned}
    0 &= \partial_t \rho + \vec{\nabla} \cdot (\rho \vec{v}) \\
      &= \partial_t \rho_P + \vec{v}_P \cdot \vec{\nabla} \rho_P + \left(\rho_B + \rho_P \right) \vec{\nabla} \cdot \vec{v}_P \\
      &= \partial_t \rho_P + \rho_B \vec{\nabla} \cdot \vec{v}_P,
\end{aligned}
\end{equation}

where we ignored all terms higher than first order in the perturbations (e.g. $\vec{v}_P \cdot \vec{\nabla} \rho_P$).

We continue with the momentum equation and make simplifications in the same spirit as before.
By $\vec{v}_P^2 \ll e_{th}$ we can ignore the nonlinear term.

\begin{equation}
    \rho_B \partial_t \vec{v}_P + \vec{\nabla} P_P = 0.
\end{equation}

Finally, we consider the energy equation. By $\vec{v}_P^2 \ll e_{th}$

\begin{equation}
    e = e_{th} + \frac{1}{2} \vec{v}^2_P = e_{th},
\end{equation}

and therefore

\begin{equation}
    \rho e = \frac{P}{\gamma - 1}, \quad \rho e + P = \frac{\gamma}{\gamma - 1} P.
\end{equation}

We obtain (applying simplifications of the same kind as before)

\begin{equation}
    \begin{aligned}
        0 &= \partial_t P_P + \gamma P_B \vec{\nabla} \cdot \vec{v}_P \\
          &= \partial_t P_P + c_s^2 \rho_{B} \vec{\nabla} \cdot \vec{v}_P \\
          &= \partial_t P_P - c_s^2 \partial_t \rho_P,
    \end{aligned}
\end{equation}

where in the last step we substituted in $\rho_B \vec{\nabla} \cdot \vec{v}_P = -\partial_t \rho_P$ from the
continuity equation. Integration with initial condition $P_{P,0} = c_s^2 \rho_{P,0}$ yields

\begin{equation}
    P_P = c_s^2 \rho_P.
\end{equation}

We therefore obtain the linearized Euler equations

\begin{equation}
    \begin{gathered}
        \partial_t \rho_P + \rho_B \vec{\nabla} \cdot \vec{v}_P = 0 \\
        \partial_t \vec{v}_P + \frac{c_s^2}{\rho_B} \vec{\nabla} \rho_P = 0.
    \end{gathered}
    \label{eq:euler_lin}
\end{equation}

For brevity, let us work in 1D from now on. We can write

\begin{equation}
    \partial_t \vec{U} + \mat{A} \partial_x \vec{U} = 0, \quad \mat{A} = \begin{pmatrix}
        0 & \rho_B \\
        \frac{c_s^2}{\rho_B} & 0
    \end{pmatrix}, \quad \vec{U} = \begin{pmatrix}
        \rho_P \\
        v_P
    \end{pmatrix}
\end{equation}

\subsubsection{Solution to the linearized Euler equations}

We diagonalize $\mat{A}$

\begin{equation}
    \mat{A} = \mat{R} \mat{\Lambda} \mat{R}^{-1}, \quad \mat{\Lambda} = \begin{pmatrix}
        \lambda_1 & 0 \\
        0 & \lambda_2
    \end{pmatrix}
\end{equation}

and define the characteristic variables

\begin{equation}
    \vec{W} = \mat{R}^{-1} \vec{U}
\end{equation}

to obtain the decoupled advection equation

\begin{equation}
    \partial_t \vec{W} + \mat{\Lambda} \partial_x \vec{W} = 0,
\end{equation}

with solution

\begin{equation}
    W_i(x, t) = W_i(x - \lambda_i t, 0).
\end{equation}

In our case

\begin{equation}
    \mat{\Lambda} = \begin{pmatrix}
        c_s & 0 \\
        0 & -c_s
    \end{pmatrix}, \quad
    \mat{R} = \begin{pmatrix}
        \rho_B & \rho_B \\
        c_s & -c_s 
    \end{pmatrix}, \quad
    \mat{R}^{-1} = \frac{1}{2} \begin{pmatrix}
        \frac{1}{\rho_B} & \frac{1}{c_s} \\
        \frac{1}{\rho_B} & -\frac{1}{c_s}
    \end{pmatrix}
\end{equation}

such that

\begin{equation}
    \vec{W} = \mat{R}^{-1} \vec{U} = \frac{1}{2} \begin{pmatrix}
        \frac{\rho_P}{\rho_B} + \frac{v_P}{c_s} \\
        \frac{\rho_P}{\rho_B} - \frac{v_P}{c_s}
    \end{pmatrix}, \quad 
    \vec{W}(t,x) = \frac{1}{2} \begin{pmatrix}
        \frac{\rho_P(x-c_s t, 0)}{\rho_B} + \frac{v_P(x-c_s t, 0)}{c_s} \\
        \frac{\rho_P(x+c_s t, 0)}{\rho_B} - \frac{v_P(x+c_s t, 0)}{c_s}
    \end{pmatrix}
\end{equation}

so

\begin{equation}
    \begin{aligned}
        \vec{U}(x,t) &= \mat{R} \vec{W}(t,x) \\
        &= \frac{1}{2} \begin{pmatrix}
            \rho_P(x-c_s t, 0) + \rho_P(x+c_s t, 0) +   \frac{\rho_B}{c_s} \left( v_P(x-c_s t, 0)    -   v_P(x+c_s t, 0)   \right) \\
            v_P(x-c_s t, 0) + v_P(x+c_s t, 0)       +   \frac{c_s}{\rho_B} \left( \rho_P(x-c_s t, 0) -   \rho_P(x+c_s t, 0)\right)
        \end{pmatrix} \\
        &=: S_t \vec{U}(x,0),
    \end{aligned}
\end{equation}

where we have defined the operator $S_t$, which one can easily verify to be linear. We will drop
the $P$ subscripts in $\rho_P$ and $v_P$ from now on.

\subsubsection{Analytical gradients for the linearized Euler equations}

Our fluid state $\vec{U}$ is from the
Hilbert space $H = L^2(\mathbb{R}) \times L^2(\mathbb{R})$
with inner product

\begin{equation}
    \langle \vec{U}, \vec{W} \rangle = \int_{-\infty}^{\infty} U_1 W_1 + U_2 W_2 \, dx.
\end{equation}

We are interested in the analytical gradient of $J: H \to \mathbb{R}$

\begin{equation}
    J(\vec{U}_0) = \frac{1}{2} \langle S_t \vec{U}_0, S_t \vec{U}_0 \rangle.
\end{equation}

Based on the previous example, we can state

\begin{equation}
    \partial J(\vec{U}_0)[\delta \vec{U}_0] = \langle S_t \vec{U}_0, S_t \delta \vec{U}_0 \rangle = \langle \vec{U}_t, \delta \vec{U}_t \rangle = \int_{-\infty}^{\infty} \rho(x,t) \delta \rho(x,t) + v(x,t) \delta v(x,t) \, dx,
\end{equation}

where we have defined $\delta \vec{U}_t := S_t \delta \vec{U}_0$ and $\delta \rho(x,t), \delta v(x,t)$ are the 
forwards solutions of the linearized Euler equations starting from $\delta \rho_0, \delta v_0$.

Note that the following derivation might be more compactly achieved in characteristic variables.

We can plug in our analytical results for $\delta \rho(x,t)$ and $\delta v(x,t)$

\begin{equation}
\begin{aligned}
\partial J(\vec{U}_0)[\delta \vec{U}_0]
= {} &
\frac{1}{2} \int_{-\infty}^{\infty} \rho(x,t)\,\delta \rho(x-c_s t, 0)\,dx \\
&+ \frac{1}{2} \int_{-\infty}^{\infty} \rho(x,t)\,\delta \rho(x+c_s t, 0)\,dx \\
&+ \frac{1}{2} \int_{-\infty}^{\infty} \rho(x,t)\,\frac{\rho_B}{c_s}\,\delta v(x-c_s t, 0)\,dx \\
&- \frac{1}{2} \int_{-\infty}^{\infty} \rho(x,t)\,\frac{\rho_B}{c_s}\,\delta v(x+c_s t, 0)\,dx \\
&+ \frac{1}{2} \int_{-\infty}^{\infty} v(x,t)\,\delta v(x-c_s t, 0)\,dx \\
&+ \frac{1}{2} \int_{-\infty}^{\infty} v(x,t)\,\delta v(x+c_s t, 0)\,dx \\
&+ \frac{1}{2} \int_{-\infty}^{\infty} v(x,t)\,\frac{c_s}{\rho_B}\,\delta \rho(x-c_s t, 0)\,dx \\
&- \frac{1}{2} \int_{-\infty}^{\infty} v(x,t)\,\frac{c_s}{\rho_B}\,\delta \rho(x+c_s t, 0)\,dx.
\end{aligned}
\end{equation}

By definition of the gradient

\begin{equation}
    \partial J(\vec{U}_0)[\delta \vec{U}_0] = \langle \vec{\nabla} J, \delta \vec{U}_0 \rangle = \int_{-\infty}^{\infty} \frac{\partial J}{\partial \rho_0} \delta \rho(x,0) + \frac{\partial J}{\partial v_0} \delta v(x,0) \, dx.
\end{equation}

We can apply changes of variables to our previous expressions to bring them to this form, for instance

\begin{equation}
    \frac{1}{2} \int_{-\infty}^{\infty} \rho(x,t)\,\delta \rho(x-c_s t, 0)\,dx = \frac{1}{2} \int_{-\infty}^{\infty} \rho(x + c_s t,t)\,\delta \rho(x, 0)\,dx.
\end{equation}

The full regrouped expression reads

\begin{equation}
\begin{aligned}
\partial J(\vec{U}_0)[\delta \vec{U}_0]
= {} &
\frac{1}{2} \int_{-\infty}^{\infty}
\Big[
\rho(x+c_s t,t) + \rho(x-c_s t,t) \\
&\qquad
+ \frac{c_s}{\rho_B}
\big( v(x+c_s t,t) - v(x-c_s t,t) \big)
\Big]\,
\delta \rho(x,0)\,dx \\
&+
\frac{1}{2} \int_{-\infty}^{\infty}
\Big[
v(x+c_s t,t) + v(x-c_s t,t) \\
&\qquad
+ \frac{\rho_B}{c_s}
\big( \rho(x+c_s t,t) - \rho(x-c_s t,t) \big)
\Big]\,
\delta v(x,0)\,dx.
\end{aligned}
\end{equation}

so that we can identify

\begin{equation}
    \frac{\partial J}{\partial \rho_0} = \frac{1}{2} \left( \rho(x+c_s t,t) +  \rho(x-c_s t,t) + v(x+c_s t,t)\,\frac{c_s}{\rho_B} -  v(x-c_s t,t)\,\frac{c_s}{\rho_B}\right),
\end{equation}

and 

\begin{equation}
    \frac{\partial J}{\partial v_0} = \frac{1}{2} \left( v(x+c_s t,t) + v(x-c_s t,t) + \rho(x+c_s t,t)\,\frac{\rho_B}{c_s} - \rho(x-c_s t,t)\,\frac{\rho_B}{c_s} \right),
\end{equation}

which completes our analytical derivation of the gradient.

\yellowbox{Note that if we replace the $L^2$-based inner product on $H$
by the physical acoustic energy inner product

\begin{equation}
\langle\vec{U}, \vec{V}\rangle_{\mathrm{ac}}=\int \frac{c_s^2}{\rho_B} \rho_U \rho_V+\rho_B v_U v_V d x,
\end{equation}

and the functional $J$ by

\begin{equation}
J_{\mathrm{ac}}\left(\vec{U}_0\right)=\frac{1}{2}\left\|S_t \vec{U}_0\right\|_{\mathrm{ac}}^2.
\end{equation}

$S_t$ is a unitary operator and the gradient collapses to the trivial

\begin{equation}
    \nabla J_{\mathrm{ac}}\left(\vec{U}_0\right) = S_t^* S_t \vec{U}_0 = \vec{U}_0
\end{equation}

(independent of $t$)
as the acoustic energy is conserved.
}

Generally, the continuous adjoint method is used to calculate such gradients.

\subsubsection{Generalization to 3D}

\paragraph{Analytical forward solution}
\label{app:analytical_forward}

As we are dealing with a linear PDE, the superposition principle applies,
and we can solve a general initial value problem in terms of a plane wave
decomposition, i.e. by Fourier transform.

The linearized Euler equations transform to 

\begin{equation}
    \begin{aligned}
        \partial_t \hat{\rho} + i \rho_B \vec{k} \cdot \hat{\vec{v}} &= 0 \\
        \partial_t \hat{\vec{v}} + i \frac{c_s^2}{\rho_B} \vec{k} \hat{\rho} &= 0,
    \end{aligned}
\end{equation}

leading to the wave equations

\begin{equation}
    \begin{aligned}
        \partial_t^2 \hat{\rho} + \omega^2 \hat{\rho} &= 0 \\
        \partial_t^2 \hat{\vec{v}} + \omega^2 \hat{\vec{k}} \, \hat{\vec{k}} \cdot \hat{\vec{v}} &= 0,
    \end{aligned}
\end{equation}

where $\omega = \left \| \vec{k} \right \| c_s$ and $\hat{\vec{k}} = \frac{\vec{k}}{\left \| \vec{k} \right \|}$.

For the density this directly motivates the ansatz

\begin{equation}
    \hat{\rho} = A \cos{(\omega t)} + B \sin{(\omega t)},
\end{equation}

where $A = \hat{\rho}_0$ by our initial conditions.

For the velocity it is useful to split into a component parallel and one perpendicular to $\vec{k}$

\begin{equation}
    \hat{\vec{v}} = \hat{\vec{v}}_\parallel + \hat{\vec{v}}_\perp, \quad \hat{\vec{v}}_\parallel = (\hat{\vec{k}} \cdot \hat{\vec{v}}) \hat{\vec{k}}, \quad \hat{\vec{v}}_\perp = \hat{\vec{v}} - \hat{\vec{v}}_\parallel,
\end{equation}

such that we obtain

\begin{equation}
    \begin{aligned}
        \partial_t \hat{\vec{v}}_\perp &= 0 \\
        \partial_t^2 \hat{\vec{v}}_\parallel + \omega^2 \hat{\vec{v}}_\parallel &= 0,
    \end{aligned}
\end{equation}

thus

\begin{equation}
    \begin{aligned}
        \hat{\vec{v}}_\perp &= \hat{\vec{v}}_{0,\perp} = \hat{\vec{v}}_{0} - (\hat{\vec{k}} \cdot \hat{\vec{v}}_0) \hat{\vec{k}} \\
        \hat{\vec{v}}_\parallel &= C \hat{\vec{k}} \cos{(\omega t)} + D \hat{\vec{k}} \sin{(\omega t)},
    \end{aligned}
\end{equation}

where $C = \hat{\vec{k}} \cdot \hat{\vec{v}}_0$ by our initial conditions.

Next, we solve for $B$ and $D$ by plugging in our ansätze into the original density equation in Fourier space

\begin{equation}
    \begin{aligned}
        (i \rho_B k D -\omega  A) \sin{(\omega t)} + (B \omega + i \rho_B k C) \cos{(\omega t)} &= 0,
    \end{aligned}
\end{equation}

so $B = -i\frac{\rho_B}{c_s} \hat{\vec{k}} \cdot \hat{\vec{v}}_0$ and $D = -i \frac{c_s}{\rho_B} \hat{\rho}_0$.

Therefore, the full solution reads

\begin{equation}
    \begin{aligned}
        \hat{\rho}(\vec{k}, t) &= \hat{\rho}_0(\vec{k}) \cos(\omega t) - i \frac{\rho_B}{c_s} (\hat{\vec{v}}_0(\vec{k}) \cdot \hat{\vec{k}}) \sin(\omega t) \\
        \hat{\vec{v}}(\vec{k}, t) &= \hat{\vec{v}}_0(\vec{k}) - \hat{\vec{k}}(\hat{\vec{v}}_0(\vec{k}) \cdot \hat{\vec{k}}) \Big( 1 - \cos(\omega t) \Big) - i \frac{c_s}{\rho_B} \hat{\rho}_0(\vec{k}) \hat{\vec{k}} \sin(\omega t),
    \end{aligned}
\end{equation}

which we can then Fourier-transform back to real space. We can also write this in the form

\begin{equation}
\begin{pmatrix}
\hat{\rho}(\vec{k}, t) \\
\hat{\vec{v}}(\vec{k}, t)
\end{pmatrix}
=
\underbrace{
\begin{pmatrix}
\cos(\omega t) & -i \frac{\rho_B}{c_s} \sin(\omega t)\,\hat{\vec{k}}^\top \\
- i \frac{c_s}{\rho_B} \sin(\omega t)\,\hat{\vec{k}} & I - (1-\cos(\omega t))\,\hat{\vec{k}}\hat{\vec{k}}^\top
\end{pmatrix}
}_{=: S(t)}
\begin{pmatrix}
\hat{\rho}_0(\vec{k}) \\
\hat{\vec{v}}_0(\vec{k})
\end{pmatrix}
\end{equation}

\paragraph{Analytical gradients}
\label{app:analytical_gradients}

By Sec. \ref{sec:quad_form_grad}, we can calculate

\begin{equation}
    \vec{\nabla} \hat{\vec{U}} = S^* S \hat{\vec{U}}_0,
\end{equation}

where

\begin{equation}
S^* S = \begin{pmatrix}
\cos^2(\omega t) + \left(\frac{c_s}{\rho_B}\right)^2 \sin^2(\omega t) & -i \sin(\omega t) \cos(\omega t) \left( \frac{\rho_B}{c_s} - \frac{c_s}{\rho_B} \right) \hat{\vec{k}}^\top \\
i \sin(\omega t) \cos(\omega t) \left( \frac{\rho_B}{c_s} - \frac{c_s}{\rho_B} \right) \hat{\vec{k}} & I + \left( \left(\frac{\rho_B}{c_s}\right)^2 - 1 \right) \sin^2(\omega t) \hat{\vec{k}}\hat{\vec{k}}^\top
\end{pmatrix}
\end{equation}

so

\begin{equation}
\begin{aligned}
\nabla_{\hat{\rho}_0} J &= \left[ \cos^2(\omega t) + \left(\frac{c_s}{\rho_B}\right)^2 \sin^2(\omega t) \right] \hat{\rho}_0 - i \sin(\omega t) \cos(\omega t) \left( \frac{\rho_B}{c_s} - \frac{c_s}{\rho_B} \right) \left(\hat{\vec{k}} \cdot \hat{\vec{v}}_0\right) \\
\nabla_{\hat{\vec{v}}_0} J &= i \sin(\omega t) \cos(\omega t) \left( \frac{\rho_B}{c_s} - \frac{c_s}{\rho_B} \right) \hat{\rho}_0 \hat{\vec{k}} + \hat{\vec{v}}_0 + \left( \left(\frac{\rho_B}{c_s}\right)^2 - 1 \right) \sin^2(\omega t) \left(\hat{\vec{k}} \cdot \hat{\vec{v}}_0\right) \hat{\vec{k}},
\end{aligned}
\end{equation}

and based on Sec. \ref{sec:grad_fourier}

\begin{equation}
    \vec{\nabla} \vec{U} = \mathcal{F}^{-1}(\vec{\nabla} \hat{\vec{U}}).
\end{equation}

\subsection{Discretized gradients}
\label{app:discretized_gradients}

Consider we want to discretely approximate gradients on $L^2([0,L])$
where $[0,L]$ is our domain of interest.

For simplicity, for a function $f \in L^2([0,L])$ we will consider its
piecewise-constant discretization $\mathfrak{f}$ on $N$ intervals with boundaries
$x_{0},\dots,x_{N}$ where $x_{0} = 0$ and $x_{N} = L$.

\begin{equation}
    \mathfrak{f}(x) = \sum_{i=1}^{N} f_i\, I_{[x_{i-1},x_{i}]}(x), \quad f_i = f\left( \frac{x_{i-1} + x_i}{2} \right), \quad I_{[a, b]}(x) := \begin{cases}
        1, & x \in [a,b] \\
        0, & \text{else}
    \end{cases}
\end{equation}

For two such discretizations $\mathfrak{f}, \mathfrak{g}$, the scalar product discretizes to
\begin{equation}
    \langle \mathfrak{f}, \mathfrak{g} \rangle_{L^2}
    = \int_0^L \mathfrak{f}(x)\,\mathfrak{g}(x)\,dx
    = \sum_{i=1}^{N} \int_{x_{i-1}}^{x_i} f_i g_i \, dx
    = \sum_{i=1}^{N} h_i\, f_i g_i,
    \qquad h_i := x_i - x_{i-1},
\end{equation}
where all cross terms vanish because the indicators $I_{[x_{i-1},x_i]}$ have
(essentially) disjoint supports.

Collecting the coefficients into vectors $\vec{f} = (f_1,\dots,f_N)^\top$ and
$\vec{g} = (g_1,\dots,g_N)^\top \in \mathbb{R}^N$, this is a weighted Euclidean
inner product
\begin{equation}
    \langle \mathfrak{f}, \mathfrak{g} \rangle_{L^2}
    = \vec{f}^\top M\, \vec{g}
    =: \langle \vec{f}, \vec{g} \rangle_M,
\end{equation}
with the diagonal \emph{mass matrix}
\begin{equation}
    M = \operatorname{diag}(h_1, \dots, h_N) \in \mathbb{R}^{N \times N},
    \qquad M_{ij} = h_i \,\delta_{ij}.
\end{equation}
Since all $h_i > 0$, $M$ is symmetric positive definite, so
$\langle \cdot, \cdot \rangle_M$ is a genuine inner product on $\mathbb{R}^N$
that approximates the $L^2$ product. For a higher-order approximation, e.g.
piecewise linear or parabolic instead of piecewise constant, the mass matrix will
generally not be diagonal anymore.

\paragraph{From Euclidean to proper gradient.}

Now consider a loss $J: L^2([0,L]) \to \mathbb{R}$ and
its discretization $\mathcal{J}: \mathbb{R}^N \to \mathbb{R}$
such that for our piecewise constant function $\mathfrak{f}$
with representation $\vec{f} \in \mathbb{R}^N$ we have
$J(\mathfrak{f}) = \mathcal{J}(\vec{f})$.

$\nabla J$ will also be piecewise constant, and we are interested
in its representation $\vec{g}$.

For any piecewise constant function $\mathfrak{h}$ with representation
$\vec{h}$, the directional derivatives agree because
$J(\mathfrak{f}) = \mathcal{J}(\vec{f})$ for all representatives, i.e.
$DJ(\mathfrak{f})[\mathfrak{h}] = D\mathcal{J}(\vec{f})[\vec{h}]$. Hence

\begin{equation}
    \langle \nabla J, \mathfrak{h} \rangle_{L^2}
    = \langle M \vec{g}, \vec{h} \rangle_{\mathrm{Eucl}}
    = \langle \nabla_{\mathrm{Eucl}} \mathcal{J}, \vec{h}\rangle_{\mathrm{Eucl}},
\end{equation}
where the first equality uses the $M$-representation of $\nabla J$ together with
$M = M^\top$, and the second is the directional-derivative identity above. It
follows that

\begin{equation}
    \vec{g} = M^{-1} \nabla_{\mathrm{Eucl}} \mathcal{J},
\end{equation}

where $\nabla_{\mathrm{Eucl}} \mathcal{J}$ is for instance provided by automatic differentiation.

\section{Gradient explosion in single shooting}
\label{app:exploding_gradients}

Chaotic systems are exponentially sensitive to the initial state. But 
while unstable modes can be traced back (they are backwards stable), 
the gradient norm explodes with the time horizon.

Consider a system described by a differential equation, 
$\partial_t u = F(u)$, with solution map $\Phi_t$. We are interested 
in solving a PDE-constrained optimization problem

\begin{equation}
    u_0^* = \argmin_{u_0} L(u_T) \quad \text{where} \quad u_T = \Phi_T(u_0).
\end{equation}

Consider we try to find $u_0^*$ by simple gradient descent on a 
scalar loss $J(u_0)=L(\Phi_T(u_0))$. For a given $u_0$, we denote

\begin{equation}
    A(t) := \partial_{u} F(u=\Phi_t(u_0)).
\end{equation}

Given the terminal value $p(T) = \partial_{u_T} L$ and the adjoint equation

\begin{equation}
    \partial_t p =  -A(t)^T p,
\end{equation}

our gradient of interest is given by integration backward in time

\begin{equation}
    \boxed{\nabla J = p(0) }
\end{equation}

The adjoint equation is a homogeneous 
linear equation, such that

\begin{equation}
    \left\Vert p(0) \right\Vert \sim e^{\lambda_{max}T} \left\Vert p(T) \right \Vert,
\end{equation}

where $\lambda_{max}$ is the largest Lyapunov exponent (given that $p(T)$ has a
non-zero projection onto the corresponding Lyapunov direction).

The nonlinear saturation that keeps the state bounded does not apply to perturbations 
or adjoints; tangent and adjoint variables evolve under linear time-dependent equations 
and can grow exponentially even when the primal trajectory remains bounded.

Thus, while tracing growing modes backwards is a well-posed problem (i.e. a small 
perturbation of the forward unstable mode does not dramatically change the initial 
state one should find), single shooting, i.e. optimization based on the gradient 
with respect to the initial state, might not be possible.

\yellowbox{Note that in our failure classification the gradient magnitude explosion
by itself is a type 1 failure and might in principle be tamed by rescaling or
preconditioning, while the problematic ill-conditioning or nonconvexity of the
reduced problem is retained.}

\end{document}